# Thermal Energy Transport in Oxide Nuclear Fuel

David H. Hurley[1]*, Anter El-Azab[2], Matthew S. Bryan[3], Michael W. D. Cooper[4], Cody A. Dennett[1], Krzysztof Gofryk[1], Lingfeng He[1], Marat Khafizov[5], Gerard H. Lander[6], Michael E. Manley[3], J. Matthew Mann[7], Chris A. Marianetti[8], Karl Rickert[9], Farida A. Selim[10], Michael R. Tonks[11], Janelle P. Wharry[2]

[1] Idaho National Laboratory, 1955 N Fremont Ave, Idaho Falls, ID 83415, USA
[2] School of Materials Engineering, Purdue University, 701 W Stadium Ave, West Lafayette, IN 47907, USA
[3] Materials Science and Technology Division, Oak Ridge National Laboratory, 1 Bethel Valley Road, Oak Ridge, TN 37831, USA
[4] Materials Science and Technology Division, Los Alamos National Laboratory, P.O. Box 1663, Los Alamos, NM 87545, USA
[5] Department of Mechanical and Aerospace Engineering, The Ohio State University, 201 W 19th Ave, Columbus, OH 43210, USA
[6] European Commission, Joint Research Center, Postfach 2340, D-76125 Karlsruhe, Germany
[7] US Air Force Research Laboratory, Sensors Directorate, 2241 Avionics Circle, Wright Patterson AFB, OH 45433, USA
[8] Department of Applied Physics and Applied Mathematics, Columbia University, 500 West 120th Street, New York, New York 10027, USA
[9] KBR, 2601 Mission Point Boulevard, Suite 300, Dayton, OH 45431, USA
[10] Department of Physics and Astronomy, Bowling Green State University, 705 Ridge Street, Bowling Green, OH 43403, USA
[11] Department of Materials Science and Engineering, University of Florida, 158 Rhines Hall, Gainesville, FL 32611, USA

**Abstract**

To efficiently capture the energy of the nuclear bond, advanced nuclear reactor concepts seek solid fuels that must withstand unprecedented temperature and radiation extremes. In these advanced fuels, thermal energy transport under irradiation is directly related to reactor performance as well as reactor safety. The science of thermal transport in nuclear fuel is a grand challenge due to both computational and experimental complexities. Here, we provide a comprehensive review of thermal transport research on two actinide oxides: one currently in use in commercial nuclear reactors, uranium dioxide ($UO_2$), and one advanced fuel candidate material, thorium dioxide ($ThO_2$). In both materials, heat is carried by lattice waves or phonons. Crystalline defects caused by fission events effectively scatter phonons and lead to a degradation in fuel performance over time. Bolstered by new computational and experimental tools, researchers are now developing the foundational work necessary to accurately model and ultimately control thermal transport in advanced nuclear fuel. We begin by reviewing research aimed at understanding thermal transport in perfect single crystals. The absence of defects enables studies that focus on the fundamental aspects of phonon transport. Next, we review research that targets defect generation and evolution. Here, the focus is on ion irradiation studies used as surrogates for damage caused by fission products. We end this review with a discussion of modeling and experimental efforts directed at predicting and validating mesoscale thermal transport in the presence of irradiation defects. While efforts into these research areas have been robust, challenging work remains in developing holistic tools to capture and predict thermal energy transport across widely varying environmental conditions.

Table of Contents

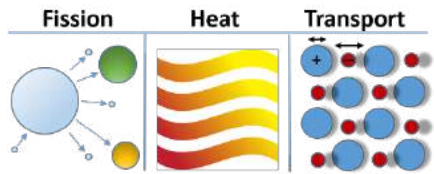

## 1 Introduction

Currently, about 73 percent of the electricity in the world is generated using heat engines [1]. This includes heat generated by burning fossil fuels (coal, gas, oil), by concentrated solar energy, and by nuclear fission. On the surface, the anatomy of a heat engine power plant looks similar regardless of the heat source. Heat extracted by passing a coolant over the heat source is used to create steam, which is then converted to electricity using a steam turbine. For a fossil fuel-powered plant, the heat source (i.e. exhaust gas) moves, allowing additional heat to be extracted by increasing the length of the heat exchanger. A nuclear plant is fundamentally different in this regard; heat is generated by nuclear fission in a crystalline solid and must be *conducted* through the fuel and transferred to the coolant, as shown in Figure 1. This perspective illustrates the close connection between conduction of thermal energy (heat) and electricity generation in nuclear power plants.

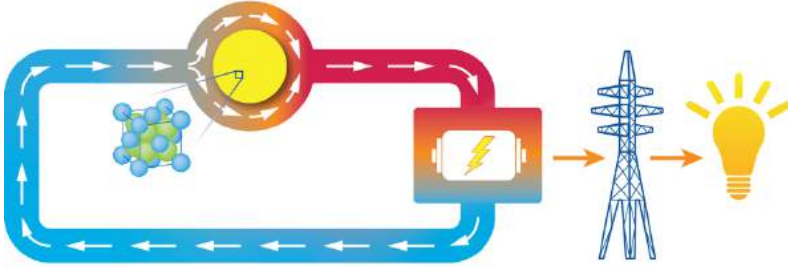

**Figure 1.** Schematic of a nuclear energy plant. The heat extracted from the solid fuel is used to drive a steam turbine and generate electricity.

Nuclear reactors operating in the world today primarily use actinide oxides (uranium dioxide, $UO_2$, and mixed oxides containing $UO_2$ and plutonium dioxide, $PuO_2$) as the fuel material. These oxide fuels have a large energy density (a single commercial reactor fuel pellet weighing about 10 g stores the same amount of energy as a 1.4 tons of coal (~1300 kg), 285 gallons of oil (~1 m$^3$), or 38,800 cubic feet of natural gas (~150 m$^3$) [2]) and electricity generation does not involve the release of greenhouse gases. In nuclear fuel, a fissile nuclide absorbs a neutron, becomes unstable, and splits, creating at least two new atoms (referred to as fission fragments), producing gamma rays and neutrons, and imparting about 200 MeV of kinetic energy into those fission fragments (as illustrated in Figure 2) [3]. The transfer of this kinetic energy to the crystalline lattice of the fuel generates thermal energy that must be conducted through the fuel and transferred to the coolant for eventual conversion to electricity. The thermal energy is conducted through the fuel almost entirely by lattice waves or phonons, due to the scarcity of free electrons in oxide fuels [4]. The rapid transfer of the kinetic energy from the fission event to the lattice also produces defects (vacancies, interstitials and extended defects) with fission fragments depositing as impurity atoms. These defects effectively scatter phonons, reducing the capacity of the fuel to conduct heat. Over the lifetime of the fuel in a reactor, the thermal conductivity decreases by as much as 70 percent [5].

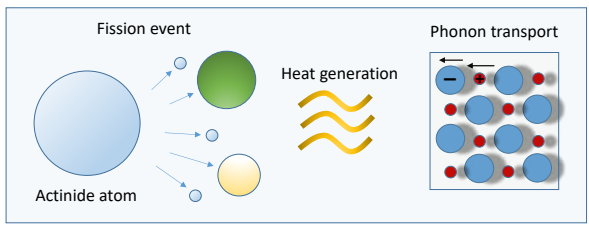

**Figure 2.** Nuclear fission is used to generate heat. Fission fragments create damage and deposit as chemical impurities. Heat is carried by phonons in oxide nuclear fuels.

However, in some cases, a cooperative influence between multiple defect types has a positive effect by increasing the lattice thermal conductivity over that expected if the defects acted in isolation. For example, in the high burnup structure (HBS) of nuclear fuels, where considerable irradiation-driven restructuring greatly increases the grain boundary surface area [6], and decreases the lattice defect concentration due to segregation to grain boundaries [7], the thermal conductivity is larger than in similar fuel that has not formed HBS [8, 9]. The myriad of defect types and interactions in nuclear fuel under irradiation has motivated researchers to consider improving function by controlling defect evolution and structure. Examples include using dopants to control defect concentrations by modifying grain size [10, 11, 12, 13], simulating HBS using nanocrystalline $UO_2$ to improve fission gas retention and improve mechanical properties [14, 15], and directly adding other materials to $UO_2$ to raise the thermal conductivity [16, 17, 18, 19, 20, 21, 22, 23, 24]. Expediting progress towards tailoring defects and microstructure to bring about desired thermal properties will require further development of fundamental, predictive models of phonon transport in actinide oxide fuels containing irradiation-induced defects [25, 26, 27, 28, 29].

This desire to understand and tailor the thermal properties of fuels to improve the efficiency, safety, and reliability of electricity generation has driven a great deal of fundamental research over the past 60 years. Our current understanding of thermal transport in actinide oxides can be traced back to early work involving phonon transport in semiconductors and insulators. This work, using analytical expressions to represent conductivity, provided a mechanistic understanding of phonon thermal transport in the presence of defects [30, 31, 32, 33]. More recently, a first-principles theoretical approach has produced accurate results for simple semiconductors with no adjustable parameters [34, 35, 36]. Similar work using the Green's function T-matrix approach has been used to calculate phonon scattering rates for simple lattice defects from first principles [37, 38, 39]. On the experimental side, this effort has included phonon structure measurements using bright neutron and X-ray sources and spatially resolved thermal conductivity measurements [40, 41, 42]. Extending these approaches to actinide oxides containing irradiation-induced defects remains a grand challenge due to several factors. These include challenges associated with fabricating high-quality single crystal samples for phonon structure measurements, complexities encountered in accurately treating electron correlation in 5$f$ electron systems, seeding samples with specific defect populations found in reactor fuel, accurately characterizing the spectrum of defects, and measuring thermal transport on length scales commensurate with defect accumulation.

Here, we review the current understanding of phonon thermal transport in oxide nuclear fuel, highlighting the above-mentioned challenges. Specifically, we provide a comprehensive review of previous work on two actinide oxides: one currently in use in commercial nuclear reactors, $UO_2$, and one advanced fuel candidate material, thorium dioxide ($ThO_2$). $ThO_2$ additionally acts as a surrogate material for $UO_2$ as it lacks the complexity of electron correlation effects. We also provide a narrower review of other oxide materials relevant to $UO_2$ and $ThO_2$ research. Our focus is primarily on defects produced by ion irradiation as ions provide a convenient and expedient approach to inject defects that are found in real fuel specimens. Emphasis is placed on identifying remaining knowledge gaps and revealing new opportunities to improve thermal transport in reactor fuel.

## 1.1 Origins of Phonon Thermal Conductivity

In nuclear fuel, thermal transport can be viewed from both a macroscopic and an atomistic viewpoint. On the macroscopic scale, thermal transport is treated within a continuum framework to obtain the temperature distribution in a fuel pellet. On an atomistic scale, thermal transport is treated within a quantum mechanical framework to obtain an expression for the thermal conductivity, which controls the temperature distribution. Phonon thermal conductivity is most often calculated using one of two methods: (1) Molecular dynamics simulation using empirical interatomic potentials (EIP) or potentials derived from first principles (FPIP); (2) Seeking solutions to the linearized Boltzmann transport equation for phonons (see Figure 3).

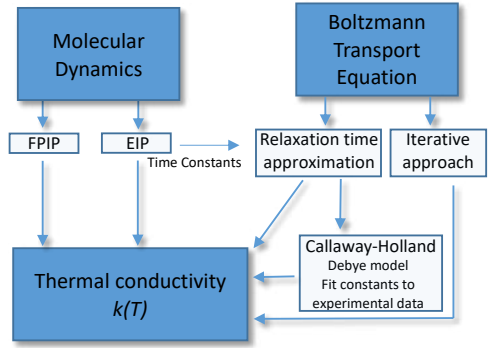

**Figure 3.** Different ways to calculate thermal conductivity.

While this review will cover studies using both methods, this section provides an overview of the Boltzmann framework as this best illustrates the motivation for the organization of this review. In crystals, vibrational energy is exchanged between phonon modes. In thermal equilibrium, the average number of modes with energy $\hbar\omega$ is given by the Bose-Einstein relation:

$$N(\omega) = \frac{1}{exp(\hbar\omega/k_B T)-1}. \quad (1)$$

The number of modes of a particular energy increases with increasing temperature. Another way of stating this is that phonons are created by increasing the temperature, and destroyed by decreasing the temperature (i.e. the number of phonons is not conserved). The strong temperature dependence of thermal conductivity in insulators and semi-conductors results partially from this lack of conservation. A thermal gradient imposed by an external heat source will force the system away from equilibrium. The thermal conductivity depends on the extent to which the phonon distribution can deviate from equilibrium for a specific thermal gradient.

The time-space evolution of $N(\omega)$ is governed by the Boltzmann transport equation (BTE) for phonons. This equation is in general difficult to solve. The two solution methods that will be discussed in this review are the exact iterative solution [43, 44] and the relaxation time approximation (RTA) [45]. The iterative solution to the BTE is well suited to calculate thermal conductivity of perfect single crystals but remains too computationally expensive to model crystals with nontrivial defects. For the RTA, lattice dynamics is not required to compute relaxation times as they can also be calculated using molecular dynamics. It is this aspect that makes the RTA well suited for treating thermal transport in crystals containing a range of defects from point defects to extended defect clusters.

Under the RTA, the conductivity can be represented as:

$$\kappa = \frac{1}{3}\sum_{qs} C_{qs} v_{qs}^2 \tau_{qs}, \quad (2)$$

where $s$ enumerates the number of possible phonon modes for a given phonon wave vector $q$. The right-hand side of this equation contains the branch-specific phonon lifetime ($\tau_{qs}$), phonon velocity ($v_{qs}$), and specific heat ($C_{qs}$), with the latter two being a function of only phonon dispersion (the relation between frequency and wavevector, $\omega(q)$). The lifetime is determined by phonon scattering, which ultimately limits thermal conductivity. Above cryogenic temperatures, in nearly perfect single crystal materials, phonon scattering is dominated by scattering with other phonons, mediated by lattice anharmonicity (3rd and higher order terms in the interatomic potential).

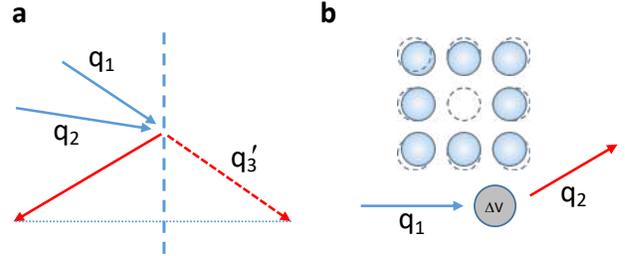

**Figure 4.** Phonon scattering: (a) anharmonic 3-phonon Umklapp scattering, does not conserve total momentum and plays an outsized role in limiting conductivity and (b) phonon defect scattering.

Structural defects caused by irradiation also influence the lifetime by introducing phonon scattering sites. Both scattering processes are illustrated in Figure 4. Using Matthiessen's rule, the total phonon lifetime can be written as:

$$\tau_{qs}^{-1}(\omega) = \tau_{anharmonic}^{-1} + \tau_{defect\ 1}^{-1} + \tau_{defect\ 2}^{-1} + \cdots, \quad (3)$$

to account for scattering from multiple mechanisms, both intrinsic (anharmonic) and extrinsic (defects). This Boltzmann transport framework provides a template that is used to organize the material presented in this review:

Section 2: Thermal transport in perfect single crystals

This section addresses the input to Equation 2 in terms of phonon dispersion and lifetime, discussing previous work aimed at understanding thermal transport in perfect single crystals.

Section 3: Defect generation evolution and characterization

This section discusses the body of work involving the generation and evolution of defects in actinide oxides using energetic ion irradiation.

Section 4: Thermal conductivity under irradiation

This section reports on the output of Equation 2, reviewing modeling and experimental efforts directed at predicting and validating thermal conductivity in the presence of defects.

## 2 Thermal transport in perfect single crystals

The absence of defects enables studies that focus on the fundamental aspects of phonon transport including the impact of strong electron correlation. Experimentally, prior efforts ranging from synthesis of single crystals to inelastic neutron and X-ray scattering are discussed. Computationally, previous work aimed at the application of advanced electronic structure calculations to systems that exhibit strong electron correlation is reviewed. In keeping with the scope delineated in the Introduction, the emphasis of this section and the sections to follow is focused on providing a review of key discoveries, highlighting fundamental mechanisms, and identifying remaining knowledge gaps.

### 2.1 Inelastic neutron and X-ray scattering

#### 2.1.1 Phonon dispersion and thermal conductivity

Measurements of the phonon dispersion curves in a material offer a unique window into the microscopic mechanisms controlling phonon thermal transport (see Figure 5). These dispersion curves relate phonon wave vector, **q**, to frequency, $\omega$, through the interatomic force constants. In principle, thermal conductivity can be represented using the measured dynamical structure factor [46]. The mode partial phonon density of states, $g(\omega)$ is evaluated by differentiating the phonon branches [47] and can be used to evaluate the heat capacity and vibrational entropy per mode [48]. The slope of a measured dispersion curve determines the phonon group velocities, $v_g$, and the linewidth of the curve is inversely related to the phonon lifetime. As illustrated in Figure 5b, for a typical acoustic phonon, the partial density of states available to carry heat becomes very small for states near the zone center (**q**=0). Consequently, even though the velocities (see Figure 5c) and lifetimes are large in this region, the contributions to the thermal conductivity can still be small. At high temperatures, with full thermal occupation ($k_B T >> \hbar\omega_{max}$), the heat capacity term in Equation 2 follows the density of states, and the contributions to thermal conductivity are more broadly distributed over the curve (Figure 5d), peaking in importance near the middle of the curve. Similar arguments can be made for dispersive optical phonons. For this reason, understanding the thermal conductivity at high temperatures requires measurements of the full set of dispersion curves, including linewidths, and this requires neutron or X-ray scattering techniques at large user facilities.

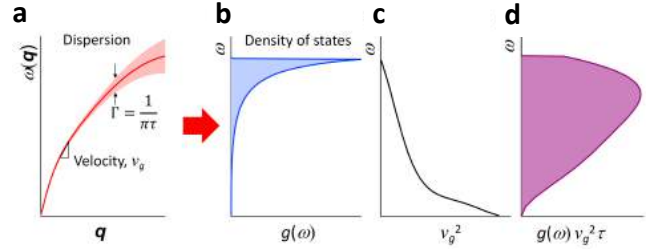

**Figure 5.** Relationship between phonon dispersion and thermal conductivity. (a) Phonon dispersion curve for a typical acoustic mode. The slope provides the phonon group velocity, $v_g$; the linewidth, $\Gamma$, is inversely related to the phonon lifetime, $\tau$, as indicated. (b) The mode partial phonon density of states scales as $g(\omega) \propto q^2(\omega)(dq/d\omega)$. (c) Phonon group velocity squared, $v_g^2 = (d\omega/dq)^2$. (d) Product of the density of states, group velocity squared, and lifetimes give a distribution for the contributions to the thermal conductivity tensor assuming full thermal occupation ($k_B T >> \hbar\omega_{max}$), where each mode contributes equally to the heat capacity. At low temperatures the higher frequency phonons thermally depopulate and no longer contribute as much heat.

#### 2.1.2 Techniques for measuring phonons

The earliest measurements of the phonon dispersion curves of actinide oxides $UO_2$ [49] and $ThO_2$ [50] were performed using the reactor-based triple-axis neutron scattering technique (Figure 6a) that was first established for measuring phonon dispersion curves by Brockhouse [51, 52]. However, the early work on actinide oxides primarily focused on determining the phonon dispersion curves or density of states and left out the phonon lifetimes, which are more difficult to measure but crucial to understanding the thermal conductivity. Additionally, because of large neutron absorption cross-sections and crystal size requirements (>0.2 cm$^3$), certain actinide oxides (e.g. plutonium dioxide) remained impractical to measure with neutrons.

The situation changed with the advent of high-energy resolution synchrotron-based inelastic X-ray scattering (IXS) spectrometers [53, 54, 55, 56] (Figure 6b), which have no isotope requirements and crystal sizes can be as small as a few micrometers [57]. This allows the measurements of individual crystal grains within a polycrystal sample. The first measured phonon dispersion curves for plutonium metal [58] and the first phonon density of states of $PuO_2$ [59] were both obtained using IXS rather than neutron scattering. IXS also opened up measurements of phonons at high pressures and in thin films using grazing incidence. Actinides are particularly well suited to IXS measurements in small geometries since the X-ray scattering per atom increases as the square of the atomic number. However, there are a limited number of facilities available for these measurements and, like triple-axis neutron scattering, it takes considerable beamtime to map a full set of phonon dispersion curves. A full set of phonon dispersion curves takes about two weeks to map, which may require multiple experimental cycles. An example is the recent phonon dispersion curves for $NpO_2$ taken from a very small crystal with X-rays at the European Synchrotron Radiation Facility (ESRF) [60].

Another important advance in measuring phonons in actinide oxides came with the wide-angle range time-of-flight spectrometers at next-generation spallation neutron sources such as the Spallation Neutron Source (SNS) at Oak Ridge National Laboratory [61], which came online in the mid to late 2000s (Figure 6c). These instruments and sources enable measurements of higher energy modes than are normally accessible with reactor-based triple-axis measurements and provide a sampling of a large volume of energy-momentum space in a single measurement – allowing the full mapping of the dispersion curves. These two advantages have proven useful in the discovery of new high energy dynamical features [62]. These measurements, however, also require large samples.

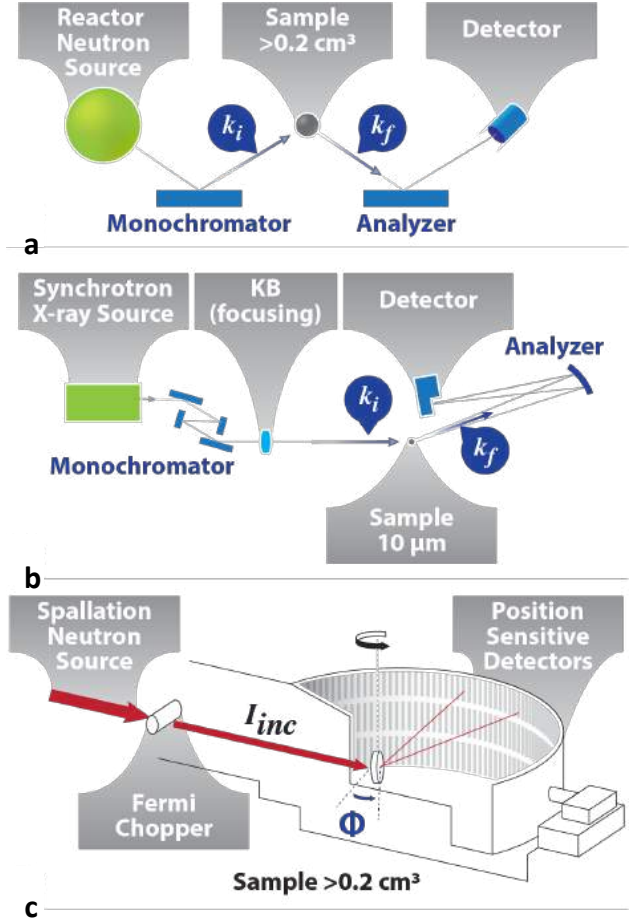

**Figure 6.** Important scattering techniques for measuring phonon dispersion curves in actinides. (a) Triple-axis neutron scattering spectrometer at a reactor source. A monochromator fixes the incident neutron wavevector, $k_i$, and an analyzer crystal selects the final wavevector, $k_f$, from which momentum transfer, $Q = k_i - k_f$, and energy transfer, $\hbar\omega = \hbar^2(k_i^2 - k_f^2)/2m$, can be selected to measure phonon dispersion curves. (b) Inelastic X-ray scattering at a synchrotron. Similar to a triple-axis instrument in operation but with a more selective monochromator to extract ~1 meV resolution from a ~20 keV beam. (c) Time-of-flight direct geometry inelastic neutron scattering spectrometer. Incident energy is selected using a chopper and final energies and wavevectors are determined by the time of flight of the scattered neutrons arriving at a wide-angle detector bank and allows for a broad sampling of energy-momentum space.

### 2.1.3 Triple-axis neutron measurements of the phonon lifetimes in UO$_2$

Building on the pioneering work of Dolling *et al.* [49], Pang *et al.* [27, 63] have shown using triple-axis inelastic neutron scattering (Figure 7) that the principal contribution to the thermal conductivity of UO$_2$ comes from the acoustic phonons together with a steep and relatively low energy longitudinal optic phonon, LO$_1$ (see Figure 7a).

In most crystalline insulators the majority of thermal heat transport is carried by the acoustic phonons, since they typically have the largest group velocities and the longest lifetimes. However, the triple-axis measurements [27] indicate that the steep longitudinal optic mode carries the largest amount of heat by branch at 295 K in UO$_2$ (see Figure 7b), which was not expected. It was also not predicted in previous simulations [64] because they underestimated the group velocities and lifetimes of the LO$_1$ phonon compared to experiment [27], rendering it less important in the simulated thermal conductivity. The steeply dispersing LO$_1$ phonon is a general feature of the fluorite crystal structure and is thus likely important in the thermal conductivity of all of the actinide oxides considered herein.

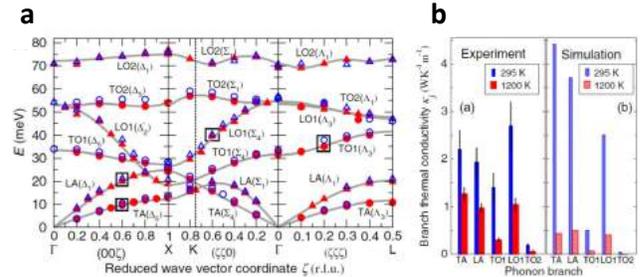

**Figure 7.** Phonon dispersion curves and branch thermal conductivity of UO$_2$, after Pang *et al.* [27]. (a) Phonon dispersion curves measured using triple-axis neutron scattering at 295 K (blue) and 1200 K (red). (b) Phonon branch thermal conductivity derived from the measured phonon dispersion and lifetimes ('Experiment') and from calculations ('Simulation'). Adapted with permission from [27]. Copyright 2013 American Physical Society.

### 2.1.4 Inelastic X-ray measurements of phonons in irradiated epitaxial thin films of UO$_2$

The focus of this section thus far has been on neutron scattering measurements on perfect or nearly perfect single crystals. However, as mentioned previously, IXS can provide phonon structure information on thin films using grazing incidence methods. Here we extend our focus momentarily to consider IXS on ion irradiated UO$_2$ thin films, as this gives a phonon-level description on the impact of irradiation. An experiment by Rennie *et al.* [65] involved epitaxial films (thickness 300 nm) of UO$_2$ that were irradiated with 2.1 MeV He$^{2+}$ ions to 0.15 dpa, resulting in homogeneous damage and a lattice swelling of $\Delta a/a \sim 0.6\%$. The film was then examined by grazing-incidence X-ray inelastic scattering using synchrotron radiation (Figure 6b) to measure whether the acoustic phonons showed any change under these conditions.

The results showed that the frequency of the phonons was not changed, but their widths Γ (Figure 5a) were substantially increased. Although it is difficult to measure optical phonons with this technique, the results with acoustic phonons are consistent with the drop in the thermal conductivity for this level of radiation damage [66]. These experiments confirm that the effects of irradiation in UO$_2$ are intrinsic to the microscopic structural changes caused by the damage, and result in shorter lifetimes of the acoustic phonons. The negligible impact of irradiation damage on phonon velocity will provide important information for studies, discussed in subsequent sections, aimed at the forward problem of predicting the thermal conductivity in defective oxides using the BTE formalism.

### 2.1.5 Time-of-flight neutron measurements reveal nonlinear modes in ThO$_2$ and UO$_2$

The vibrational motions in crystalline solids are typically phonons, equivalent to the normal modes of classical mechanics. The total number of phonon modes in a system with N atoms in the primitive cell is 3N. In the case of the fluorite structured ThO$_2$ and UO$_2$, N=3 and there is a total of 9 phonon branches. Recent inelastic neutron scattering measurements observe all 9 branches [67]. However, in addition to the 9 phonon modes expected, a dispersing mode has been observed in this system at energies above the highest energy phonon mode, which is referred to as the Nonlinear Propagating Mode (NPM). This mode is observed in UO$_2$ and ThO$_2$, as well as in natural fluorite, CaF$_2$. It is observable at temperatures ranging from 5 K to 1200 K. The observed dispersion and linewidth are comparable to a typical phonon dispersion and linewidth in this system. It also repeats with the same symmetry as the phonon modes, evidence that the NPM propagates with the same periodicity as the phonons.

Additional degrees of freedom can lead to additional modes but given the wide range of observations and no evidence of a phase transition leading to the absence or appearance of the NPM, this is unlikely. Nonlinear terms in the equations of motion, as a result of the interatomic potential not being perfectly harmonic, can also produce additional modes as observed previously [68]. It has been argued that this is the source of the mode in these materials with support from first-principles calculations [68].

The high group velocity of the NPM suggests that it may make a positive contribution to thermal conductivity. However, the high energy of the mode indicates that this could only be the case at high temperatures (80 meV corresponds to about 930 K). The mode may also act as an additional scattering channel for normal phonons, decreasing thermal conductivity, so the impact of NPMs on thermal conductivity remains an open question. A full accounting of the phonons and NPMs from first principles is needed to resolve this question.

### 2.2 First-principles thermal transport in perfect crystals

ThO$_2$, a band insulator, does not have ground state *f*-shell electrons, while UO$_2$ has a partially filled *f* shell and is a Mott insulator. Thus, in addition to being of interest to the nuclear fuels community, these materials are model systems for understanding the role of electron correlation. We begin by generically considering the state of first-principles computations of electronic and lattice properties of crystals as compared to experimental measurement on high purity single crystals. After the general discussion, we critically review the literature concerning application of first-principles computation to thermal transport in ThO$_2$ and UO$_2$, highlighting main findings and identifying open issues.

### 2.2.1 General considerations

The foundation of any first-principles prediction of thermal transport in crystals begins with the computation of the various quasiparticles which will be carriers of heat (e.g. electrons, phonons, etc.), requiring the approximate solution to the many-body Schrödinger equation for electrons interacting with some symmetric arrangement of nuclei. Fortunately, we have elegant mathematical formalisms, such as density functional theory (DFT) [69, 70, 71, 72, 73], which have allowed us to make substantial progress in obtaining approximate solutions to this many-body problem that represents the forefront of mathematical and computational physics. Further, first-principles computations of various sorts will then be needed to calculate the intrinsic scattering mechanisms of the respective quasiparticles, and we will focus on phonon-phonon interactions in the subsequent sections.

### 2.2.2 First-principles approaches for crystals beyond DFT

#### 2.2.2.1 Hybrid Functionals and DFT+U

Hybrid functionals mix standard DFT exchange-correlation functionals with some fraction of Hartree-Fock exchange [74]. Generically speaking, this approach has been very successful for correcting ground and excited state properties in atoms and molecules [75, 76], weakly correlated band insulators [74], defect energetics [77], and strongly magnetic insulators [78, 79, 80, 81]. However, hybrid functionals can be unreliable for metals [82] and can fail completely for strongly-correlated electron materials. For example, the Heyd-Scuseria-Ernzerhof hybrid functional [83] is qualitatively wrong for the classical material VO$_2$ [84], where the undistorted rhombohedral phase is predicted to be insulating [85], while it is known to be metallic [84]; the dimerized monoclinic phase is predicted to be higher in energy than the rhombohedral phase in addition to being a magnetic insulator [85], while it is experimentally known to be the ground-state structure [84] and nonmagnetic (i.e. no local moments observed) [86]; and the predicted band gap of 2.23 eV is in poor comparison with the experimental value of 0.6-0.8 eV [87]. A further limitation of hybrid functionals is the rather large computational cost when using a plane wave basis, scaling approximately as the cube of the number of atoms in the unit cell [88].

DFT+U [89] bears many similarities to hybrid functionals, in that DFT+U can be viewed as a sort of local hybrid functional. The connection between DFT+U and hybrid functionals has indeed received attention from a methodological perspective [90, 91]. While DFT+U does not contain the non-local aspects of a hybrid functional, it does account for the important local aspects present in many strongly correlated materials. More importantly, the computational cost of DFT+U is considerably smaller than hybrid functionals. Additionally, while DFT+U has more empirical parameters than hybrid functionals (i.e. U, J, and double counting in DFT+U vs. the alpha mixing parameter in hybrids), this is not entirely a disadvantage because it facilitates an empirical exploration of the various components of the functional. Altogether, DFT+U is a very efficient theory, which can operate in the same spirit as DFT, performing full structural relaxations and exploring a wide phase space of possibilities.

## 2.2.2.2 DFT + Dynamic mean-field theory

All known implementations of DFT break down in certain strongly-correlated electron materials, illustrating the difficulty of developing sufficiently robust approximations for the exchange-correlation functional. Instead of creating functionals of the density, one can choose to create functionals of other observables, which might be more sensitive to capturing the physics of strong correlations, such as the single particle Green's functions [92, 93, 94]. In such approaches, there is an analogy to the exchange-correlation functional, which embodies the unknown portion of the energy as a functional of the Green's function, and the functional derivative of this quantity with respect to the Green's function is the well-known self-energy. The dynamical mean-field theory (DMFT) can be viewed as a tool which allows for the approximate computation of the self-energy [95].

DMFT was created in the context of the Hubbard model [96, 97, 98], which is one of the simplest models of interacting electrons on a lattice; consisting of an on-site Coulomb repulsion and a nearest neighbor hopping parameter to embody both the kinetic energy and some external potential. The Hubbard model cannot be solved in general, but it can be solved in one dimension using the Bethe ansatz [99, 100] and in infinite dimensions in conjunction with DMFT [101]. DMFT maps the Hubbard model to an effective Anderson impurity model (AIM), where the bath is determined via the DMFT self-consistency condition. Some other approach is then needed to solve the AIM, and most prevalent approaches to the many-body problem have been applied [101, 102]. Given that the AIM is still a tractable many-body problem, DMFT allows for the infinite dimensional Hubbard model to be solved numerically exactly, using techniques such as quantum Monte Carlo [103, 104]. These numerically exact techniques, however, are computationally expensive, limited to finite temperatures, and may face minus sign problems [105].

While DMFT can clearly capture the physics of the Mott transition in the infinite dimensional Hubbard model, it is not tractable for treating all the electrons in a strongly correlated electron material. Fortunately, it is only a subset of electrons in real materials that give rise to strongly correlated electron behavior – typically $d$ and $f$ electrons. It is then natural to combine the best aspects of both DFT and DMFT by constructing an energy functional of two variables: the density and the local Green's function of the correlated orbitals [102]. This functional is then approximated using some exchange-correlation functional for DFT, along with DMFT, yielding the so-called DFT+DMFT approach. DFT+DMFT has been placed on the same footing as DFT, allowing for detailed total energy calculations of complex materials where DFT sometimes qualitatively breaks down [106, 107, 108, 109, 110, 111].

Upon executing a DFT+DMFT calculation, one needs some approach to solve the effective AIM of the DMFT portion of the calculation, and there is always a balance between accuracy and computational cost. The impurity solver with the smallest computational expense is Hartree-Fock, and this leads to the well-known DFT+U approach. It is worth noting that DFT+U [112] was developed around the same time as DMFT, and therefore DFT+U greatly preceded DFT+DMFT. Therefore, it is often not appreciated that DFT+U is a special case of DFT+DMFT [102].

### 2.2.3 Computing phonons and their interactions from first principles

Computing phonons and their interactions from first principles amounts to computing the Taylor series expansion of the Born-Oppenheimer potential of a given first-principles approach with respect to the nuclear displacements of the crystal. A crystal is infinite in extent, and therefore the computed Taylor series at each order must be truncated at some maximum range in real space, or density in reciprocal space. A sufficient resolution needs to be obtained at each order such that the observable at hand is converged to within a sufficient tolerance, and therefore the necessary resolution will vary depending on which observable is being considered. For a comprehensive review of computing phonons and their interactions from first-principles we refer the reader to [113].

There are two basic approaches for computing derivatives of a Born-Oppenheimer potential [114, 115]: perturbative approaches and finite displacements. We use the latter term broadly to include any approach which explicitly displaces the nuclear coordinates and solves the Schrödinger equation of the displaced structure, including usual finite difference approaches or more complicated fitting procedures based on finite displacements. The nature of the perturbative approach and the order to which it can be carried out will depend on the particular first-principles approach that is being used. Finally, these two approaches are naturally combined, using a perturbative approach to obtain some low order derivatives (e.g., Hellman-Feynman forces) and finite displacements for higher order derivatives, allowing many derivatives to be simultaneously measured in a given finite displacement calculation [113, 116, 117].

### 2.2.4 Computing thermal transport from phonons and their interactions

The anharmonic vibrational Hamiltonian poses a nontrivial many-body problem of interacting bosons, and exactly evaluating the thermal conductivity is highly nontrivial. The most common approach is the linearized BTE [118]. The inputs to the BTE are the phonon eigenvalues and eigenvectors, the phonon velocities, and typically the cubic phonon interactions. The BTE may be numerically solved without approximation using iterative techniques [44] and variational approaches [119], where the latter guarantees numerical convergence. Solutions to the BTE based on phonons and phonon interactions have been executed over a large number of material systems [120, 121], often resulting in impressive agreement with experiment, though most systems that have been studied are band insulators.

Typical approaches to the Boltzmann equation have numerous limitations. For example, most studies only account for scattering via cubic phonon interactions and do not account for thermal expansion of the crystal, and both of these assumptions are only valid at sufficiently low temperatures. Several recent studies have attempted to go beyond these limitations, by including quartic phonon interactions [122, 123, 124, 125] and thermal expansion [125] when solving the Boltzmann equation. If one only needs results in the classical regime, another approach is to use Kubo-Green linear response in conjunction with classical molecular dynamics [126, 127]. A potential advantage of this route in conjunction with first-principles molecular dynamics is to retain phonon interactions to all orders and include all possible allowed scattering process.

### 2.2.5 First-principles calculations of UO$_2$ and ThO$_2$

The ground state and thermodynamic properties of ThO$_2$, including the phonons, are reasonably well described by DFT [128, 129, 130, 131], as expected for a band insulator. A recent study [132] compared the accuracy of three different exchange-correlation functionals to experimental results (see Figure 8). Overall, the Strongly Constrained and Appropriately Normed (SCAN) functional and the Local Density Approximation (LDA) show better agreement with experiment than the Generalized Gradient Approximation (GGA). UO$_2$ nominally contains two electrons in the uranium $f$ shell, given that the remaining four uranium valence electrons will be transferred to the oxygen atom. However, in reality, hybridization substantially alters this nominal charge counting. The partially filled $f$ shell gives rise to a low energy Hamiltonian, which will exhibit a competition between the strong on-site Coulomb repulsion and delocalization, resulting in the "actinide Hamiltonian" [133], which is a combination of the well-known Hubbard and periodic Anderson model. Therefore, one is faced with a challenging many-body problem, with potential for strongly correlated electron behavior. Moreover, the physics of UO$_2$ are further complicated by the relatively strong spin-orbit coupling in uranium.

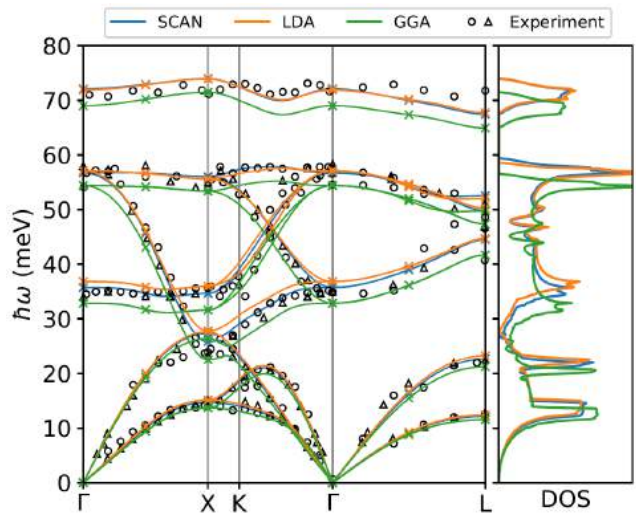

**Figure 8.** Phonon dispersion for ThO$_2$ computed using DFT with three different exchange-correlation functionals (LDA, GGA, SCAN) and measured using inelastic neutron scattering. Right panel: The phonon density of states. Reproduced with permission from [132]. Copyright 2021 Dennett et al. (Published by Elsevier Ltd.).

A large number of DFT, DFT+U, hybrid functional, DFT+DMFT, and other first-principles calculations have been performed on UO$_2$ [134]. The LDA and GGA qualitatively fail, producing a metallic state [135]. The PBE0 and HSE hybrid functionals both produce an antiferromagnetic insulator [136, 137]. DFT+U also produces an antiferromagnetic insulator and can even capture the proper 3**k** structure [138, 139, 140, 141]; reasonable agreement with the spin-wave spectra is also obtained [142]. DFT+U calculations can provide moderate agreement with the experimental phonon spectrum [143, 144]. Of course, DFT+U is confined to ordered states, and DFT+DMFT will be needed to extend to the paramagnetic state.

DFT+DMFT calculations for UO$_2$ have been executed using a variety of different approximations to solve the DMFT impurity problem, including exact diagonalization for a small bath [145], Hubbard -I approximation [146], and continuous time quantum Monte Carlo (CTQMC) [146]. It is straightforward for DFT+DMFT to capture the paramagnetic insulating state in all cases. Studying small energy details with DFT+DMFT is still challenging due to the severe tradeoff between precision and computational expense when solving the DMFT impurity problem. Precise techniques such as CTQMC will inherently hit a computational wall at sufficiently low temperatures, while inexpensive techniques such as Hubbard-I are not terribly reliable. A technique that is well suited to the paramagnetic state at zero temperature for UO$_2$ is the rotationally invariant slave boson theory [147], which finds UO$_2$ to be an orbitally-selective Mott insulator. However, the shortcoming of this approach is that it only has a semi-quantitative characterization of Mott physics, and more precise methods are needed to further investigate these findings.

With these approaches for the lattice properties of UO$_2$ and ThO$_2$, thermal conductivity has been computed for both materials in various levels of approximations. The lowest level approximation only includes the anharmonicity at the level of the mode averaged Grüneisen parameter in conjunction with the Slack equation [148], and this has been executed using various DFT functionals in ThO$_2$ [129, 130, 131, 149]; and it has been executed for UO$_2$ using DFT(GGA)+U [150] in addition to DFT+DMFT where the DMFT impurity problem is solved via exact diagonalization using a truncated bath [64]. The next level of approximation is directly computing the cubic phonon interactions from first principles, computing the phonon lifetimes from perturbation theory, and then using the RTA. This has been executed for ThO$_2$ using DFT [151, 152] and UO$_2$ using DFT+U [27], where the latter demonstrated rather substantial deviation from experiment. The next level of approximation goes beyond the RTA and fully solves the BTE, which has been executed for ThO$_2$ [153, 154]. It should be noted that quantitative differences were found in the two aforementioned studies, despite the fact that the calculations were nominally very similar, highlighting some sort of methodological shortcomings.

### 2.3 Thermophysical properties
### 2.3.1 Introduction

In addition to thermal conductivity, the heat capacity, thermal expansion, and the elastic constants are all directly determined by the phonon structure. The thermophysical properties of actinide oxides have been a subject of many studies during the last 70 years. This fundamental research, initiated by the Manhattan Project, started various systematic studies of actinide oxides that led, among others, to the discovery of phase transitions in both UO$_2$ [155] and NpO$_2$ [156] at temperatures below 50 K. Neutron diffraction experiments established the antiferromagnetism in UO$_2$ below 30 K, [157, 158], but the situation in NpO$_2$ remained a mystery for many years. The antiferromagnetsim (AF) of UO$_2$ was originally thought to be a simple type I AF single-**k** (1**k**), but work by Burlet, et al. [159] strongly suggested that the ordering was of a 3**k** nature, and this was confirmed unambiguously by Blackburn and coworkers [160]. In such a magnetic structure, the moment directions are non-collinear but point along alternate <111> directions. At the same time, resonant X-ray experiments [161] had established that the ordering at 25 K in NpO$_2$ was of the electric quadrupoles, and they were also found soon after in UO$_2$ [162]. Much of these complications and the associated theory is reviewed in [163], and the latest excitation spectra of UO$_2$, as measured with polarized neutrons, was published in 2011 [164]. Hints of the elusive quadrupolar excitations [165] were found in this study, but an unambiguous proof of their existence remains a challenge. A more recent (shorter) review on these challenges in $UO_2$ appeared in 2020 [166].

Our focus here is on studies that target low temperature (room temperature and below) thermophysical properties. This is in keeping with our later discussion of thermal transport in the presence of defects. When investigating the influence of irradiation defects, it is often advantageous to freeze out phonon-phonon interaction by conducting experiments at low temperatures.

### 2.3.2 Heat capacity

The low-temperature heat capacity of $ThO_2$ has been measured on both polycrystalline samples and single crystals [29]. As expected for a diamagnetic insulator, the heat capacity does not show any sign of phase transitions, and its temperature dependence is governed by the acoustic and optical contributions which can be described using the DFT harmonic approximation (see Figure 9). $UO_2$ orders antiferromagnetically (non-collinear 3**k** magnetic structure) below the Neel temperature, $T_N = 30.5$ K, as discussed above. In $UO_2$ the transition is accompanied by a Jahn-Teller distortion of the oxygen atoms, discovered in 1976 [167], and the electric quadrupolar ordering (3**k** type) [142, 163]. Because of that, the heat capacity shows a large first-order-type anomaly at the transition (see Figure 9).

there are much fewer studies on the specific heat of $NpO_2$ and $PuO_2$. $NpO_2$ has electric quadrupole ordering at 25.5 K [161] and a complex ordering of a higher order magnetic multipoles [163, 174, 175, 176]. Similar to $ThO_2$, the temperature dependence of $PuO_2$ (polycrystalline samples) shows no anomaly in agreement with a temperature-independent Pauli paramagnetic state [170, 171, 177], which arises from the singlet-ground state established for this $Pu^{4+}$ ($5f4$) material [178]. The similarity in the behavior of heat capacity in these oxides agrees with the similarities in phonon dispersions observed in these materials. Also, the low-temperature electronic contribution to the specific heat can be extrapolated to zero at T = 0, in perfect agreement with their Mott insulating ground state.

### 2.3.3 Thermal conductivity

Despite its importance as a nuclear fuel material, the thermal conductivity of $ThO_2$ single crystals has been measured only recently [179]. Overall, the temperature dependence of the thermal conductivity of $ThO_2$ is typical for diamagnetic insulators with a characteristic maximum below 40 K. Its behavior can be described by phonon thermal transport by taking into account boundary, defects, and Umklapp phonon-phonon scattering mechanisms [180]. Uranium dioxide, in contrarst, shows an anomalous behavior with a double-peak structure (see Figure 10) and low magnitude of $\kappa(T)$.

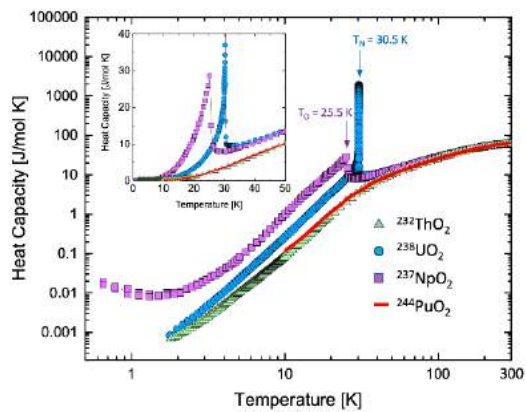

**Figure 9.** The heat capacity of $ThO_2$ (green triangles) [29], $UO_2$ (blue circles) [168], $NpO_2$ (purple squares) [169] single crystals, and polycrystal $PuO_2$ (red line) [170, 171]. In the case of $UO_2$, a large first-order type anomaly is observed at 30.5 K resulted from structural Jahn-Teller distortion and antiferromagnetic ordering. For $NpO_2$ the transition at 25.5 K is caused by the ordering of high-order magnetic multipoles. The increase of heat capacity below 3 K in various neptunium systems is usually associated with a nuclear Schottky term due to the splitting of the I = 5/2 nuclear ground level of the $^{237}$Np nuclei by the hyperfine field. In $NpO_2$, however, the origin of the anomaly is unclear since the nuclear contribution seems to be too small to explain the difference [172]. $UO_2$ data adapted with permission from [168]. Copyright 2019 American Physical Society. $NpO_2$ data adapted with permission from [169]. Copyright 2016 American Physical Society. $PuO_2$ data adapted with permission from [170]. Copyright 1976 American Institute of Physics. $ThO_2$ data adapted with permission from [29]. Copyright 2020 Dennett et al. (Published by AIP Publishing).

The anharmonic effects, first observed in diffraction experiments by Willis and colleagues [173], are important in describing the heat capacity in this oxide family. In $UO_2$, the observation of a small anharmonic specific heat contribution is the result of relatively large energy-dependent anharmonic effects which have opposite signs, leading to a total contribution near zero [168]. As might be expected,

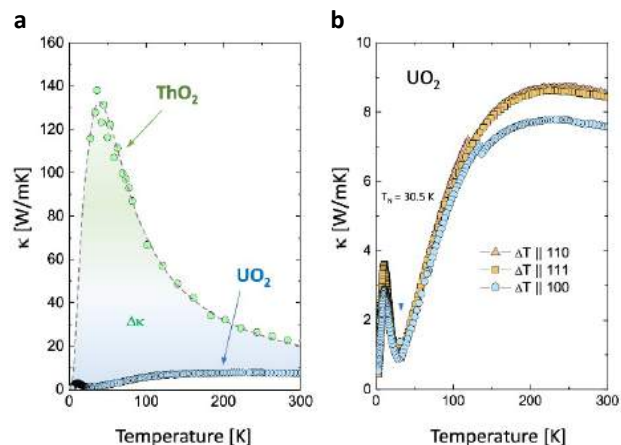

**Figure 10.** (a) The thermal conductivity of $ThO_2$ (green circles) and $UO_2$ (blue circles) single crystals measured along the <100> crystallographic direction. Note the significant differences in $\kappa(T)$ and its magnitude between $ThO_2$ and $UO_2$. (b) The temperature dependence of the thermal conductivity of $UO_2$ single crystal measured in the <100> <110> and <111> crystallographic directions. $ThO_2$ data adapted with permission from [179]. Copyright 2010 American Chemical Society. $UO_2$ data adapted with permission from [180]. Copyright 2014 Gofryk et al. (Published by Springer Nature).

That behavior cannot simply be explained by different strength of defect or grain boundary scattering processes present in different materials as the thermal conductivity of $UO_2$ is similar for stoichiometric pellets, polycrystals, and single crystals [181]. To explain that discrepancy, it has been suggested that the reduction of the thermal conductivity in $UO_2$ is due to the strong resonant scattering process of phonons and electronic degrees of freedom, such as spins, with the characteristic energy of the order of 10 meV [180]. Similar behavior has also been suggested to describe the $\kappa(T)$ behavior in other magnetic materials [182, 183, 184]. Furthermore, the thermal conductivity of $UO_2$ appears to exhibit anisotropic behavior (Figure 10b) with slightly smaller (~10%) heat conduction along the <100> direction [180]. If confirmed, this unusual behavior for the second-rank thermal conductivity tensor indicates that there is some "hidden" anisotropic behavior (in certain phonon interactions or couplings) that break the cubic symmetry in this material and lead to anisotropic thermal transport. Some microscopic evidence to support these macroscopic measurements has come from measuring the low-temperature phonon linewidths with IXS [185]. A strong anisotropy is found in that the linewidths are strongly broadened over the energy range 5–15 meV in the <100> direction, but no effects are observed in the <011> direction. Further study is needed to definitively connect these phonon features and macroscopic thermophysical properties.

### 2.3.4 Elastic constants and thermal expansion

Elastic constants have been studied experimentally using a pulse echo technique on single crystals of $ThO_2$ and $UO_2$ at room temperature [186, 187]. The thermal expansion of $UO_2$ single crystals has been measured by Brandt and Walker in 1967 [188, 189]. That was the first direct demonstration that the phase transition is first-order in the lattice properties. The volume collapse at $T_N$ was first observed in 1974 [190] and, more recently, by X-ray diffraction [191]. Furthermore, the measurements of elastic constants versus temperature have not only exhibited changes at the magnetic transition but also confirmed that the elastic properties at high temperature (up to room temperature) are much more complicated than anticipated. Similar indications have come from susceptibility measurements [192] and of polarized neutron studies [193] that showed the short-range order up to at least 200 K. Elastic constants are directly related to low frequency dispersion of the three acoustic phonon modes. In cubic systems, there are three independent elastic constants, $C_{11}$, $C_{12}$, and $C_{44}$. In $UO_2$, the temperature dependence of $C_{44}$ (shear along the <111> crystallographic direction – the cube diagonal) is highly anomalous [188, 189] and indicative of a considerable interaction between lattice and spins even well above the Neel temperature (see Figure 11). The strong spin-phonon coupling and magnetoelastic interactions have also been confirmed by inelastic neutron scattering experiments [194, 195]. This is more completely discussed in a recent neutron inelastic scattering study [164], the review article [163], and the 2020 review [166].

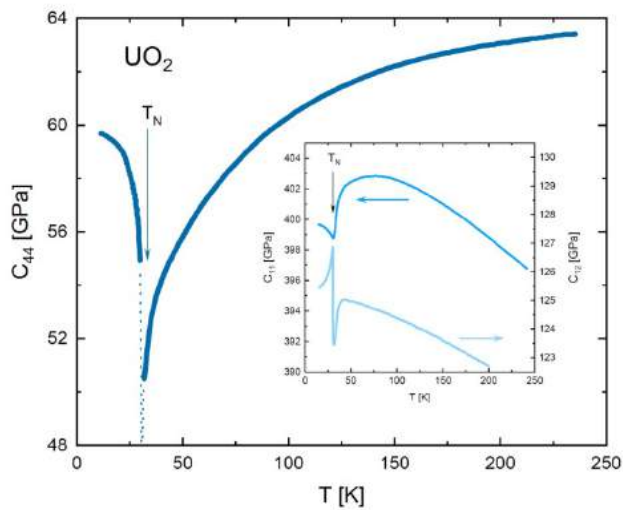

**Figure 11.** Change in elastic constants with temperature for $UO_2$. Data adapted with permission from [188]. Copyright 1967 American Physical Society.

Recently, the thermal expansion of $UO_2$ has been studied under high magnetic field. It was found that $UO_2$ becomes a piezomagnet when the magnetic field is applied along the <111> direction in the antiferromagnetic state and exhibits a peculiar memory effect due to a partial domain reorientation [196]. These effects are due to the strong magnetoelastic interactions and symmetry of the non-collinear 3**k** antiferromagnetic order in this material that allows the appearance of the piezomagnetism in $UO_2$ [197].

## 2.4 Crystal growth
### 2.4.1 Challenges in growing single crystals

Many of the experimental efforts discussed herein either require single crystals or perform best when analyzing single crystals. The $AO_2$ (A= Ce, Th, U, or Pu) family, however, is extremely challenging to grow as single crystals because the high melting points (2470-3377°C) [198, 199, 200, 201] exceed those of the containers used in melt growth syntheses and the solubilities are low in most industrially feasible solvent systems [202, 203, 204, 205]. Overviews of the successful growth techniques have been published elsewhere [206, 207, 208, 209, 210]), but herein the discussion is focused on how each successful growth technique maximizes some of the desired attributes while having shortcomings in other aspects. As a result, when approaching a study of the $AO_2$ family, or even any individual member, compromises must typically be made with what single crystal properties are of highest importance. The main areas where this tradeoff occurs is in the crystal purity, growth speed, structural quality, and size. The various oxidation states that are available for U and Pu further complicate crystal production [211].

### 2.4.2 Bulk crystal growth techniques

The cornerstone of large (>1 m), high purity (>>99.9%), high quality crystal growth in the semiconductor industry is the pulled melt crystal growth methodology, such as that employed in the Czochralski method. Maintaining a feedstock of molten $AO_2$, however, is infeasible because their melting points exceed that of widely used containment vessels. This has resulted in the $AO_2$ family being the subject of more exotic crucible methods, such as the cold crucible technique ($ThO_2$ and $UO_2$ [212]), closed capsule zone melting ($UO_2$ [213]), and using a floating zone furnace ($UO_2$ [214, 215]). An alternative route to avoiding crucible destruction is to have a much more rapid melt growth, such as in arc melting ($UO_2$ [214, 216, 217, 218] and $ThO_2$ [219, 220, 221, 222, 223]), a solar furnace ($UO_2$ [224] and $ThO_2$) [225]), or the Verneuil method ($UO_2$ [226]). Although many of these techniques yield large (>1 cm$^3$) products, they are never single crystal for the $AO_2$ family. Rather, agglomerates of multiple single crystals that are sintered together are produced [227].

Samples used for analysis are often cut from these large domains such that a single crystal is acquired. This means that all single crystals thus obtained have been mechanically altered which, in the case of A = U or Pu, could mean that the surface adopts a new cation-to-anion ratio. Furthermore, the production of large crystal domains signifies that orientation control is difficult for melt growth of these refractive oxides. The presence of these large domains indicates growth conditions are quite heterogeneous, suggesting that crystal quality might vary within a single given domain, but experimental investigations have rarely been performed to assess this.

If the melt growth is eschewed because of these assorted shortcomings, the next most common growth techniques dissolve the feedstock under conditions well below the melting point and then crystallize out the feedstock from the solution. Chemical vapor deposition (CVD) or chemical transport ($UO_2$ [228, 229, 230, 231], $U_xTh_{1-x}O_2$ [232], and $U_{1-x}Pu_xO_2$ [233]), flux ($CeO_2$ [234, 235, 236, 237, 238, 239, 240, 241], $ThO_2$ [220, 223, 234, 235, 236, 237, 238, 242, 243, 244, 245, 246, 247], $UO_2$ [217, 248, 249], $U_xTh_{1-x}O_2$ [249], and $PuO_2$ [242, 250, 251, 252, 253]), and hydrothermal or solvothermal synthesis ($CeO_2$ [254], $ThO_2$ [29, 67, 179, 255, 256, 257, 258, 259, 260, 261, 262], $U_{1-x}Th_xO_2$ [262, 263, 264], $UO_2$ [256, 257, 258, 264, 265, 266, 267], and $PuO_2$ [268]) fall into this category and have been employed for the entire $AO_2$ family. High-quality, well-faceted single crystals of $ThO_2$ and $UO_2$ produced using the hydrothermal method are shown in Figure 12. Whereas melt crystal growth is largely a physical growth method (although oxidation, reduction, and crucible contamination can become major chemical factors), solution-based growth techniques are truly chemical growth methods. This complicates matters in that the solvent-reactant, -product, and –container interactions all become variables in addition to chemical species and oxidation/reduction. The added complexity means that large, high-quality single crystals can be grown, but the development of a reliable, large-scale technique is often a cumbersome undertaking. Furthermore, solvent contamination is an inherent challenge, frequently resulting in measurable impurities in single crystals produced via these pathways. In some cases, however, ideal solvents can be found, such as how the highest purity $CeO_2$ single crystals have been fabricated by the hydrothermal technique in a KF solution with impurities of only 100 ppm of K and 10 ppm of F [254]. Despite the purity, the 100-micrometer size somewhat limits their utility. Seeded growth can be performed easily in most of these techniques (provided a suitable seed can be procured), enabling orientation control and iteratively improving crystal quality.

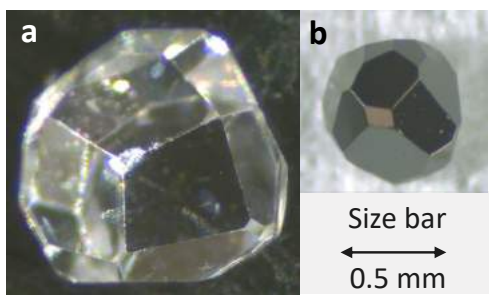

**Figure 12.** A high quality, well faceted single crystal of a) $ThO_2$ and b) $UO_2$ produced using the hydrothermal method.

There are a few techniques that fall outside of the purview of both melt techniques and transport assisted growths, including sublimation in vacuum or from a plasma ($ThO_2$ [269] and $UO_2$ [217, 269, 270]), thermal decomposition ($UO_2$ [271] and $PuO_2$ [272]), electrodeposition ($UO_2$ [273, 274, 275, 276]), and high temperature sintering ($UO_2$ [217]). Sublimation has thus far only yielded crystals (4 to 12 mm) of $AO_2$ with purities that are often worse than the starting material. Thermal decomposition, where something other than $AO_2$ (such as $AOCl_2$) is the feedstock for a flux growth, maintains the same drawbacks of flux growth with the added complication of the starting material itself being a source for impurities. Electrodeposition suspends two electrodes in a molten flux containing the feedstock material, repeating the same shortcomings of flux growth and enhancing them with the possibility of electrode contamination. High temperature sintering is essentially limited by growth time, as grains will slowly enlarge, but the atomic mobility is extremely low compared to fluid or gas growth mechanisms. Despite these challenges, high quality crystals have been produced by these techniques, but they are often small, contain impurities, or both.

In the techniques discussed here, some type of tradeoff is necessitated in the crystal purity, growth speed, structural quality, or size. Even with these compromises, however, at the present there is not a direct route to making large (> 1 cm$^3$) <u>single</u> crystals of any member of the $AO_2$ family. Growth procedures are discussed that report larger crystals, but these often have grain boundaries or large domains such that the term "single crystal" cannot readily be applied. As of now, however, there are some leaders in the crystal growth of each $AO_2$ member. Dark brown $CeO_2$ crystals with edges up to 8 mm, for example have been grown with a (100) cubic morphology using a flux mixture of $Pb_3P_2O_8$, $PbF_2$, $Al_2O_3$, NaF, $MgF_2$, and $V_2O_5$ [235]. The crystal purity might be low, however, as similar flux solutions have displayed > 1 wt. % Pb inclusions in $CeO_2$ [240]. Adjusting the flux to $Li_2WO_4$ has resulted in lower impurity concentrations (200 ppm W and <50 ppm Li), but the size decreases to maximum edge sizes of 2 mm [239]. For $ThO_2$, a 18.92 x 9.69 x 0.85 mm single crystal with (100) cubic morphology has been grown by the hydrothermal technique [67] and the bulk purity (> 99 at. % for cation content) is higher than the starting $ThO_2$ feedstock material [29]. The first 50 nm of the surface, however, have a high concentration of impurities, demonstrating that the impurities from the starting $ThO_2$ feedstock were purified out of the bulk material and then deposited on the sample during the final cooling cycle. $UO_2$ crystals with side lengths approaching 10 mm and a variety of crystal morphologies have been synthesized with electrolytic decomposition in a molten flux [274]. Altering conditions decreases the size to 3 mm per side but obtains crystals with highly defined cubo-octahedral morphology and a decrease of ~33% of the impurities present in the starting material [273]. The radiological concerns of $PuO_2$ make its crystal growth a rare endeavor, but $2 \times 2 \times 3$ mm single crystals with a (111) octahedral morphology have been grown from a $Li_2Mo_4O_7$ flux with relatively low concentration of impurities (highest was Fe at 200 ppm) [251].

### 2.4.3 Epitaxial thin films

Epitaxial thin films of $AO_2$ ($CeO_2$ [277], $ThO_2$ [278], $UO_2$ [211, 279, 280], $PuO_2$ [211]) produced via reactive magnetron sputtering, polymer assisted deposition, and sol-gel techniques represent an alternative route to the synthesis of bulk single crystals. While much remains to be done before samples of the quality used in the semiconducting industry are available, this field has started to pay dividends. For example, antiferromagnetism of $UO_2$ thin films on different substrates has been studied [281] and shows the effect of induced strain by the substrate-film interaction.

Currently, films are commonly produced using commercially available substrates such as $LaAlO_3$, $SrTiO_3$, or YSZ, which impart strain into the grown oxide layer. This explains, in part, why thin films have been shown to have broad X-ray diffraction (XRD) rocking curves [282], large deviations from bulk lattice parameters [282], and columnar grains [283]. Nonetheless, tremendous potential exists in this field as suitable single crystal $AO_2$ substrates become available. A start along this line is reported on $U/UO_2$ interfaces [284] and $UO_2/ThO_2$ interfaces [258].

## 2.5 Outlook

Foundational work on phonon transport in nearly perfect single crystals is the first step in developing a comprehensive understanding of thermal transport in oxide nuclear fuel under irradiation. On the experimental side, this portion of the review focused on inelastic particle scattering to obtain detailed information about the phonon structure and measurement of low temperature thermophysical properties. These studies revealed gaps in understanding that are the source of continued research. Examples include the origin of non-linear propagating modes found in several fluorite crystal structures, including $UO_2$ and $ThO_2$, and the possibility for anisotropic thermal conductivity in $UO_2$ due to phonon interactions that locally break cubic symmetry. On the modeling side, electronic structure calculations were reviewed with focus on strongly correlated materials. The shortcomings of DFT to accurately predict the branch specific contribution to conductivity in $UO_2$ point to new opportunities for DFT-DMFT modeling approaches. Lastly, central to all fundamental studies, a short review was presented on growth of single crystals of actinide oxides. As new high-quality single crystal substrates of actinide oxides become available, opportunities exist for synthesis of epitaxial thin films with quality rivaling that found in the semiconductor industry. Having access to interfaces with chemical precision would open new frontiers for understanding thermal transport under irradiation in nuclear fuel.

## 3 Crystalline defect generation, evolution, and characterization

In line with the main goal of this review, here we present an overview of defects in $UO_2$, $ThO_2$ and $CeO_2$ that is closely tied to thermal transport. Defects play a critical role in thermal transport because of their ability to scatter phonons. Experiments have shown that the thermal conductivity in $UO_2$ under reactor conditions is influenced by: (1) irradiation defects, (2) intragranular clustering of fission products, and (3) clustering at existing and newly formed grain boundaries [5,8] (see Figure 13). Until recently, the impact of many of these defects on thermal transport has only been treated empirically as the concentration, size, and distribution of these defects is controlled by complex mechanisms and are highly dependent upon local irradiation conditions. Accordingly, here we primarily focus on smaller-scale defects that can be introduced in a controlled ion beam environment and where recombination, saturation and clustering play an outsized role.

In the last several decades, a number of studies have focused on radiation damage in $UO_2$, including dislocation loop formation [285], radiation induced lattice expansion [286, 287, 288], HBS development [6, 289] and fission gas release [290, 291, 292, 293, 294]. The irradiation microstructure of $CeO_2$, an important surrogate material for oxide nuclear fuels, has started to receive significant attention over the last decade [28, 295, 296, 297, 298]. $ThO_2$ has received sporadic attention since the 1960s but is now receiving more consideration as it has emerged as a promising advanced nuclear fuel [299, 300, 301, 302, 303].

In this section, we discuss the nature of irradiation induced defects as revealed in experiments, select methods for characterization of defects and their evolution, off-stoichiometry and equilibrium point defect disorder in oxides, the mechanisms of defect clustering and long-term evolution in irradiated oxides, and give an outlook on required future developments in understanding defects from the perspective of thermal transport. The primary focus here will be key findings tied to ion beam irradiation studies. Ion beam irradiation has proven a flexible tool for the generation of defects found in real fuel [29, 304, 305, 306, 307, 308, 309, 310, 311, 312, 313, 314, 315, 316, 317, 318, 319, 320, 321, 322, 323, 324, 325, 326, 327, 328]. While beyond the scope of this review, it is noted here that complementary studies involving neutron [329], $\alpha$-particle [286, 287, 288, 330], or electron [331] irradiation can be used to judge the efficacy of using a particular ion irradiation conditions (e.g. ion mass, energy and fluence).

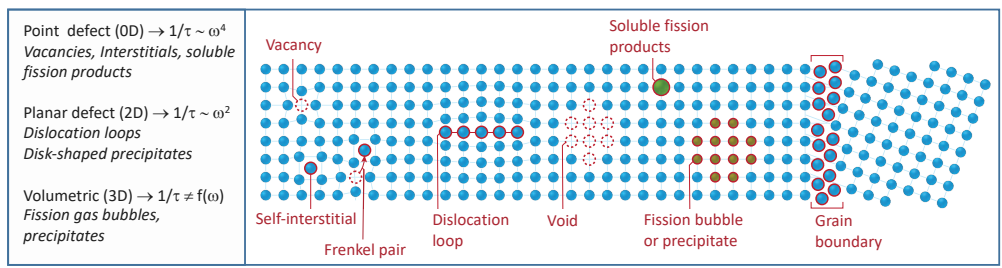

**Figure 13.** Illustration of irradiation defects found in oxide nuclear fuel. The phonon scattering rate is sensitive to the dimensionality of the defect.

As shown in the left pane of Figure 13, the frequency dependence of the phonon scattering rate, $\tau^{-1}$, is sensitive to the dimensionality of the defect with respect to the wavelength of the heat carrying phonons. Thus, to provide a distinct phonon transport perspective, this section is organized according to the size and dimensionality of the defect. In particular, we introduce intrinsic point defects first, followed by point defects and defect clusters that form under irradiation, then other extended defects such as defect decorated grain boundaries and grain boundaries formed due to restructuring.

### 3.1 Intrinsic point defects

Intrinsic point defects are present in single crystal oxides under thermodynamic equilibrium in the form of vacancies, interstitials, and native dispersed impurities. The presence of these point defects makes oxides off-stoichiometric. Off-stoichiometry is important in understanding structural properties, thermo-kinetic properties, and performance of oxide nuclear fuels. Moreover, carefully accounting for the impact of intrinsic defects on thermal transport is the first step in providing a comprehensive description of thermal transport under irradiation.

### 3.1.1 Characterization techniques

Accurate characterization of the size and distribution of intrinsic, sub-nanometer defects remains a challenge. Here we provide a brief review of two experimental techniques used to gleam information about sub-nanometer defects. Point defects in oxides such as $UO_2$, $CeO_2$, and $ThO_2$ have been extensively characterized using light source techniques [322, 332, 333, 334, 335, 336, 337, 338, 339, 340, 341, 342, 343, 344]. This field of study has recently been review by Lang, et al. [345]. However, our focus here will be on positron annihilation spectroscopy and optical spectroscopy, which have so far seen limited use in oxide fuels, but show particular promise for characterizing the spectrum of point defects and small clusters that reside in ion-irradiated oxide fuels and fuel surrogates. While each technique exhibits demonstrated potential, particular care must be paid to the characterization of point defects in ion irradiated oxides.

#### 3.1.1.1 Positron annihilation spectroscopy

Positron annihilation spectroscopy (PAS) can accurately probe individual atomic vacancies as well as small and large vacancy clusters with remarkable sensitivity of less than 1 ppm providing quantitative information about their size and density [346]. When introduced into a solid, a positron quickly loses its kinetic energy and eventually annihilates with an electron, producing two photons that are emitted in nearly opposite directions. The emitted photons are Doppler shifted by the longitudinal momentum of the electron. Measurement of this shift has been developed as an effective method known as Doppler broadening spectroscopy (DBS) to provide information about the concentration and characteristics of defects in solids, as the electron momenta sampled by the positron is strongly affected by trapping at defects.

The positron annihilation rate is governed by the overlap between the positron wave function and the electron density at the annihilation site, providing information about the local electron density which is directly related to the defect nature and size. Thus, positron annihilation lifetime spectroscopy (PALS) provides a direct method for measuring the size and density distributions of atomic scale defects. It can discriminate between vacancy clusters from single vacancies to ~50-100 vacancies until they become large enough to be resolvable by transmission electron microscopy.

While standard PAS techniques are extremely useful, the typical penetration depth is several hundreds of micrometers and as a consequence they cannot be applied to ion irradiated materials as the damage layer is only a few micrometers thick. In the 1980s, the development of slow positron beams and variable energy positron techniques made depth-resolved measurements and characterization of ion irradiated materials possible. The first studies of depth resolved PAS were performed on He ion irradiated Cu and Ni [347].

Laboratory-based slow positrons or variable energy positron beams begin with a positron source having a continuous energy distribution. The most commonly used source is the radioactive isotope $^{22}Na$. The fast positrons emanated from the source are moderated down to ~75 meV as they pass through a thin foil (e.g. W and Ni) [347, 348]. The positrons extracted from the moderator are then guided through a vacuum system to the target using magnetic or electrostatic lenses. The extracted positrons are unmoderated and of high energy, so need to be filtered through crossed electric and magnetic fields (E×B filter) or a magnetic velocity selector. The left side of the schematic of a magnetically transported positron beam in Figure 14 illustrates the extraction of slow positrons through a magnetic velocity selector. The figure also shows the moderation process. After extracting the slow positrons, the direct current (DC) slow e+ beam is then accelerated to variable energies, typically up to 30 keV, where variable energy DBS (VEDBS) can be performed. VEDBS can provide qualitative information about the defect structures and measure the overall defect level in ion irradiated materials. Such measurements were applied in the last few decades to ion irradiated actinide oxides and the knowledge learned from them will be briefly discussed in the next section. However, the full characterization of radiation induced defects in terms of size, density, and depth distributions can only be extracted from PALS measurements which require pulsed positron beams with short pulses on the order of 100-200 ps. In the last several years, reactor- and accelerator-based pulsed positron beams have been developed with sufficient temporal resolution for characterization of small defects induced by ion irradiation in some metals as a function of depth [349]. While these user facility adaptations of positron characterization provide the irradiation materials science community with tremendous new capability, multiple measurement methodologies cannot be easily accommodated. Current efforts directed at developing pulsed positron beams derived from radioactive isotopes (see Figure 14) will usher in a new era in positron characterization, by enabling in-situ measurements of materials during ion irradiation [350]. Figure 14 shows the incorporation of a bunching system to a magnetically transported beam to convert the DC slow e+ to pulses with hundred picoseconds width relevant for PALS measurements in a wide range of ion irradiated materials. Then both DBS and PALS can be simultaneously done as illustrated in Figure 14.

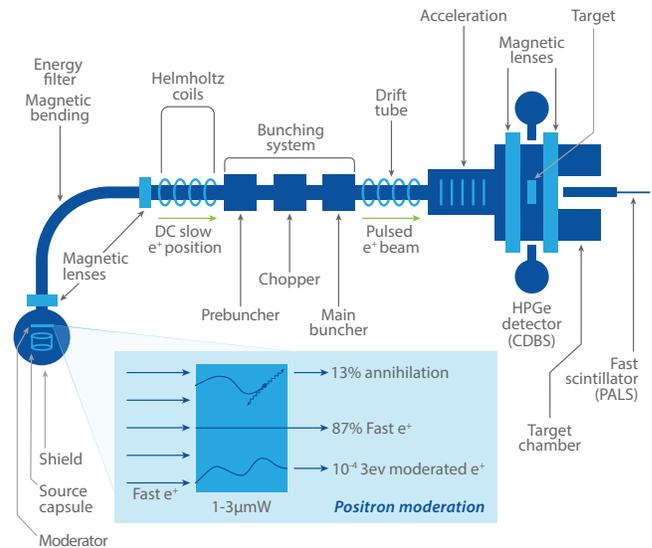

**Figure 14.** A simplified schematic of a magnetically transported slow pulsed positron beam with positron annihilation lifetime spectrometer and coincidence Doppler broadening spectrometer. Vacuum components, HV/current sources, and beam diagnostic tools are not shown.

#### 3.1.1.2 Optical spectroscopy

Optical absorption and luminescence spectroscopy are convenient methods for characterizing point defects in irradiated oxides as oxygen vacancies have prominent signatures. While these spectroscopies have been applied to characterize a large range of irradiated oxides [351, 352, 353], application to irradiated actinide oxides is limited. The optical response of oxides depends strongly on the preferred charge state of the cation. We thus start our discussion with $ThO_2$ as Th has only one preferred charge state. We then discuss $CeO_2$ and $UO_2$, both of which have multiple charge states.

In ThO$_2$, Th only exhibits a +4 charge state. Therefore, when one considers an oxygen vacancy, two electrons (in a neutral defect) and one electron (in a positively charged defect) are localized on the vacant site and create a virtual atom having specific electronic states. Optical induced transitions of an electron within this virtual atom contribute to the emergence of absorption and luminescence bands within the bandgap. Owing to their impact on optical permittivity, these defects are referred to as color centers. Figure 15(a) shows representative spectra for proton irradiated ThO$_2$ exhibiting absorption peaks at 1.8 and 2.0 eV [132]. The location of these peaks matches previous reports on irradiated ThO$_2$ [219, 354]. In contrast to Al$_2$O$_3$ and MgO [351, 352, 353], the number of studies of electronic structure of defects in ThO$_2$ has been limited [149, 355]. Accordingly, assignment of these absorption/luminescence peaks to defect type requires more systematic studies.

In CeO$_2$, the oxygen vacancy induced peaks have a different origin [356, 357]. Unlike ThO$_2$, CeO$_2$ does not have color centers as extra electrons are localized on cerium ions, causing them to change charge state from Ce$^{4+}$ to Ce$^{3+}$. The extra electron on the Ce$^{3+}$ ion occupies the 4$f$ electron state, which is empty in Ce$^{4+}$. Additionally, it has been reported that 5$d$-band electrons in Ce$^{3+}$ exhibit additional crystal field splitting. In perfect CeO$_2$, Ce$^{4+}$ has an octahedral coordination leading to a splitting of 5$d$-bands into two groups of bands, each with T$_{2g}$ and E$_g$ symmetry (Figure 15e) [358]. When defects are present, the symmetry of the crystal structure is reduced which results in further splitting of $d$-bands (Figure 15e). Overall, this leads to an emergence of additional absorption bands corresponding to electron transitions from 4$f$ bands into crystal field split 5$d$ bands as shown in the spectra of irradiated CeO$_2$ (Figure 15b), which exhibits an additional absorption peak at 2.8 eV [359]. Computationally, there is strong evidence supporting localization of electrons on Ce ions [355].

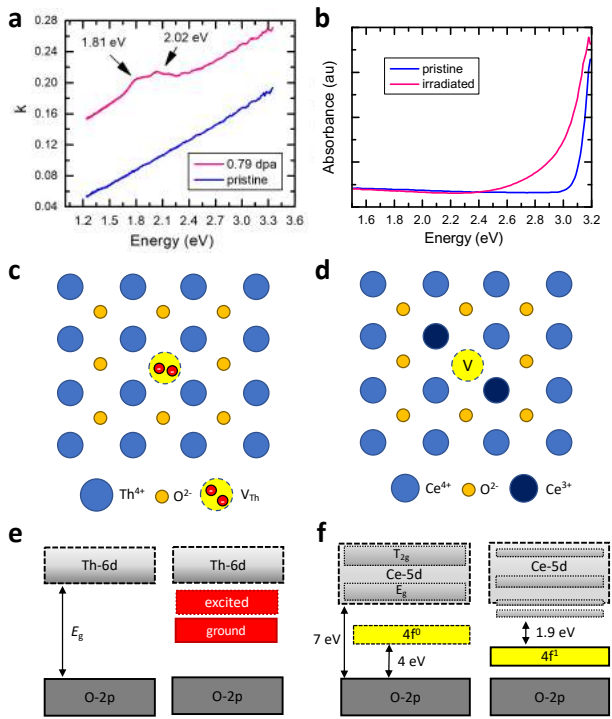

**Figure 15.** Optical characterization of defects in irradiated fluorite oxides. Optical absorption in (a) proton irradiated ThO$_2$ and (b) electron irradiated CeO$_2$. Atomic configuration of oxygen vacancies in (c) ThO$_2$ and (d) CeO$_2$, corresponding electronic band diagrams are shown in (e-f). Data in (a) adapted with permission from [132]. Copyright 2021 Dennett et al. (Published by Elsevier Ltd.). Data in (b) adapted with permission from [360]. Copyright 1994 Elsevier Sciences B.V.

Electronic band-structure for pristine UO$_2$ have been thoroughly investigated experimentally [361]; however, investigation of the electronic structure of defects has been limited to computational efforts involving DFT+U treatments of electron correlation [362, 363]. While in ThO$_2$ and CeO$_2$ the highest populated level corresponds to 2$p$-bands of oxygen, in UO$_2$ it corresponds to 5$f$-bands. In ThO$_2$, the lowest unoccupied state is 5$d$, whereas in CeO$_2$ it is 4$f$. In UO$_2$ an additional effect is present, where the $f$-band is split due to strong electron correlation effects making it a Mott insulator [364]. The challenge here involves properly treating electron correlation to accurately assign absorption/luminescence peaks to defect type.

**Raman Spectroscopy**

Raman spectroscopy allows probing of vibrational properties of fluorite crystals [365, 366, 367] and can be used to characterize point defects and small defect clusters. Raman spectra of UO$_2$ has been presented in a number of studies [258, 368], while reports on ThO$_2$ are more limited [29, 321]. An extensive body of literature exists for CeO$_2$ and ZrO$_2$ [369, 370]. A perfect crystal structure of these fluorites exhibit a mode at ~ 450 cm$^{-1}$, characterized as a triply degenerate mode with T$_{2g}$ symmetry at the Γ-point [371]. The other 3 optical modes at the Γ-point (see Figures 7 and 8) have symmetry T$_{1u}$ and are not Raman active, but infrared active. Point defects contribute to the emergence of new vibration modes, some of which are Raman active [372, 373, 374]. The vibrational structure of anion vacancies, interstitial and cation substitution have been analyzed in fluorites using the Green's function approach for lattice dynamics [370, 372, 373, 374]. Group theory considerations are used to determine the symmetry of the defect structure and its vibration modes [370].

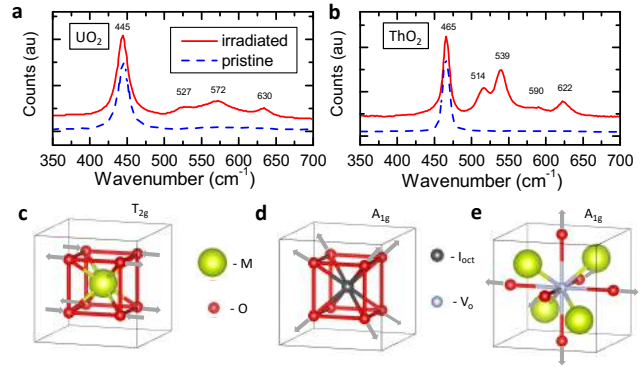

**Figure 16.** Raman spectra in pristine and irradiated (a) UO$_2$ and (b) ThO$_2$. (c-e) Cartoon representation of the vibrational modes in perfect and defected fluorite systems. Data in (a) and (b) adapted with permission from [368]. Copyright 2015 Elsevier B.V.

Figure 16a depicts the Raman spectra in proton irradiated UO$_2$ and Figure 16b depicts the same for ThO2 [29]. Observed defect peaks share features similar to ones in CeO$_2$. In CeO$_2$, the peak at 560 cm$^{-1}$ is attributed to oxygen vacancies and the peak 590 cm$^{-1}$ is associated with substitutions on the cation sublattice [371, 375]. In UO$_2$, peaks at 630 cm$^{-1}$ have been attributed to cube-octahedral clusters [376], while the peak at 570 cm$^{-1}$ corresponds to a mode having T$_{1u}$ symmetry that is Raman activated in the presence of disorder [377]. Another feature observed in irradiated fluorites is a broadening and peak shift of $T_{2g}$ owing to disorder-induced phonon confinement effect [277, 371]. Assignment of the defect peaks in irradiated $ThO_2$ is pending further studies [29].

### 3.1.1.3 Correlative Characterization

It is important to note that the positron and optical characterization methods described above have unique advantages and limitations. While PALS can size-discriminate from single vacancy-type defects through di-vacancy to larger scale clusters, it is not sensitive to interstitial-type defects. In contrast, optical spectroscopy can provide information about both vacancy- and interstitial-type defects, but its ability to size-discriminate is somewhat limited. There are thus advantages in combining these techniques. Optical, photo-luminescence, and Raman spectroscopies can successfully detect oxygen vacancies while PAS directly measures cation vacancies and the defect complexes of vacancies and impurities. Combining more than one technique such as PAS and Raman spectroscopy can thus be effective in revealing the variety of different types of defects present in actinide oxides [378]. For example, applying PAS and optical spectroscopies can solve the challenge in assigning absorption/luminescence peaks to a specific type of defect. For instance, certain emission (luminescence) peaks in oxides may be due to donor-acceptor recombination. Detecting cation vacancies by PAS can often reveal the origin of such luminescence peaks.

### 3.1.2 Intrinsic point defects: characterization and modeling

For homogenously distributed intrinsic defects, standard positron techniques can also be used to characterize vacancy-type point defects and small clusters. PAS measurements have been applied to characterize defects in oxidized uranium metal [379], $UO_2$ and $ThO_2$ powders [380, 381, 382], and doped $UO_2$ [383]. Changing U concentrations strongly modifies the density and size of U vacancies as revealed from PALS and DBS measurements. There is strong evidence that $UO_2$ doped with metal oxides can increase the grain size in fresh fuel, leading to enhanced fission gas retention [10, 11, 12, 13]. However, it is not clear what influence doping has on intergranular intrinsic point defects. From a phonon transport perspective, the presence of intrinsic point defects is important as they effectively scatter phonons. Arborelius et al. used PALS to characterize point defects in Cr and Al doped $UO_2$ [383]. The PALS measurements presented in Figure 17 reveal that doping with Cr and Al using the advanced doped pellet technology (ADOPT) leads to a large number of vacancies and complete positron trapping.

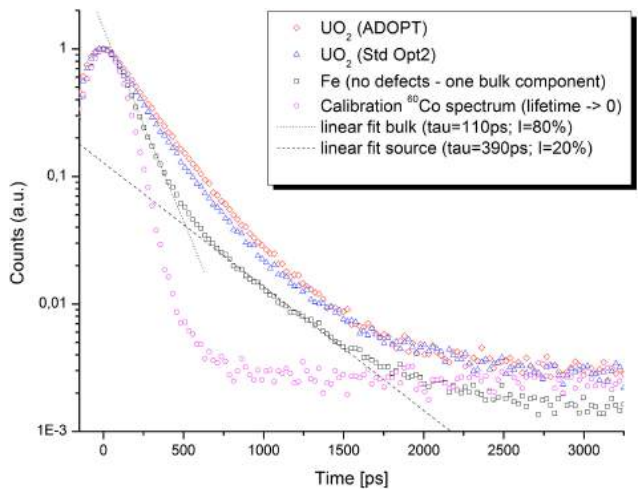

**Figure 17.** Positron lifetime spectra measured for doped and undoped $UO_2$ fuel samples, compared with the spectrometer resolution (Co-60 spectrum) and defect free Fe metal. The doped sample is referred to as the ADOPT sample (Advanced Doped Pellet Technology), the undoped sample is referred to as the Standard Optima 2 sample (Std Opt2). Reproduced with permission from [383]. Copyright 2017 Chollet et al. (Published by EDP Sciences).

X-ray diffraction (XRD) has proven to be a useful tool to shed light on the nature of point defects in fluorite oxides. Using Vegard's law, which relates the lattice constant to radius of each ion, the impact of oxygen vacancies in hypostoichiomtric $CeO_{2-x}$ and trivalent lanthanide (Ln) doped $Ce_{1-x}Ln_xO_{2-y}$ on lattice constant have been well documented [384]. Similar analysis has been applied to Ln-doped and nonstoichiometric $UO_2$ [385]. Ionic radii for different charge states of Ce, U, and Ln agree well with those tabulated [386]. However, because X-ray diffraction provides an average lattice constant, it is often difficult to tell if off-stoichiometric defects are isolated point defect or if they combine into larger clusters or even precipitate into different phases. To address these shortcomings, researchers have started to combine XRD with first principles models to delineate between different scenarios. Recently ab-initio methods have been implemented to further investigate impact of the charge state of defects on the relaxation volume of the defected cell, which is expected to be related to the radii of associated ionic representation of defects [387]. Furthermore, more careful analysis of the lattice constant in non-stoichiometric $UO_2$ considering a possibility of different defects' charge states using thermodynamics has suggested that uranium vacancies are absent in $UO_{2+x}$ [388].

The impacts of defects in nonstoichiometric and doped $UO_2$ on Raman spectra has been investigated in several studies [377, 389, 390]. Raman spectra of the pristine samples exhibit two primary peaks. The first peak at 445 cm$^{-1}$ is attributed to $T_{2g}$ symmetry phonons also visible in neutron scattering measurements. The second peak at 1150 cm$^{-1}$ is the 2$^{nd}$ order Raman peak corresponding to second harmonic of LO mode located at 575 cm$^{-1}$ and thus denoted as 2LO peak. Upon introduction of trivalent lanthanides into $UO_2$, a broad feature centered around 560 cm$^{-1}$ has been reported [377]. Detailed analysis of the broad feature suggests a presence of two closely located peaks at 540 cm$^{-1}$ and 575 cm$^{-1}$. The first has been attributed to oxygen vacancies emerging due to the trivalent nature of the dopants and the latter is attributed to Raman activation of LO mode caused by symmetry breaking. Oxygen rich $UO_2$ as well as oxidized $U_{1-x}Ln_xO_{2-y}$ exhibit defect peaks, the most prominent of which is located at 630 cm$^{-1}$ and is attributed to oxygen interstitials forming cuboctahedral clusters such as in $U_4O_9$ [389, 390]. All these studies suggest that presence of the peaks can be used as an indication for a particular defect type and peak intensity can be related to defect concentration.

The above examples illustrate that the state of defects in oxides can be more complex than just dispersed vacancies, self- or impurity-interstitials, and substitutional impurities. The presence of point defects is also often associated with other electronic carriers such as electrons and holes. Thermodynamic modeling of equilibrium defect disorder can capture these carriers simultaneously with the atomic defects themselves. However, under irradiation, this goal is far from complete. Here we review the equilibrium point defect disorder calculations focusing on $UO_2$ and $ThO_2$.

$UO_2$ can exist in the form $UO_{2\pm x}$, where $x$ denotes the deviation from the reference stoichiometric state $UO_2$. The finding that $UO_2$ can be off-stoichiometric agrees with experimental data of oxygen partial pressure against O/U ratio published by Gúeneau et al. [391] where O/U ratios are calculated up to 2.2 for temperatures ranging from 800 K to 2500 K. As for $ThO_2$, the phase diagram reported in Agarwal and Parida [392] based on published data [393, 394, 395] shows the oxide system to exist as a hypo-stoichiometric oxide above 1300 K. Past work by Murphy et al. [396] and Bergeron et al. [397] investigated the off-stoichiometric properties of $ThO_2$ by density functional theory (DFT) simulations using the CASTEP code and through thermodynamic modeling, respectively, showing $ThO_2$ to exhibit hypo-stoichiometry.

Point defect disorder models for oxides predict the concentrations of point defects, electrons, and holes under prescribed temperature and oxygen pressure conditions. A typical model is based on minimizing the Gibbs free energy of the defective oxide, subjected to the constraint of electroneutrality and prescribed temperature and oxygen pressure. The results of this approach, adapted from Schmalzried [398], are discussed below [149, 399].

Common Brouwer diagram depiction of oxides show that O vacancy defects dominate the hypo-stoichiometric regime and are charge compensated by free electrons. Meanwhile, O interstitial defects dominate the hyper-stoichiometric regime and are charge compensated by free holes in the system [400]. Findings by Hassan et al. [401] for defects in $UO_2$ show domination of U vacancy in the hyper-stoichiometric regime of $UO_2$ instead of O interstitial defects. Subsequent work by other authors then showed further investigation of the specific conditions in which this case occurs. For instance, Cooper et al. [402] found that if vibrational entropy is omitted, O interstitials are predicted to be the dominant mechanism of excess oxygen accommodation over only a small temperature range (1265 K-1350 K). Conversely, if vibrational entropy is included, O interstitials dominate from 1165 K to 1680 K (Busker potential) or from 1275 K to 1630 K (CRG potential). Below these temperature ranges, excess oxygen is predicted to be accommodated by U vacancies, while above them the system is hypo-stoichiometric with oxygen deficiency accommodated by O vacancies [402]. A similar deduction was made by Soulié et al. [403] who found that at high temperature O interstitials are dominant, either in isolated form or in clusters depending on the deviation from stoichiometry. At temperatures lower than 1300 K, they predict U vacancies to be dominant in the near-stoichiometric material. As shown in the hyper-stoichiometric regime, there is ongoing debate with respect to the defect – U vacancy or O interstitial – ultimately responsible for this stoichiometry.

For $ThO_2$, it is found that the oxide exists mostly as a hypo-stoichiometric oxide. Murphy et al. [396] predict the hypo-stoichiometric regime to be characterized by O vacancy defects, charge compensated by conduction band electrons. The simulations also highlight the importance of the poroxo-oxygen interstitial defect, which is predicted to form with a significantly higher concentration than octahedral O interstitial defects under hyper-stoichiometric conditions. $ThO_2$ is hypo-stoichiometric when subjected to high temperatures (2400 – 2655 K) [399]. It is found that the extent of hyper-stoichiometry in the $(Th,U)O_2$ system depends strongly on temperature and oxygen partial pressure [399].

In summary, extensive work has been done to study the thermodynamics of oxides. However, there is still much work ahead in this field. For instance, the study of the effect of vibrational entropy is relatively incomplete and is yet to be extended to study the thermodynamics of mixed oxides. Proper calculation of entropy of formation of defects take into consideration its temperature dependence. This, however, does not seem to be the practice followed in computing the internal energy of formation, where the defect formation energies are often computed at 0 K. Interestingly, the lattice vibrational information used to compute the entropy of formation and its dependence on temperature can be used to properly compute the same for internal energy of formation of defects.

## 3.2 Irradiation induced defects

The complexity of the radiation environment associated with nuclear fuels brings about a rich spectrum of defects. This includes defects created directly by fast neutron damage [404], by high-energy fission products [329], defects created by alpha decay of radioactive uranium isotopes [329, 405], and defects created by beta and gamma decay. However, damage caused by high energy fission fragments is the most significant damage source by orders of magnitude. After the damage event, the lattice contains isolated fission product atoms, point defects, and to a lesser degree small defect clusters [302]. Over time and at temperature, these defect populations evolve to form a rich spectrum of defect from point defects to large clusters. As a surrogate to damage caused by neutrons, ion beam irradiation has proven to be a convenient approach to separate the effects of specific defects on thermal transport. Proton (and He) beam irradiations provide a relatively flat damage profile over tens of micrometers (and close to 10 µm), respectively, using ion energies of a few MeV, which is sufficiently thick for modern thermal transport measurements [29, 42, 315, 406, 407]. Moreover, proton irradiation has been shown to be an effective means for producing irradiated microstructures comparable to those produced by neutron irradiation [408]. This is a very rich field of research and providing a comprehensive, historical review is well beyond the scope of this article. Instead, in this section we present a review of main finidings and open issues, with the intent of identifying new opportuinites for impactful research.

### 3.2.1 Characterization techniques

In section 3.1.1, we discussed two under-utilized techniques for characterization of point defects. Here we focus on two techniques, analytical TEM and atom probe tomography (APT), that show promise for characterizing of larger scale defects that form due to clustering. For irradiated metals, these two characterization techniques are starting to be used in a correlative fashion. Extending correlative studies to oxides is an area of research with tremendous opportunities.

#### 3.2.1.1 Analytical transmission electron microscopy

Conventional TEM has been widely used to study extended defects, dislocation loops and cavities in irradiated actinide oxides and their non-radioactive surrogates. A variation of conventional TEM, scanning transmission electron microscopy (STEM), involves raster scanning a tightly focused (~0.1 nm) electron beam over a sample to form an image. This configuration makes STEM suitable for analyzing local chemistry using spectroscopy methods. Electron-energy loss spectroscopy (EELS) and energy dispersive X-ray spectroscopy (EDS), are powerful tools for imaging radiation induced defects, phase changes, and small-scale changes in composition. EDS, a relatively simple technique to apply, is well suited to measure elemental composition and has been used extensively to characterize phases and fission products in nuclear fuel [328, 409, 410]. However, EDS is not capable of measuring changes in stoichiometry which is required for understanding how interfaces, such as grain boundaries, influence local stoichiometry. It is thought that interfaces that are decorated with defects can have an important role in determining thermal conductivity of fuels.

EELS has been successfully used in identifying oxidation state of U and Ce in their oxides. Phonon scattering at interfaces such as grain boundaries may significantly affect the thermal transport of materials. The grain boundary chemistry in $CeO_2$ has been revealed using STEM/EELS mapping of the Ce $M_{4,5}$ edge, which reflects the transitions from the Ce 3$d$ → 4$f$ states [407, 411, 412, 413, 414]. The relative intensities of $M_4$ and $M_5$ are characteristic of the 4$f$-shell occupancy and have been used to determine oxidation state of Ce and O vacancy concentration. The O vacancy concentration at grain boundaries depends on the boundary character, and the more distorted boundary, the more oxygen vacancies it forms. Similarly the U $M_{4,5}$ edge, which results from the transitions from the U 3$d$ → 5$f$ states has been used to characterize the oxidation state of U [415, 416]. Compared to the Ce $M_{4,5}$ edge, the U $M_{4,5}$ edge at higher energy loss has a lower signal, thus requiring long collection times to obtain an appropriate signal-to-noise ratio and subsequently presenting experimental difficulties due to beam drifting. This leads to a reduced spatial resolution, and the overall ability to measure chemical changes at the nanometer and atomic scale is affected. Recently, atomic-scale EELS for U using the U $M_{4,5}$ edge has been achieved thanks to large probe convergence angles, high currents and small probe sizes in modern aberration corrected microscopes [417].

*3.2.1.2   Atom probe tomography*

APT relies on field evaporation of the specimen to produce a three-dimensional (3D) reconstruction containing atom-by-atom position-sensitive composition information. The use of a laser (as opposed to the conventional voltage APT mode) is necessary to induce field evaporation in semiconducting and insulating materials [418]. Effective field evaporation in oxides requires uniform laser absorption and efficient dissipation of thermal energy, which are highly dependent upon the optical properties, thermal properties, and surface chemistry of the oxide [419]. Field evaporation parameters must therefore be tailored or optimized in each oxide for accurate chemical quantification [420, 421, 422, 423, 424, 425, 426]. Poorly optimized field evaporation parameters can produce thermal artifacts, surface diffusion, and preferential evaporation that ultimately degrade the accuracy of APT quantitative compositional analysis [427].

There is no clear trend in optimal APT field evaporation parameters for nuclear fuel oxides and their surrogates. For example, $CeO_2$ can yield sufficient field evaporation with low energy laser pulses [422], while $UO_2$ requires moderate laser energies [420], and $ThO_2$ requires high laser energies with a low specimen temperature [255]. In $CeO_2$, higher laser energy and low base temperature results in degradation of mass spectrum resolution [422]. Recent work has used Bayesian expectation-maximization statistical methods to assign ion identities to data convoluted in thermal tails, enabling accurate quantification of the chemical composition of thermally-affected mass spectra in $ThO_2$ [255]. While challenges remain for APT characterization of oxides, there exists considerable opportunity for using APT in a correlative fashion with analytical transmission electron microscopy.

*3.2.1.3   Correlative TEM-APT characterization*

While TEM and APT have several advantages for characterizing composition at the nanoscale, each characterization technique has well known limitations [428, 429, 430, 431, 432]. TEM/STEM is excellent in crystallographic/structural analysis. Grain orientations, phase structure, and defects can be identified or visualized using electron diffraction and high-resolution imaging techniques. The grain orientations may also be measured by APT, but the spatial resolution of APT datasets is determined by the regularity of the field-evaporation sequence. The dependence of field-evaporation on grain orientations and on the local bonding state may reduce the spatial resolution in the reconstructed 3D atom maps, especially at locations of defects or interfaces [428, 433].

STEM/EDS has been routinely used to quantify local chemical compositions at the nanoscale. The elemental sensitivity of EDS depends on the atomic number of the elements. The quantification of light elements, such as O and N is based on low-energy X-rays, K peaks and their *k*-factors depend on X-ray absorption. Because of inaccuracy of the absorption correction due to self-absorption of low energy X-rays, their poor detection efficiency, and the low fluorescence yield, the quantification of light elements is very challenging. Compared to EDS, EELS tends to work better for relatively low atomic numbers. However, quantification of EELS spectra relies on obtaining single scattering profiles, which can only be realized when very thin samples are used. In thicker samples, multiple inelastic scattering changes edge shapes and intensities and therefore quantification errors can be introduced [434]. When the size of the target defect is smaller than the sample thickness, the measured composition is an effective composition that is an average over the sample thickness and may not be representative of the target defects.

The quantitative element analysis at the near-atomic scale using APT is rather robust because ionic species, in most cases, can be simply identified according to their mass-to-charge ratio, while there are intrinsic challenges associated with quantification of O especially from nano-sized features. It is well known that the concentration of O in materials is strongly dependent on the analysis parameters of APT, such as laser wavelength, laser pulse energy, and field condition of the specimen [435, 436]. For example, at high laser energies, metal atoms are preferentially evaporated where O atoms diffuse and desorb, due to inhomogeneous thermal absorption of the laser, leading to inaccuracies in the measured composition. In addition, APT only measures chemical composition in a very localized volume ~$10^6$ nm$^3$, in comparison to the volume ~$10^{10}$ nm$^3$ used for STEM/EDS/EELS measurement in TEM [430].

Spatially correlated TEM and APT studies are of significance to understand the radiation effects in nuclear materials. Recently researchers have investigated combining APT and STEM, with the advantages of one offsetting the limitations of the other technique. Wang et al. [432] quantify chemical compositions around nano-voids in NiCoCr solid solution alloys. Figure 18 shows an example of correlative imaging using STEM and APT. Lach et al. [431] characterized the radiation-induced segregation around grain boundaries and dislocation loops, together with precipitation in a 304 stainless steel. These correlative studies thus far have been applied to metals, thus while there is great potential for extending correlative studies to oxides, there exist considerable challenges.

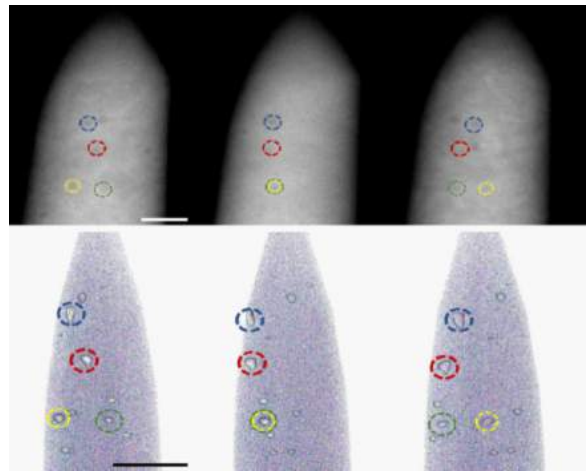

**Figure 18.** STEM and APT images acquired for a NiCoCr needle-shaped specimen at different rotation angles. The scale bar for both sets of images is 40 nm. Adapted with permission from [432]. Copyright 2020 Wang et al. (Published by Springer Nature).

### 3.2.2 Irradiation-induced point defects

At near room temperature and low damage ranges (<0.1 dpa), the formation of extended defects is limited due to limited diffusion. Several reports have applied Raman spectroscopy to characterize point defects in irradiated fluorite oxides [437, 438, 439]. One notable study performed depth resolved radiation damage upon irradiation of $UO_2$ by 25 MeV $He^{2+}$ ions [438]. The ion impacted region of the cross-sectional scan exhibited a reduction in intensity of the $T_{2g}$ peak at 445 $cm^{-1}$ and emergence of new defect peaks at 527, 574, and 634 $cm^{-1}$. The peaks at 527 and 634 $cm^{-1}$ peaks were attribute to oxygen vacancies and interstitials, respectively. The peak at 574 $cm^{-1}$ has been attributed to activation of the Raman-forbidden LO peak caused by symmetry breaking. A cross-sectional line scan along the incident direction of the implanted He ions was performed, and it was shown that the intensity of $T_{2g}$ peak decreases and the intensity of the defect peak increases proportional to the amount of depth-dependent displacement damage predicted using SRIM [438]. In a more recent report, it was postulated that the LO peak is due to defects on the uranium sublattice and associated with changes of the charge state of uranium from $U^{4+}$ to either $U^{3+}$ or $U^{5+}$ [439].

Similar measurements have been performed in light ion irradiated $ThO_2$ [29, 368]. In 2 MeV proton irradiated $ThO_2$, 4 prominent peaks at 515, 535, 585 and 630 $cm^{-1}$ were identified [29]. The latter 3 are similar to the defect peaks in $UO_2$, whereas the peak at 515 $cm^{-1}$ has not been reported in other fluorite oxides. Similar peaks were observed in 21 MeV $He^{2+}$ irradiated $ThO_2$ at 514, 539, 590 and 622 $cm^{-1}$, but were not identified as Raman peaks by the authors and instead were attributed to luminescence [368].

XRD diffraction has been applied to monitor microstructure evolution in irradiated $UO_2$ [287, 440]. During isochronal annealing of neutron and alpha particle irradiated $UO_2$ different stages of defect recovery were identified. These stages were attributed to mobility of different defect types. These results enabled the determination of migration barriers and relaxation volume of Frenkel pairs, without clear differentiation between uranium and oxygen sublattices. Similar measurements have been applied to alpha irradiated $CeO_2$ and $PuO_2$ [286].

For self-irradiation studies involving doping with a radioactive isotope, traditional PAS can be used to measure vacancy type defect distribution throughout the bulk. One such example used PALS to characterize the damage generated by alpha self-irradiation in $UO_2$ [441]. This study suggested that the point defects observed using PALS correspond to uranium vacancies. From a phonon transport perspective, defects on the uranium and oxygen sublattices can influence phonon scattering in significantly different ways. In the case of high energy electron and ion irradiation, defects are generated in the bulk of the samples and traditional PALS was successfully applied to study defects in $CeO_2$ and $UO_2$ using this approach [442]. Additionally, coupling experimental results to a density functional theory description revealed detailed information about the structure of vacancy type defects in $UO_2$. Wiktor *et al.* [443] used this approach to reveal that the dominant irradiation induced defect in 45 MeV alpha-irradiated $UO_2$ is a neutral bound Schottky defect of neutral $V_U+2V_O$ tri-vacancies.

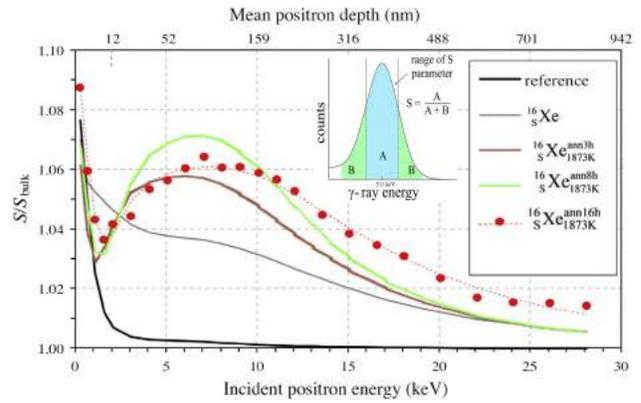

**Figure 19.** Defect parameter S as a function of depth in $UO_2$: Reference sample, Xe implanted sample, and post Xe-implantation annealing samples. (Inset) Schematic illustrating how S-parameter is extracted from the 511 keV peak in DBS. Adapted with permission from [444]. Copyright 2013 Elsevier B.V.

In the early 1990s, researchers first demonstrated the applicability of using slow positrons to perform depth resolved studies of ion implanted $UO_2$ [445]. More recent studies applied variable energy DBS (VEDBS) to characterize vacancy type defects in electron and light ion irradiated $UO_2$ [446, 447, 448]. In one study involving electron irradiation of $UO_2$, results reveal that of U-related vacancies are formed when stepping from 1 MeV irradiation to 2.5 MeV irradiation [446]. In another study involving 1 MeV He implantation, it was found that the nature of the irradiation induced vacancy type defects does not change with fluence [447]. VEDBS was also applied to investigate the depth distributions of defects in Xe implanted $UO_2$. As an example, Figure 19 [449] depicts the defect parameter S extracted from DBS measurements as a function of depth for unirradiated $UO_2$ (reference sample), a Xe implanted sample, and a post-Xe irradiated annealing sample. The large nonlinear increase in S indicates the formation of Xe-bubbles after annealing.

### 3.2.3 Dislocation loops

At intermediate temperatures and low damage levels, ion irradiation may create both point defects and dislocation loops [28, 29, 315]. Detailed TEM and atomic resolution STEM imaging shows the dislocation loop characteristics of single crystal $ThO_2$ (Figure 20) [303]. They are interstitial faulted loops or Frank loops with Burgers vector, $\vec{b}$ of $1/3\langle 111 \rangle$ and a habit plane of $\{111\}$. The large, extended strain field around faulted loops scatter phonons significantly, leading to a larger reduction of thermal conductivity compared to perfect loops [315]. In neutron and ion-irradiated $UO_2$, $ThO_2$ and $CeO_2$, both faulted loops with $\vec{b}$ of 1/3<111> and perfect loops with $\vec{b}$ of 1/2 <110> are found. MD simulations [298,303,326] showed that mainly 1/3<111> Frank loops are energetically favorable when the loop size is small (several nm), and that 1/2<110> perfect loops are preferred when the loops grow. Thus, there must be an unfaulting process of loops under irradiation and the formation of perfect loops can be ascribed to a reaction between a Frank partial and a Shockley partial [298, 303, 326].

However, this unfaulting process during loop growth has not been directly observed in actinide oxides and their surrogate, $CeO_2$.

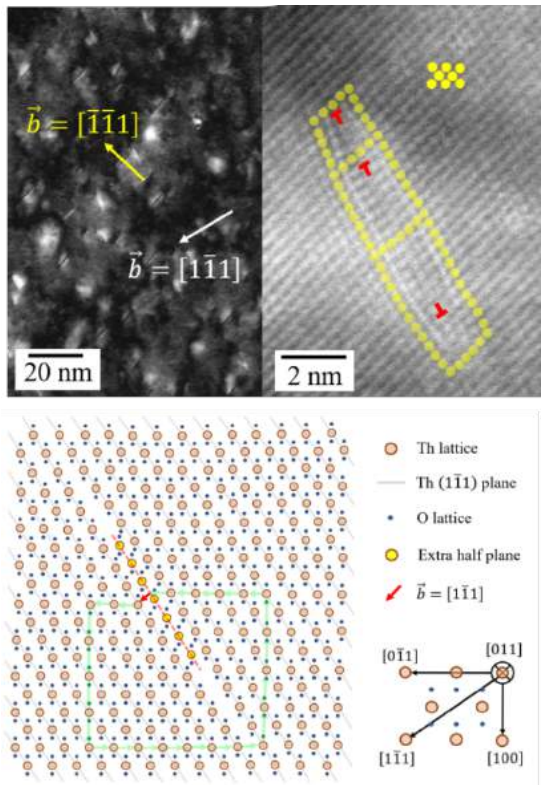

**Figure 20.** Weak beam dark field TEM image (top left), atomic resolution STEM image (top right) dislocation loops in ThO$_2$ and schematic representation (bottom) of $1/3\langle 111\rangle$ Frank partial dislocation loop showing extra half plane and Burgers vector $\vec{b}$ when seen along [011] zone axis. Adapted with permission from [303]. Copyright 2021 Elsevier B.V.

In the past decade, the formation and evolution of extended defects, dislocation loops and bubbles under heavy ion irradiation have also been well studied using in situ ion irradiation facilities such as the Intermediate Voltage Electron Microscope (IVEM) at Argonne National Laboratory [295, 296, 312, 314] or JANNuS-Orsay in France [309, 317, 318, 324, 325, 326]. The size and density of dislocation loops increase with the irradiation dose, before the density saturates and then dislocation loops start to transform to dislocation segments and tangled networks at higher doses through coalescence and coarsening mechanisms [295, 296, 312, 313]. Both dislocation loops and cavities can nucleate under heavy ion irradiation at room temperature [311, 314, 318], which is different with light ion (e.g. proton) irradiation. At a low irradiation temperature, some dislocation loops can nucleate directly at the clusters produced in cascades and such a process does not require U interstitials and U vacancies to be diffusive. Electrons or light ions such as protons will produce damage as isolated Frenkel pairs or in small clusters while heavy ions and neutrons produce damage in large clusters. The direct nucleation of extended defects at small clusters is more difficult compared to large clusters, which may explain why loops do not form under proton irradiation at room temperature. For in situ irradiation, it is noted that the TEM samples are very thin (∼100 nm) and the surface is a strong sink for irradiation induced defects. For those samples prepared by focused ion beam (FIB) milling, FIB damage may contribute to the defect evolution.

### 3.2.4 Cavities and Precipitates

Defects that are large in comparison to the wavelength of the phonons responsible for thermal transport can have a significant impact on thermal conductivity. While voids and bubbles can be formed under heavy ion irradiation, precipitates can form due the coalescing of impurities, doping atoms and fission products. Formation of these large-scale defects typically requires long range diffusion. While heavy ion irradiation has been used to study fission gas diffusion in UO$_2$ [310, 311, 314, 317], ThO$_2$ [299] and Th$_{1-x}$U$_x$O$_2$ [300], fission gas bubble nucleation, growth, migration, and self-organization in ThO$_2$ and Th$_{1-x}$U$_x$O$_2$ is has yet to be studied in detail.

#### 3.2.4.1 Cavities

Cavities, including voids and fission gas bubbles, have been characterized extensively in UO$_2$ and CeO$_2$. Generally TEM has been used to characterize bubble sizes and distributions [308, 450, 451, 452, 453], while techniques such as secondary ion mass spectroscopy (SIMS) [454] and XAS [455, 456, 457] have been used to quantify gas pressures.

Michel et al. [452] performed one of the earliest in situ TEM studies of bubble evolution UO$_2$ irradiated using 390 keV Xe ions. Their results confirm the rapid formation of a small (less than 1 nm) bubble population. The novel result lies in the fact that the bubble concentration reaches a near asymptotic value. While providing invaluable data against which cluster dynamics models may be assessed, no clear physical explanation was found for saturation of the bubble concentration. Theoretical and experimental work is currently ongoing to elucidate this longstanding conundrum.

Using Kr irradiation, He *at al.* [311] showed the bubble size to be a weak function of ion dose but strongly dependent on the temperature. Kr bubble formation at room temperature was observed for the first time, indicating that either U vacancies are mobile at room temperature, or that Kr bubbles may directly nucleate at the vacancy clusters produced during cascades with such a process not requiring Kr and U vacancies to be diffusive. In a similar study involving Xe irradiation of UO$_2$ single crystals, it was shown that small bubbles (1-2 nm) were formed at room temperature indicating the bubble formation does not require long-range diffusion of Xe atoms nor U vacancies [314]. These studies involving Kr irradiation indicating bubble formation at room temperature are to be contrasted with another study involving gas atom precipitation in CeO$_2$ at room temperature and 600°C [295]. In this study, no gas bubbles were observed at room temperature.

In a study by Ye *et al.* [296] that looked at the effect of Xe irradiation on cavity formation in UO$_2$ and CeO$_2$, it was suggested that implanted Xe ions were trapped in vacancy clusters and aggregated into bubbles at elevated temperature. The size and density of those features are comparable to the fission gas bubbles in low burnup UO$_2$ fuel in a similar temperature range. Kinetic Monte Carlo analysis revealed that O mobility in UO$_2$/CeO$_2$ is substantially higher in hypo-stoichiometric conditions (MO$_{2-x}$) than hyper-stoichiometric conditions (MO$_{2+x}$).

Void formation has been studied using Au irradiation of thin UO$_2$ foils. Because most of the Au passes through the foil, this irradiation regime only creates ballistic damage. In one study, Onofri et al. [327] found that void nucleation is a heterogeneous process since new voids do not involve long-range diffusion of point defects and gas atoms. In a related study by Sabathier et al. [458] it was suggested that heterogeneous nucleation of voids was induced by energetic cascade overlap. Such nanovoids are likely to act as sinks for mobile fission products during reactor operation.

Researchers are now looking beyond TEM methods to characterize fission gas bubbles. One of the early studies in this area involved using APT to map the Kr profile in ion irradiated $UO_2$ as a function of depth [311]. The Kr concentration measured by APT was considerably smaller than predicted by SRIM. This was attributed to the gas being lost to the vacuum within the analysis chamber and not registered by the detector. However, using APT to directly characterize bubbles can be problematic. For instance, aberrations are associated with tomographic reconstructions near surfaces (internal and external) and may change the evaporative characteristics of atoms. As mentioned in a previous section, correlative studies that utilize STEM and APT are beginning to demonstrate promising results tied to characterization of bubbles in metallic samples. A recent study by Perrin-Pellegrino [459], extended this approach to oxides. Using APT in concert with TEM revealed the existence of nanoclusters enriched in Xe with a size and density comparable to bubble populations observed with TEM.

Coincidence Doppler broadening spectroscopy (CDBS) is another powerful tool to characterize the chemistry of gas bubbles, identify the chemical environments around vacancies and probe atomic scale precipitates [460]. Nagi et al., [461] employed it to detect the ultrafine precipitation of Cu impurities in nuclear reactor pressure vessel (RPV) steel. The work revealed that positrons can be trapped by embedded ultrafine particles forming a quantum dot-like positron states, thus enabling the detection of atomic scale precipitates. In general, correlative studies using PAS and TEM will prove useful to characterize the size, size distribution and local chemistry of defects ranging from single vacancies to clusters and large voids [462, 463]. This will be crucial to understanding defect formation at the early stages and the mechanisms of defect accumulation and recombination.

### 3.2.4.2 *Precipitates*

Precipitation in irradiated oxide fuel can have significant impact on physical properties including swelling, melting point, and thermal conductivity. Moreover, the irradiation-induced aggregation of solute species into precipitates can bring about a cooperative effect wherein the integrated impact on thermal conductivity of impurity atoms in solution differs from that of precipitates composed of the same elemental species. Characterization of precipitates in oxide fuels and surrogates is commonly performed using TEM, APT, and light source techniques.

In a recent study involving emulated spent fuel, Jiang *et al.* [464] used TEM and APT to study thin films of $CeO_2$ doped with metallic fission products. They found a uniform distribution of doped metals within the as-grown film. Pd particles of ~3 nm appeared near dislocations after He irradiation. Annealing at 1073 K in air leads to precipitates of Mo and Pd at grain boundaries and further annealing results in precipitate coarsening.

The structure of the precipitate (e.g. crystalline versus amorphous) can have a significant impact of phonon scattering and transmission through the precipitate. In addition to studies that investigate the composition of precipitates, there have been several studies that have addressed the structure of precipitates. In one study, the local atomic structure of chromium bearing precipitates in chromia doped uranium dioxide was investigated by combining micro-beam X-ray diffraction and absorption spectroscopy [465]. The XRD data indicated that the chromia precipitates contain structural disorder and μ-XAS results provide insight into the oxidation state of the chromium. In a study targeting fission gas bubble formation, Yun *et al.* [466] found solid Xe precipitates in irradiated 5% La doped $CeO_2$. The fact that these precipitates only form at very high doses, indicated a dose threshold in their formation.

Oxide fuels are also known to exhibit steep radial gradients in burnup, which can have significant implications on chemical clustering and segregation. Bachhav and coworkers [467] have recently demonstrated a novel APT-based approach to determine local burnup in irradiated fuels. This method, which relies upon isotopic ratios of $^{235}U$, $^{236}U$, $^{238}U$, $^{239}Pu$, and $^{237}Np$, provides localized understanding of U enrichment and local burnup. In a follow-on study [468], the team linked burnup to local fission product chemistry. By conducting APT at the center and on the edge of an irradiated $UO_2$ pellet, they observed burnup-related differences in the irradiation-induced nucleation of metallic fission product Mo-Tc-Ru-Rh-Pd precipitates. Precipitates were larger at the pellet edge than in the center.

### 3.3 Modeling of defect clustering

There are several methods used to model defect evolution. Molecular dynamics (MD) can be used to investigate defect evolution at very short time scales and provide useful information on initial conditions used for modeling approaches that target long time evolution. For scales of interest to the materials discussed in this review, there are two main modeling methods, kinetic Monte Carlo and rate theory models. While kinetic Monte Carlo retains spatial correlation, all possible events and their rates must be known in advance. Conversely, rate theory does not retain spatial correlation but \ exhibit predictive capability. Rate theory models have been used extensively for metallic systems; however, more development is required for application to oxide nuclear fuel. Accordingly, our focus in this section is on reviewing recent work and identifying future opportunity associated with rate theory models.

Two types of rate theory models exist in the literature to model point defect and cluster evolution in irradiated oxides. The first is a simplified model designed to yield the time evolution of point defects interacting with a mean density of features such as loops and voids at an average evolving size. The second, termed cluster dynamics, is one in which the density of clusters of all sizes evolve simultaneously with point defects. These models are briefly reviewed below.

### 3.3.1 Simple rate theory model of defect evolution

Simple rate theory models consider three processes: point defect generation by irradiating particles, mutual recombination of interstitial-vacancy pairs, and defect clustering into dislocation loops and voids [297, 328, 469]. However, for low dose proton irradiation at intermediate temperatures, void nucleation may be suppressed for kinetic reasons in some oxides, e.g., $ThO_2$, and loop nucleation dominates due to the relatively higher mobility of interstitials. A rate theory for such a situation consists of a set of ordinary differential equations [328]:

$$\frac{\partial C_i}{\partial t} = G - S_{ij} C_i C_j, \tag{5}$$

describing the time rate of change for defect type (*i,j*), including monomers $O_i$, $V_O$, $Th_i$, $V_{Th}$ and loops (L). $G$ is the generation rate for monomers and di-interstitial formation for loop nucleation. Mutual interactions between defects ($S_{ij}$) represent recombination for each ($O_i$, $V_O$) and ($M_i$, $V_M$) cation pair and di-interstitial formation for two cation interstitials. These di-interstituals act as the only source of dislocation loop nucleation. Terms $S_{iL}$ capture absorption of monomers by loops and contribute to loop size evolution [328], which is given by:

$$\frac{d}{dt}\left(\frac{\pi b}{3\Omega_0}R_L^2 C_L\right) = S_{iL}C_i C_L, \qquad (6)$$

with $R_L$ being the loop radius, b the Burgers vector, and $\Omega_0$ the volume per atom. The $S_{ij}$ rate constants depend on monomer diffusion coefficients and geometrical factors capturing atomic structure of defects. Expressions of these coefficients can be found in [328]. Figure 21 compares the predicted evolution of loop diameter and loop density in proton irradiated $CeO_2$, $ThO_2$ and $UO_2$ [29, 315, 328] to experimental measurement. It can be seen that the loop growth rate and loop concentration are inversely correlated and are proportional to the mobility of cation interstitials in this system. Faster cation mobility promotes faster loop growth, however, leads to lower supersaturated concentration of monomers [470, 471]. These results indicate that cation interstitial migration is fastest in $CeO_2$ and slowest in $ThO_2$.

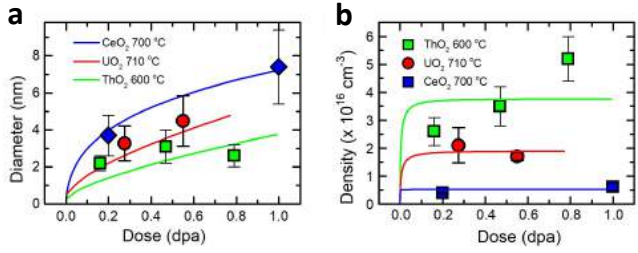

**Figure 21.** Comparison of dislocation loop evolution in proton irradiated $CeO_2$, $ThO_2$, and $UO_2$ using rate theory modeling (solid lines) to experimental measurement (symbols). $UO_2$ data adapted with permission from [28]. Copyright 2020 Elsevier Ltd. $ThO_2$ data adapted with permission from [132]. Copyright 2021 Dennett et al. (Published by Elsevier Ltd.). $CeO_2$ data adapted with permission from [315]. Copyright 2019 The American Ceramic Society.

### 3.3.2 Cluster dynamics approach to defect evolution

The formation of extended defects such as dislocation loops and voids under irradiation results from the diffusion and interaction of vacancies and interstitials produced by atomic displacements. This dynamic generation and interaction process may be studied using cluster dynamics (CD), an approach based on chemical reaction rate theory [472,473]. In this approach, the temporal evolution of the density of defect clusters in size space is controlled by the emission and absorption of mobile point defects generated in the collision cascade. The CD approach assumes a uniform distribution of defect clusters of varying sizes and considers that the above-mentioned interactions take place continuously in space and time. CD approaches have been used extensively in the study of radiation-induced defect generation and evolution, including spatially-dependent approaches which take into account the local configuration of defects of different sizes [474, 475, 476, 477]. However, these spatially-correlated approaches have, so far, found wide use in metallic systems, while mean-field approaches lacking spatial correlation have been applied more readily to oxide fuel materials [478, 479, 480]. In either case, the rate at which clusters of a specific size accumulate in the material is obtained as a sum of their generation and depletion rates due to the absorption and emission of point defects by similar or different-sized clusters. This approach has been used extensively to study defect evolution in materials like Fe [481], stainless steel [482, 483], and Zr [484], as a robust multiscale method that takes input from atomistic simulations and successfully predicts the evolution of defect densities over extended length and time scales. In multicomponent systems, the CD framework has been used to study precipitation kinetics in binary [485, 486] and ternary alloys [485].

Irradiation of oxides generates charged point defects of different species in the displacement cascade; for instance, in irradiated $UO_2$ the point defects obtained are O-self interstitial atoms (SIAs), U-SIAs, O-vacancies and U-vacancies denoted with their nominal charge states as $O_i^{-2}$, $U_i^{+4}$, $V_O^{+2}$, and $V_U^{-4}$, respectively. The point defects can also have non-nominal charges; for instance, $U_i^{+3}$, $V_U^{-3}$, etc. depending on the stability of individual ionic species in the irradiated oxide [401]. To make a CD approach tractable, some limits must be placed on the charges and spectrum of cluster sizes considered. Skorek et al. [487] studied the evolution of fission gas bubbles in $UO_2$, assuming Schottky defects, i.e, $U_i - 2O_i$ and $V_U - 2V_O$ defect complexes are generated in the cascade rather than isolated SIA and vacancies. They further assumed that Schottky defect complexes are mobile and hence migrate and agglomerate into clusters having stoichiometric compositions, preserving the charge neutrality of the matrix. The clusters trap fission gas atoms, which are otherwise highly mobile, and impede their rate of escape to grain boundaries. Though the simplified CD model was able to predict the experimentally observed release rate of the fission gas, it did not allow for the matrix to become hyper-stoichiometric under irradiation. A more detailed mean-field approach was later taken by Khalil et al. by allowing for the asymmetric (non-stoichiometric) generation of point defects of different species in the displacement cascade due to their varying displacement energies, and also allowing for the possibility of point defect agglomeration into clusters having non-stoichiometric compositions [488]. They defined the cluster composition space (CCS) for loops, as shown in Figure 22a, by counting the respective point defects i.e, $O_i^{-2}$, $U_i^{+4}$ for SIA loops and $V_O^{+2}$, $V_U^{-4}$ for vacancy loops, in the fluorite crystal structure of $UO_2$. The mesh points in Figure 22a represent all possible defect compositions, where each point $(m,n)$ indicates vacancy clusters with $m$ $V_U$ and $n$ $V_O$ and can either move horizontally or vertically by interaction (absorption or emission) with U and O defects, respectively. Figure 22b gives the time evolution of SIA clusters at 800°C and a dose rate of $9.2\times10^{-4}$ dpa/s, showing off-stoichiometric loop growth at higher doses. Vacancy loops evolve similarly, favoring off-stoichiometric compositions with larger loop sizes. This modified CD model gives detailed insight into the wide range of stoichiometry of defect clusters and their evolution with time. However, a significant source of error in this model may lie in the assignment of binding energies for each defect species. These energies are routinely assigned for defect clusters in pure materials using molecular dynamics, although in reality these too may change as a function of stoichiometry.

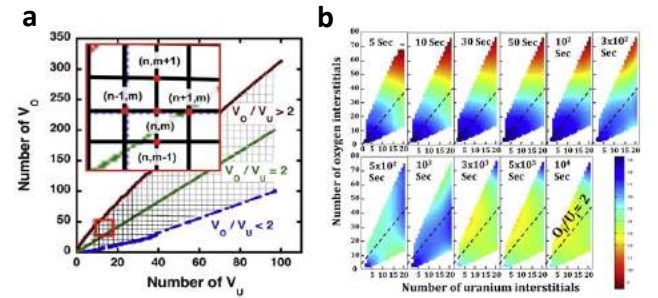

**Figure 22.** (a) CCS for vacancy loops. The blue and brown lines represent the minimum and maximum number of $V_O$ that can exist in a cluster with a given number of $V_U$. The middle green line represents the 2:1 ratio of $V_O$ to $V_U$. (b) Time evolution of SIA cluster density. Adapted with permission from [488]. Copyright 2017 Elsevier B.V.

While the approach described above relies on the input of defect monomers only, MD simulation of primary radiation damage in the mixed $Th_{1-x}U_xO_2$ oxide system shows that small vacancy and SIA clusters may be produced directly in the damage cascade [302]. Thus, in modified next-generation CD models for oxides, the production rate of small clusters should be provided as input parameters similar to monomers. Also, the strict assumption of monomers being the only mobile species in the general CD model [478, 488] needs to be discarded. Instead, the spatial and temporal evolution of loops and voids, driven by the diffusion of small clusters should be targeted directly. Such a modified CD model can be extended to introduce new terms in the cluster evolution equations that govern the migration of vacancy and SIA clusters. As reviewed by Kohnert and coworkers [489], spatially resolved cluster dynamics has proven useful in modeling the evolution of complex metallic microstructures resulting from defect diffusion. Such spatially-correlated methods are, indeed, the next logical progression for the CD approach applied to oxide fuel materials.

## 3.4 Grain Boundaries

Here we first look at microstructure evolution in the presence of isolated grain boundaries. Characterization of isolated boundaries is a notably easier task than characterizing boundaries in nanocrystalline materials. Moreover, new experimental tools have enabled studies that target the impact of isolated boundaries on thermal transport. These studies of thermal transport can then be compared with prediction in a one-to-one fashion. The ultimate goal for materials scientists will be to translate understanding of isolated boundaries to nanocrystalline restructured oxide fuel.

### 3.4.1 Intrinsic Grain Boundaries

Grain boundaries can significantly influence microstructure evolution in the environments in which nuclear fuels operate. In addition to providing sinks and sources for defects, grain boundaries can accelerate mass diffusion. Numerous studies have identified accelerated mass transport pathways along grain boundaries in $UO_2$ for self-diffusion (i.e. U and O species) [490, 491, 492, 493, 494] and He [495]. In general, grain boundaries in $UO_2$ have a higher concentration of mobile charge carriers than the bulk, leading to enhanced oxygen transport along grain boundaries [493]. The result is highly anisotropic ionic diffusion localized along grain boundaries, dependent upon temperature, local structure, and grain boundary character [493, 496].

The impact of grain boundary structure in $UO_2$ on segregation has also been studied with MD simulations [497, 498]. It was found that high-angle grain boundaries are energetically favored to accommodate Xe versus low angle boundaries. Motivated by this computational study, Valderrama et al. [499] investigated segregation of Kr at grain boundaries in ion irradiated $UO_2$. It was found that high angle grain boundaries contain substantially more Kr than low angle grain boundaries. They also found that excess Kr grain boundary segregation leads to bubble formation. In another study targeting bubble evolution in Kr-irradiated $UO_2$ during annealing [310], it was found that the size of intergranular bubbles increased more rapidly than intragranular ones, and bubble denuded zones near grain boundaries formed in all the annealed samples at 1000 to 1600°C. The area density of strong segregation sites in the high-angle grain boundaries is much higher than that in the low angle grain boundaries. The geometric constraints for intergranular bubbles, restricted to planar surfaces, results in the spacing between these large bubbles being much smaller than spacing between intragranular bubbles [500]. This deviation from a random distribution due to the cooperative behavior of pre-existing grain boundaries and mass transport can have a profound impact on thermal transport properties [501].

Computational studies have confirmed that the grain boundary character has a large impact on both intergranular bubble nucleation and growth. In a study involving the segregation of Xe to a select group of grain boundaries in $UO_2$ [496], it was suggested that segregation properties of grain boundaries may result in different nucleation rates for fission gas bubbles. This was investigated in more detail using random-walk Monte Carlo simulations [502], showing that the segregation energy of a grain boundary has a large impact on its intergranular bubble nucleation rate. It was also shown that the interconnection rate of intergranular bubbles is dependent on the grain boundary character, since the grain boundary energy dictates the contact angle of the bubble with the grain boundary [503].

Similarly, grain boundary segregation of dopants in $CeO_2$, as a model material or fuel cell component, is dependent upon grain boundary character [504]. Li et al. [505] used APT to investigate Y segregation in doped $CeO_2$. They found that Y segregates at grain boundaries but is uniformly distributed within the grains. In a similar study involving Gd-doped $CeO_2$ [506], it was found that Gd strongly segregates at the grain boundary and within nano-sized domains within the grains. $CeO_2$ accommodates aliovalent dopants such as Gd and Mn through a dopant-defect complex. First principles studies from Dholabhai et al. [504] suggest that the arrangement of these complexes at grain boundaries strongly influences the stability of oxygen vacancies. However, Sun et al. [507] used atomistic simulations to show that dopant arrangements at dislocations in $CeO_2$, in concert with the strain fields around the dislocation, can slow oxygen ion transport along the dislocation. The reduced ionic conductivity in Gd-doped $CeO_2$ is supported by experimental evidence of grain boundary segregation of Gd and the formation of nanosized Gd-rich domains within $CeO_2$ grains, as observed by laser-assisted APT [506].

### 3.4.2 Restructured grain boundaries

The formation of HBS in commercial reactor fuel was first observed in the late 1950s and early 1960s [508, 509]. This area of research remained relatively untouched until the mid-1980s when researchers readdressed the subject in a response to the industrial trend towards increasing fuel burn-up. This structure, characterized by the formation of submicrometer-sized grains and micrometer-sized intergranular fission gas bubbles, forms near the rim of the fuel pellet where the temperature is relatively low, and the damage accumulation is relatively high. The high damage rate is due to creation of fissile $^{239}$Pu caused by resonant absorption of epithermal neutrons by $^{238}$U. In this region, the fission product concentration and damage levels are significantly higher than average values. Originally, it was thought that the formation of the rim structure would negatively impact thermal transport, leading to high centerline temperatures and enhanced fission gas transport [510]. In the early 2000s, Spino and Papaioannou [511] speculated that formation of the rim structure would have a net positive impact on thermal conductivity by removing intragranular, uniformly distributed defects (point defects, dislocation loops, and fission products). In 2004, Ronchi et al.[5] provided a systematic study of the impact of the HBS on thermal conductivity using a combination of standard commercial reactor fuel and fuel samples irradiated in a test reactor. They found that the formation of the HBS has a net beneficial impact on thermal conductivity, with an intrinsic conductivity gain of 6-7% over fuel irradiated at intermediate temperatures where the HBS did not form. Bai et al. used simulations to show that the likely cause of this thermal conductivity gain is the reduction in fission gas within the matrix due to grain boundary segregation [512]. There have been several research efforts aimed at understanding the formation mechanisms and the impact on physical properties using samples exposed to neutron irradiation. Comprehensive reviews of these studies are provided by Ronndinella and Wiss [6] as well as by Wiss et al. [513].

Here we focus on ion irradiation studies aimed at investigating the formation mechanisms responsible for the formation of the HBS [510, 514, 515, 516, 517]. There are two broad groups of theories for the formation of the HBS [6]. The first involves nucleation and growth either at highly defective sites (e.g. dislocation tangles) or at amorphous regions near fission tracks. The second involves polygonization, where grains subdivide along dislocation networks produced by radiation damage. It is generally thought that nucleation and growth will lead to high angle grain boundaries and polygonization will lead to low angle grain boundaries. In an effort to isolate the formation mechanism, researchers in the 1990s started to investigate dislocation produced by heavy ion irradiation. One of the earlier studies involved Xe irradiation of single crystal $UO_2$ samples to a dose of $1\times10^{17}$ ions/cm$^2$ [510]. Using Rutherford channeling and X-ray diffraction measurements, it was found that a fine grained polycrystal was formed having a misalignment between grains of a few degrees. Another study in support of the polygonization mechanism involved investigation of ion tracks in $UO_2$ irradiated with 100 MeV Zr and Xe ions [516]. In this study it was argued that overlapping fission tracks led to the formation of subgrains at relatively low temperatures. A more recent study used 84 MeV Xe ion irradiation and found that grain polygonization due to accumulation of radiation-induced dislocations caused the fine grains to form [518]. There are many opportunities for new research involving ion studies aimed at unraveling the formation mechanisms of the HBS. For instance, the role that fission products play in the formation of the HBS could be addressed by using doped samples. Another area that may provide rich physics involves studies that target the role of existing grain boundaries. This is especially interesting in light of a recent study [519] using modern electron microscopy techniques that suggest that the formation of low-angle boundaries initiate at existing high angle boundaries.

### 3.5 Outlook

Irradiation produces a wide range of defects in oxide fuels. A clear understanding of the role of these defects requires isolating the individual defects and studying them individually. While proton irradiation at low temperature can generate point defects and very small clusters, intermediate temperature irradiation can generate loops and point defects. Higher temperature annealing, however, can be used to eliminate point defects and control the size distribution of loops via coarsening. In the presence of gaseous species such as He, bubbles may also evolve in specimens that have loops. As such, a hierarchy of irradiation experiments can be used to investigate the different defects one by one, for comparison with computational modeling. At high temperature, the unfaulting of Frank loops could happen in proton irradiated samples. Since the effects of faulted and unfaulted loops on thermal transport are quite different due to strain fields, it is of significance to distinguish loop types and quantify their size and density and also measure the strain fields around them. In situ TEM visualization under irradiation could be used to capture the unfaulting process.

There is a common understanding that a nonequilibrium point defect concentration is established under an irradiating particle flux. Most post-irradiation measurements are not able to capture this supersaturated state, as by the time the ex-situ characterization is done a different distribution of point defects is probed. Evidence for this effect has been demonstrated in several recent in-situ measurements of various properties including thermal conductivity [520, 521], elastic properties [522, 523, 524], vibrational spectra observed using Raman spectroscopy [525] and optical properties [526, 527]. Similar approaches can be applied to the material systems considered in this review.

The ongoing efforts in developing in-situ PAS techniques during ion irradiation would offer great opportunities to detect the formation of defects at the very early of stages of irradiation and monitor their evolution to larger clusters and voids in actinide oxides. PAS can detect atomic size vacancies with sensitivity better than 1 part per million, thus providing new capabilities not attained by in-situ TEM. Capabilities for the indirect measurement of defect concentrations through their impact on thermophysical properties under ion irradiation are also rapidly being developed through the use of in-situ transient grating spectroscopy and other laser-based techniques [522, 528]. As thermal transport in oxides is impacted by the in-situ defect content which cannot be predicted from ex-situ measurements, gaining insight into defect concentrations while active generation is ongoing is of highest priority.

Computational modeling of defects is important to conduct simultaneously with experiments. Thermodynamics modeling requires ab initio underpinning, especially in fuels containing U and Pu. This type of modeling is not just useful in understanding the dominant defects in the oxide in both the hypo- and hyper-stoichiometric regimes, but it is also important in understanding and interpreting self-diffusion data for both cations and anions near thermal equilibrium conditions. Cluster dynamics modeling, on the other hand, is important in yielding the cluster size and composition distributions, which represents a critical input to thermal transport models. While the recent trends in cluster dynamics of oxides paid attention to the cluster composition issues, more research in this area is still required to account for the variability of cluster shapes and the charge effects in both the energetics and kinetics of clustering in oxides.

## 4 Thermal conductivity under irradiation

In this section, we review key studies aimed at modeling and measuring thermal conductivity in the presence of irradiation defects. From the modeling side, we focus on studies that use both the Boltzmann transport framework as well as molecular dynamics to predict thermal conductivity in $UO_2$, $ThO_2$, and oxide fuel surrogates, each containing crystalline defects found in reactor fuels. From the experimental side, we focus on work that uses ion beams to either directly emulate the multi-feature microstructures found in real fuels or to perform single-effects studies to isolate individual populations of microstructural features [304, 529]. The use of ion beam studies has been necessitated by the extreme difficulty associated with measuring conductivity of reactor-irradiated fuel, as well as the challenge of connecting measurements of conductivity to specific microstructural features in multi-feature systems (e.g. point defects and defect clusters) [530]. Moreover, when paired with new experimental methods for measuring thermal conductivity of thin damaged layers, ion irradiation experiments can be conducted rapidly, enabling researchers to probe a large parameter space of exposure and test conditions. The diagram shown in Figure 23 exemplifies a diverse set of methods that are needed to conduct a comprehensive study of phonon transport in defective crystals.

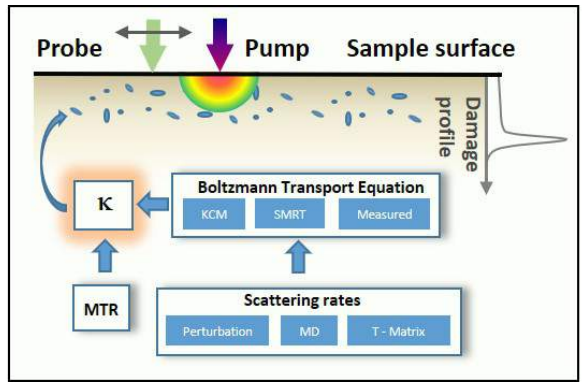

**Figure 23.** Illustration of the methods discussed in this section. Modulated thermal reflectance is used to extract the thermal conductivity of the thin damage layer caused by ion irradiation. Various levels of solutions to the Boltzmann transport equation for phonons are then compared to measured values of conductivity.

We begin by discussing the limitations of current fuel performance codes and highlighting the promise of advanced codes currently under development, with an emphasis on thermal conductivity. We then discuss recent experimental efforts that have enabled measurements of thermal conductivity on length scales commensurate with damage accumulation associated with ion irradiation. This is followed by a discussion of atomistic and atomistically-informed models of thermal transport in the presence of irradiation-induced microstructure. We end this section by reviewing recent work directed at comparing model prediction to experimental measurement of thermal conductivity in defective single crystals of $ThO_2$, $UO_2$, and surrogate oxides.

### 4.1 Fuel performance codes and thermal conductivity

The safe and efficient operation of commercial nuclear reactors relies on analysis provided by fuel performance codes [531, 532, 533]. The temperature distribution within the fuel, governed by the fuel thermal conductivity, plays a central role in fuel performance. Fuel performance codes must consider thermal conductivity as a time- and space-dependent property that changes dynamically during operation due to fission damage, the buildup of fission gases, and a myriad of other parameters [5].

Current light water reactor fuel performance codes based on $UO_2$, such as FRAPCON [531], TRANSURANUS [532], and, more recently, BISON [534, 535] rely on empirical correlations between conductivity, burnup, and temperature [5,9]. Within these codes, the fuel thermal conductivity, $\kappa$, is modeled using a simple expression derived from the kinetic theory of gasses that approximates a broad phonon spectrum by a narrow phonon band [536, 537]:

$$\kappa = \frac{1}{A+BT}. \quad (4)$$

In Equation 4, $A$ captures any conductivity reduction due to defects and $B$ parametrizes intrinsic conductivity. Fuel performance codes use the functional form of Equation 4 to define the thermal conductivity of the fuel by defining $A$ and $B$ as a function of fuel burnup and irradiation temperature [5]. In particular, $A$ can be represented as $A = A_0 + \sum x_i A_i'$, where $i$ is a summation over defects, $x_i$ is their concentration, $A_i'$ is the scattering parameter for individual defects, and $A_0$ is the residual value for a pristine material [25, 530]. Recommended values of $A_0$ and $B$ for pristine samples of $UO_2$, $ThO_2$, and $CeO_2$ are summarized in Table 1 and representative conductivity profiles as a function of temperature are shown in Figure 24. The simplified formulation, and minor variations thereof, has been used to describe the thermal conductivity of fresh and irradiated $UO_2$ fuel [538, 539], mixed uranium-plutonium dioxide (MOX) fuel [540, 541, 542], and rare-earth doped $UO_2$ [543].

**Table 1.** Parameters for high temperature conductivity of pristine fluorite oxides.

| Material | References | B [m/W] | $A_0$ [m·K/W] |
|---|---|---|---|
| $UO_2$ | [544] | $2.165 \times 10^{-4}$ | 0.0375 |
| $ThO_2$ | [545] | $2.25 \times 10^{-4}$ | $4.2 \times 10^{-4}$ |
| $CeO_2$ | [407] | $1.89 \times 10^{-4}$ | $7.84 \times 10^{-4}$ |

In these fuel performance codes, the temperature distribution is used to predict important physical properties that impact fuel performance, including the behavior of fission products, swelling, creep, and the release of fission gases [533]. However, the approximate treatment of thermal conductivity leads to increased uncertainty [546] and, correspondingly, to larger safety margins. Thus, microstructure-based fuel performance codes will be required to expand the application of these codes to higher burnups and/or alternative fuel utilization schemes proposed by new reactor concepts [547].

In the 1990s, researchers started to develop more advanced mechanistic descriptions of thermal conductivity of $UO_2$. One of the more notable studies [530] used measurements on simulated fuel (SIMFUEL) to extract the impact of dissolved fission products and large fission gas bubbles on thermal conductivity. Recently, several effective medium models have been developed to capture the impact of large, intergranular fission gas bubbles on the fuel thermal conductivity [26, 501, 548]. General models that predict the impact of randomly distributed bubbles on the thermal conductivity, such as the Maxwell-Eucken equation [549], do not accurately describe the impact of the intergranular gas bubbles because they are not randomly distributed; they are aligned on grain boundaries that also provide some thermal resistance [501]. Initial models were developed by fitting to mesoscale finite element method predictions of the thermal conductivity [501, 548]. Later, a model was developed that represented the impact of the grain boundary and intergranular bubbles on the fuel thermal conductivity using a network of thermal resistors that predicted an effective thermal resistance of the grain boundary as a function of the fraction of the grain boundary area that was covered by gas bubbles [26]. A later sensitivity analysis showed that the thermal resistor model did not demonstrate the same sensitivities to the average bubble radius as a mesoscale finite element model [550], indicating that it may be an over-simplification of the heat transport through fuel with intergranular fission gas bubbles.

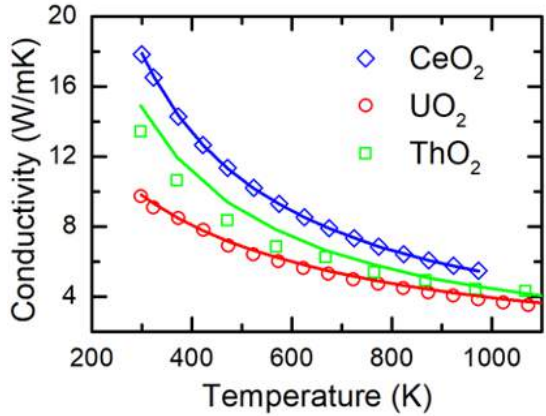

**Figure 24.** High temperature thermal conductivity of selected fluorite oxides. Solid lines are predicted values using Equation 4 and the open symbols are data taken from the literature. $CeO_2$ data (model and experiment) adapted with permission from [407]. Copyright 2014 The American Ceramics Society. $UO_2$ experimental data adapted with permission from [551]. Copyright 2013 Elsevier B.V. $ThO_2$ model data adapted with permission from [545]. Copyright 1997 Elsevier Science B.V. $UO_2$ model data adapted with permission from [544]. Copyright 2000 Elsevier Science B.V. $ThO_2$ experimental data adapted with permission from [154]. Copyright 2019 Elsevier B.V.

Most of the effective medium modeling approaches have focused on fission gas bubbles, which are typically large in comparison to the wavelength of the phonons responsible for transporting thermal energy. However, it has been recognized for some time that smaller radiation-induced defects such as dislocation loops, and sub-TEM-resolution defect clusters, contribute significantly to a reduction in thermal conductivity. Until recently, the impact of smaller defects has been neglected as they are more difficult to characterize and require a more detailed accounting of phonon structure [29, 66, 315, 552, 553, 554, 555].

## 4.2 Experimental measurement of thermal conductivity

While our focus here is on measuring the thermal conductivity of fuel surrogates, it is important, contextually, to mention the large amount of work that has been devoted to measuring thermal conductivity of reactor fuel. Since the early 2000s, researchers primarily in Europe and Japan have sought to investigate thermal conductivity in real fuel specimens that have been reinstrumented with thermocouples and tested in research reactors. These studies illustrate the challenge of truly understanding thermal transport in nuclear fuel under irradiation. Most studies performed used two measurements of temperature (fuel centerline and cladding temperature) along with estimates of the magnitude and distribution of the heat source to extract information on the average thermal conductivity of the fuel across the pellet radius.

Using these tools, irradiation effects have been measured directly on neutron exposed samples [8, 543] and simulation fuel - where the irradiation induced microstructure was emulated through addition of fission products [556, 557] and porosity controlled via conventional ceramic processing methods [558, 559, 560, 561]. Outside of reactor environments, the thermal conductivity of fresh fuels and their surrogates has been measured extensively using laser flash methods [407, 551, 558, 559, 560, 562, 563], direct heat-flow conductivity measurements [179, 180], and recently with femtosecond and continuous wave (CW) laser-based approaches [320,564]. In the following, we review several efforts used to understand thermal conductivity specifically in ion irradiated fuel and surrogate materials with a special emphasis on recent efforts targeting single effects studies employing ion beam irradiations.

In the last decade, to overcome challenges associated with the small damage volumes created by ion beam irradiation, several pump-probe optical techniques relying on photothermal material excitation have been developed. A brief description of the three which have been utilized most frequently - transient grating spectroscopy (TGS), time domain thermoreflectance (TDTR), and modulated thermoreflectance (MTR) – will be provided. For each of these methods, the key parameter which must be matched to the ion irradiation conditions is the thermal wave penetration depth, $L_{th}$. Importantly, the range of $L_{th}$ values available through these methods spans from ~0.5 to ~30 μm. This length scale is commensurate with the damage profile associated with ion irradiation. Schematics of each of these methods at the sample surface are shown in Figure 25. These methods are non-destructive, allowing for thermal transport characteristics to be captured while leaving as-irradiated material available for detailed microstructure characterization when necessary.

TGS utilizes periodic laser excitation generated by crossing two laser pulses from the same source, with the fringe pattern spacing determined by the optical geometry. The resulting periodic thermal and elastic excitation is monitored by the diffraction of a CW probe from the sample [565]. The resulting thermal wave penetration is commonly given as $L_{th}=\Lambda/\pi$, where $\Lambda$ is the fringe spacing projected onto the sample surface [566]. The most detailed work to date using TGS to study thermal transport in ion irradiated materials has focused on metals irradiated with both light and heavy ions [566, 567]. *In situ* ion irradiation capabilities with concurrent TGS property monitoring have also recently been developed [522, 564].

The TDTR methodology most often relies on a train of ultrafast laser pulses which are split into pump and probe beam paths [406,568]. The pump injects high frequency thermal waves into the sample and the probe measure the temperature field evolution by monitoring small changes in temperature induced optical reflectivity (thermoreflectance). The simplest form for the thermal wave penetration for the TDTR geometry is given as $L_{th} = \sqrt{D/(\pi f)}$, where $D$ is the diffusivity of the substrate and $f$ is the pump modulation frequency [406, 569]. TDTR has been used to study a variety of ion irradiated materials including nuclear fuel [570], nuclear fuel cladding [571], silicon carbide [572] metallic multi-layers [571], silicon [571, 573], sapphire [574], and diamond [575] across orders of magnitude in thermal conductivity.

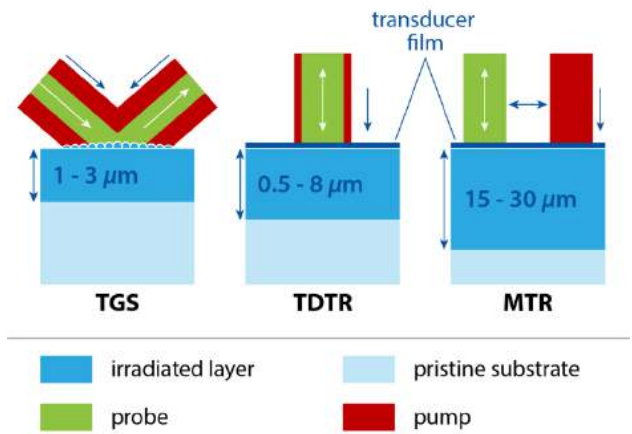

**Figure 25.** Schematics of the TGS, TDTR, and MTR optical geometries at the sample surface. For each case, irradiated layer thicknesses, and corresponding thermal wave depths, are given as examples of those used in the literature and do not describe an exhaustive range.

MTR, also referred to as spatial domain or beam-offset thermoreflectance [576, 577, 578, 579, 580, 581], has been used in the majority of studies of ion irradiated actinide oxides and their surrogates. MTR utilizes modulated continuous wave (CW) lasers to inject thermal waves into a sample under investigation. Pump and probe beams generated through either free-space or fiber-coupled CW lasers are focused onto a sample of interest coated with a transducer film [529]. The spatial extent of the thermal wave is measured by scanning either the pump or the probe along the surface of the sample [578, 582]. In this geometry, $L_{th}$ is again controlled by the applied modulation frequency of the pump and given by the same expression shown above for TDTR. MTR has also been applied on a variety of materials following ion irradiation including silicon [583], sapphire [584], fuel cladding alloys [585], and a variety of oxide nuclear fuels and surrogates including $UO_2$ [586], $CeO_2$ [315], and $ThO_2$ [29]. Recent work has focused on measurements of defect-bearing oxides at cryogenic temperatures in order to discriminate between phonon scattering resulting from multiple types of defects [132].

Experimental studies focusing on thermal transport in ion irradiated oxide nuclear fuel and surrogates have found the most success to date using MTR. Of these methods, MTR has the combination of highest spatial resolution – using spot sizes on the order 1-2 μm – and highest sensitivity to low-conductivity materials. Both of these features are key when interrogating the actinide oxide fuels as detailed surface preparation is often challenging due to material handling constraints, leaving surfaces to be measured which are not optically smooth over hundreds of micrometerss. MTR also retains theoretical sensitivity to both in-plane and cross-plane thermal transport relative to the free surface [584], useful for systems such as $UO_2$ where intrinsic or imposed thermal anisotropy has been identified [180]. In contrast, the other methods described retain primary sensitivity to only a single direction of transport (TGS, in-plane; TDTR, cross-plane). MTR has a final advantage of requiring simple, compact, CW laser sources. This has allowed MTR to be deployed in complex physical environments such as a highly radioactive glove box for the direct study of spent nuclear fuel [587].

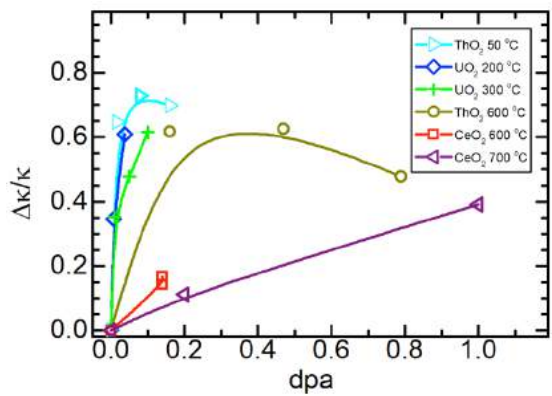

**Figure 26.** Summary of thermal conductivity measurement in ion irradiated fluorites. For $T_{irr}/T_m < 0.2$ microstructure is point defect dominated, whereas for $T_{irr}/T_m > 0.3$ it is loop dominated ($T_{irr}/T_m$ is the ratio of irradiation temperature to melting temperature). 200°C $UO_2$ data adapted with permission from [320]. Copyright 2014 Elsevier B.V. 300°C $UO_2$ data adapted with permission from [586]. Copyright 2018 Elsevier B.V. 700°C $CeO_2$ data adapted with permission from [315]. Copyright 2019 The American Ceramic Society. $ThO_2$ data adapted with permission from [29]. Copyright 2020 Dennett et al. (Published by AIP Publishing). 600°C $CeO_2$ data adapted from [297]. Copyright 2021 Acta Materialia Inc.

Example MTR-measured values of thermal conductivity reduction in ion irradiated $UO_2$, $CeO_2$, and $ThO_2$ are shown in Figure 26 [29, 297, 315, 320, 586]. A major trend evident in Figure 26 is the inverse correlation between irradiation temperature and thermal conductivity reduction; as irradiation temperature increases, more conductivity is retained in these fluorite materials. Irradiations performed at lower temperatures cause degradation of thermal conductivity primarily due to point defects (vacancies, interstitials, and charged defects) [29]. By increasing irradiation temperature, point defects become more mobile and mutually recombine at higher rates. This recombination results in a lower density of point defects even at higher irradiation doses. In parallel, however, at higher temperatures the increased point defect mobility also allows dislocation loops to nucleate and grow; these loops tend to dominate the microstructure of fluorite oxides irradiated at elevated temperatures [29, 297, 315]. A qualitative investigation of data in Figure 26 suggests that dislocation loops contribute to the observed conductivity reduction for irradiations at 600°C and higher.

### 4.3 Boltzmann transport framework

In this section, we review efforts to model thermal transport in actinide fuels and surrogate oxides using the Boltzmann transport equation. Our discussion of Boltzmann transport applied to oxide fuels and surrogates only considers the RTA as it is best suited for treating thermal transport in crystals containing a range of defects from point defects to extended defect clusters. We first discuss phonon scattering by both intrinsic 3-phonon processes and defects as the cross-sections defined for these processes can be used for any level of approximation. Within the RTA formalism, individual phonon scattering mechanisms are independently treated and added together using Matthiesen's rule defined previously in Equation 3. We pay close attention to the form of the cross section for point defects as this model has been used frequently to gauge the impact of irradiation on thermal conductivity in oxide fuels. This is followed by a short discussion of early work aimed at simple analytical solutions to the RTA. We end this section by discussing the single mode relaxation time approximation (SMRT) that more accurately captures phonon structure.

### 4.3.1 Phonon scattering

Three-phonon scattering processes shown in Figure 4 provide an intrinsic limitation to phonons' ability to conduct heat in anharmonic crystals. They are considered as perturbations to a harmonic oscillation resulting in finite phonon lifetimes. As a result the 3-phonon interactions shown in Figure 4a can be represented by [588]:

$$\tau_{qs,\text{anh}}^{-1} = P_{qs}^{(\text{anh})} N_3(qs, \omega) \quad (5)$$

where $P_{qs}^{(\text{anh})}$ is average anharmonicity parameter and $N_3(qs, \omega)$ is the occupation weighted density of states that captures the 3-phonon scattering phase space, satisfying energy and momentum conservation laws.

Defects act as phonon scattering centers, reducing phonon lifetimes and mean free paths. Scattering by the core of the defect is due to atomic mass mismatch and changes to interatomic force constants. Additionally, defects introduce a strain field which scatters phonon through anharmonic interactions. The general form for scattering rate due to phonon interactions with the defect depicted in Figure 4b can be written in a form similar to Equation 5:

$$\tau_{qs,\text{def}}^{-1} = V_{qs}^{(\text{def})} D_2(\boldsymbol{q}, \omega), \quad (6)$$

where $V_{qs}^{(\text{def})}$ is the defect induced perturbation to the interatomic potential energy and $D_2(\boldsymbol{q}, \omega)$ is the density of states that captures phonon-defect scattering phase space [536]. In the Debye approximation, this perturbation may be carried out to define simple functional forms for defect scattering rates for various types of defects. A list of these are provided in Table 2.

**Table 2.** Phonon scattering rates with various defects.

| Type | Core | Strain field |
|---|---|---|
| Point defect 0D [30] | $n_x S_c^2 \dfrac{V_0}{4\pi N v_s^3} \omega^4$ | $n_x S_s^2 \dfrac{V_0}{4\pi N v_s^3} \omega^4$ |
| Line defect 1D (screw dislocations) [30,589] | $N_d \dfrac{V_0^{4/3}}{v_s^4} \omega^3$ | $0.06 \gamma^2 b^2 N_d \omega$ |
| Planar defect 2D (stacking faults) [590] | $\propto \omega^2$ | $N_f \dfrac{0.7 a^2 \gamma^2}{v_s} \omega^2$ |
| Platelets -disk shaped precipitates [591,592] | $N_p \dfrac{24\pi h^2 R_p^2}{v_s} \omega^2$ | N/A |
| Voids 3D [593,594] | $\pi v_s R_v^2 N_v$ | N/A |

$n_x$ are the point defect concentrations of a given species, $S_c^2 = \dfrac{\Delta M_i^2}{M^2}$ and $S_s^2 = 2\left(\dfrac{\Delta K_i}{K} - 2Q\gamma \dfrac{\Delta R_i}{R}\right)^2$ are point defect scattering cross-sections for that species, $N_d$ is dislocation line density, $N_f$ is stacking fault density, $N_p$ and $R_p$ are platelet density and radius, $N_v$ and $R_v$ are spherical void density and radius. Note that technically $1^{st}$ term in $S_s^2$ can be treated as a part of core scattering.

It should be noted that the power exponent for frequency in core scattering scales with defect dimensionality, which is determined by the density of available states that a phonon can scatter into subject to applicable conservation laws, including energy and the tangential component of the momentum assuming a specular reflection occurs [30]. Larger scale defects, including grain boundaries, large voids and bubbles, precipitates, and macroscale porosity are most frequently accounted for using effective medium approaches [26, 530]. Phonon interactions with point defects require a defect-type-dependent scattering cross-section to be defined.

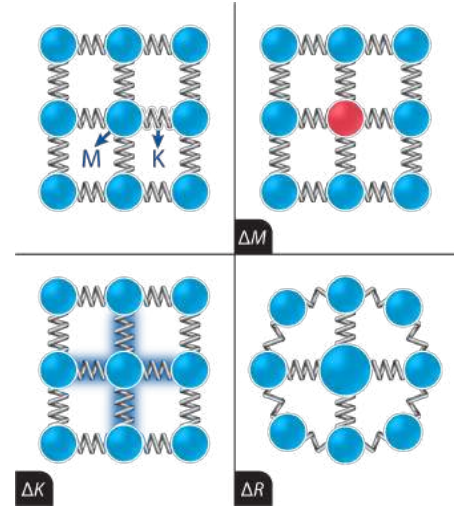

**Figure 27.** Schematic illustrating a change in mass, force constants, and atomic radius resulting from a point defect present in a crystal lattice. Each change is modeled by Klemens' expression (Equation 7) for the scattering strength or scattering cross-section for phonon-point defect scattering [30].

For point defects specifically, the perturbation to the local potential has several components, including changes in atomic mass, force constants, and strain field surrounding the defect. Using a second-order perturbation theory, Klemens derived a general expression for a "scattering parameter" or "scattering strength" for phonon-point defect interactions, $\Gamma$, in terms of the mass difference of lattice points, harmonic force constant difference, and the elastic strain scattering due to atomic radius change [30] (Figure 27). This parameter $\Gamma$ is given by:

$$\Gamma = \sum_i n_i S_i^2 = \sum_i n_i \left( \frac{\Delta M_i^2}{M^2} + 2 \left( \frac{\Delta K_i}{K} - 2Q\gamma \frac{\Delta R_i}{R} \right)^2 \right) \quad (7)$$

where $n_i$ is the fractional concentration of the $i$th defect type, $\gamma$ is Grüneisen anharmonicity parameter, and $Q$ approximates the number of distorted nearest-neighbor bonds surrounding the point defect (3.2 for substitutional defects and 4.2 for vacancies in the fluorite oxide lattice). However, for significant densities of point defects, the permutations of defect configurations within a unit cell must be accounted for [407, 558, 595]. Furthermore, Klemens' original formula assumes a single atom per unit cell, and the treatment of compound materials requires further consideration. Gurunathan et al. provided a detailed description of this and showed that refined Klemens' general formulation may still be used by careful accounting of the average unit cell mass around defects of different types [596, 597].

### 4.3.2 Klemens-Callaway-Debye approximation

One convenient approximation to BTE approach is a method generally referred to as the Klemens-Callaway method (KCM) [536, 598], which approximates phonon branches by a simple Debye model. This allows the summation over $q$-space to be replaced by an integral over frequency, $\omega$, where scattering rates are also expressed in terms of frequencies. In the KCM, thermal conductivity is defined as:

$$\kappa = \int_0^{\omega_D} C(\omega, T) v_s^2 \tau^{-1}(\omega) D(\omega) d\omega \quad (8)$$

where $D(\omega) = 3N\omega^2/V_0\omega_D^3$ is the Debye density of states, $\omega_D^3 = 6\pi^2 v_s^3 N/V_0$ defines the Debye frequency, $C(\omega, T)$ is specific heat defined in terms of Bose-Einstein distribution $C(\omega, T) = \hbar\omega \frac{\partial N^0(\omega, T)}{\partial T}$, and for fluorite oxides $N = 3$ (the number of atoms per primitive cell). Other parameters are defined and are listed for UO$_2$, ThO$_2$ and CeO$_2$ in Table 3. Phonon scattering rates are calculated using perturbation theory assuming all phonons are represented by a single band. In this approximation the 3-phonon scattering rate is given by $\tau_{3ph}^{-1}(T, \omega) = BT\omega^2 \exp(-T_D/3T)$ where $B$ is fit from experimental data [32].

If only point defects are considered in the high temperature limit, such that they and intrinsic 3-phonon scattering are the sole scattering mechanisms present, Equation 8 can be analytically integrated when $\tau_{3ph}^{-1}(T, \omega) = \frac{2\gamma^2 k_B}{Mv_s^2 \omega_m} T\omega^2$ is defined in terms of reduced Debye frequency $\omega_m^2 = \frac{6\pi^2 v_s^2}{V_0}$ to obtain:

$$\kappa_p = \kappa_i \left(\frac{\omega_0}{\omega_m}\right) \operatorname{atan}\left(\frac{\omega_m}{\omega_0}\right), \quad (9)$$

where $\kappa_i$ is intrinsic conductivity in the absence of defects:

$$\kappa_i = \left(\frac{Mv_s\omega_m^2}{4\pi^2\gamma^2 N}\right) \frac{1}{T}. \quad (10)$$

**Table 3.** Parameters for calculation of thermal conductivity following KCM approach.

| Property | UO$_2$ [320] | ThO$_2$ [179] | CeO$_2$ [407] |
|---|---|---|---|
| Velocity, $v_s$ (m/s) | 2644 | 3145 | 3270 |
| Unit cell volume, $V_0$ (Å$^3$) | 163.6 | 175.2 | 159.2 |
| Debye Frequency, (THz) | 43 | 50.2 | 53.8 |
| Debye temperature, $T_D$ (K) | 328 | 381 | 409 |
| B, (s/K) ×10$^{-18}$ | 1.6 | 1.2 | 1.2 |
| Grüneisen parameter | 1.8 | — | 2.5 |

Here, the correction due to defects is:

$$\left(\frac{\omega_0}{\omega_m}\right)^2 = \left(\frac{4N\gamma^2 k_B T}{Mv_s^2}\right) \frac{1}{\Gamma}, \quad (11)$$

where $\Gamma$ is computed via the expression given in Equation 7. These analytical forms are useful comparison tools for estimating contributions from individual microstructural features, however more care in accounting for all contributions to phonon-defect scattering must normally be taken when comparing the results of a computation to measured values of conductivity.

### 4.3.3 Single Mode relaxation time approximation

The simple nature of Equation 4 as used in performance codes makes it practical for engineering application, however, its validity is limited to high temperatures where phonon-phonon scattering dominates. The KCM approach described above is a first step in making more accurate calculations of low temperature conductivity. However, the Debye approximation is known to underrepresent the role of long mean free path phonons, requiring a more accurate representation of phonons across the Brillouin zone [42, 599]. As discussed in Section 1, numerous phonon modes exist in the crystalline structure of solids, each having its own vibration frequency, velocity, and lifetime. All these details influence the thermal conductivity of a given material [536, 600, 601]. Different approaches for calculating thermal conductivity using first principles in defect-free systems have been reviewed in Section 2. Since it is computationally expensive to follow similar methods for systems having non-trivial defects, we consider here several levels of simplification, which bring both intrinsic anharmonic and defect perturbations to similar levels of complexity.

A simplification in the calculation of the 3-phonon scattering phase may be made based on the single mode relaxation time (SMRT) approximation [600]. In the SMRT, the scattering phase-space term in Equation 6 retains exact phonon dispersion, but the anharmonic interaction term is represented by a single Grüneisen parameter. This method is especially useful for systems like UO$_2$, where the accuracy of current DFT methods is questionable [27, 602]. While dispersion curves can be easily validated by neutron scattering experiments, the data on 3-phonon scattering is limited. This approach has been applied recently to ThO$_2$ [555] to predict bulk, single-crystal thermal conductivity in comparison to measured values [179]. Figure 28 compares some of the notable results from previous studies that considered thermal conductivity of ThO$_2$.

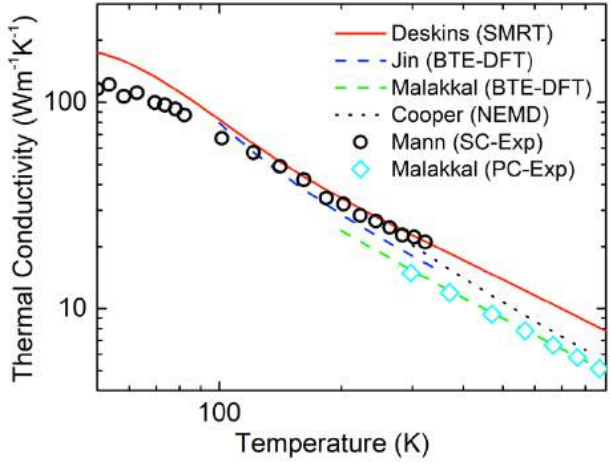

**Figure 28.** Comparison of different methods for calculation of thermal conductivity in $ThO_2$. SMRT by Deskins et al. [555] (red curve) and bulk $ThO_2$, NEMD results by Cooper et al. [603] (dotted line), and BTE-DFT results by Malakkal et al. [154] (green dashed line) and Jin et al. [152] (blue dashed line). The black circles are experimental results by Mann et. al. [179] for 96.7 TD $ThO_2$ single crystal and cyan diamonds for polycrystalline $ThO_2$ [154]. Data by Deskins et al. adapted with permission from [555]. Copyright 2021 AIP Publishing. Data by Cooper et al. adapted with permission from [603]. Copyright 2015 Elsevier B.V. Data by Malakkal adapted with permission from [154]. Copyright 2019 Elsevier B.V. Data by Jin et al. adapted with permission from [152]. Copyright 2021 IOP Publishing Ltd. Data by Mann et al. adapted with permission from [179]. Copyright 2010 American Chemical Society.

The associated phonon-point defect scattering rate of a phonon with wave number $q$ and polarization $s$, accounting for the actual dispersion curves, can be written in terms of the scattering parameter as [596]:

$$\tau_{PD,qs}^{-1} = \frac{\Gamma V_0}{4\pi v_{p,qs} v_{g,qs}^2} \omega^4, \qquad (12)$$

where $v_{p,qs}$ is the phonon phase velocity. In the Debye limit, this form reduces to that listed in Table 2 for point defects. Equation 12 takes into account the dispersive nature of the phonon and distinguishes between phonon group and phase velocities. The expansion of phonon scattering rates in Table 2 to SMRT requires recasting them into a form that accounts for differences between $v_p$ and $v_g$. Using dispersion curves fit to inelastic neutron scattering data [50], Deskins et al. [555] have implemented the SMRT approximation in the case of uranium substitutional defects in $ThO_2$ and compared the resulting temperature dependence of thermal conductivity to measured values [604] and those from MD simulation [603] (Figure 29). This comparison shows that in these fluorite oxide systems, making the SMRT approximation still produce a reasonable description of thermal conductivity in defect-bearing systems across a wide temperature range through the BTE. As more accurate first principles treatments of thermal transport in $UO_2$ become available, researchers will be able to revisit assumptions regarding lattice anharmonicity and phonon-defect scattering that are tied to the SMRT.

However, the Boltzmann framework, at any level of approximation discussed here, typically only considers 3 phonon processes (cubic phonon interactions) and may be inadequate to describe phonon thermal transport at high temperature where four-phonon scattering can become significant [123, 605]. An additional limitation of the standard implementation of BTE at higher temperature may arise from the fact that materials undergo thermal expansion resulting in phonon softening. A detailed accounting for this effect would require recalculation of phonon dispersion and relevant anharmonic terms as a function of temperature. Nevertheless, while these higher order effects are important to the broader community, they do not come into play for work reviewed in this section for two reasons. First, in an effort to isolate the role of defects, experiments are typically carried out at room temperature and below to reduce the obscuring influence of phonon-phonon interactions. Second, measurements at elevated temperature of ion irradiated material introduce significant complications due to kinetic defect processes such as recombination and agglomeration.

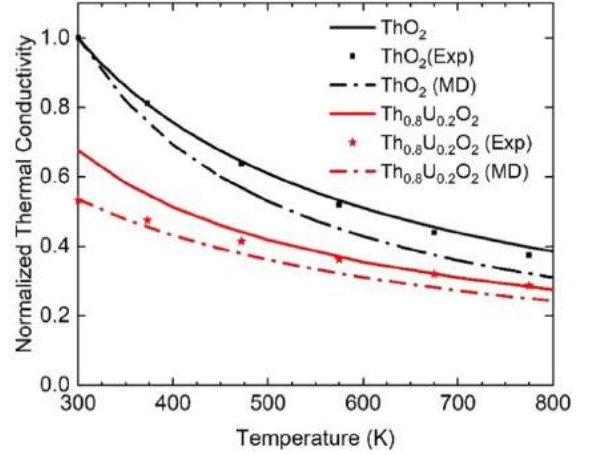

**Figure 29.** Reduction of thermal conductivity presented by Deskins et al. (solid lines) compared to experimental and MD results containing 20% uranium substitutional defects. Each value is normalized by the respective pure $ThO_2$ value at 300 K from each study. Adapted with permission from [555]. Copyright 2021 AIP Publishing.

### 4.4 Molecular dynamics simulations of thermal conductivity

Molecular dynamic simulations offer an alternative way of calculating thermal conductivity in lattices where defects and other features can be directly specified. It can be applied to defective structures and captures higher order anharmonic effects, which are too computationally expensive to be implemented within the BTE. The application of additional post-processing can reveal detailed mechanisms of phonon scattering.

#### 4.4.1 Bulk thermal conductivity

Classical molecular dynamics (MD) involves solving Newton's equations of motion to determine the trajectories of atoms within a crystal lattice [606]. To approximate an infinite lattice, a structure of many thousands of atoms can be replicated periodically in all dimensions, creating a supercell. The behavior of the system is dictated by the description of interatomic forces using an interatomic potential, derived empirically by fitting to experimental material properties or to forces calculated from DFT. During MD simulations, thermal transport within such an ensemble can be evaluated using either the non-equilibrium MD (NEMD) method or the Green-Kubo equilibrium method [607].

During NEMD, the system is driven out of its equilibrated state by application of a heat flux (Figure 30), which creates a thermal gradient [608], or by application of a thermal gradient, which creates a heat flux [609]. In either case, following a certain amount of time the system will reach steady state, whereby the heat flux, $\vec{q}$, and thermal gradient, $dT/dx$, become constant, and the thermal conductivity is defined by Fourier's Law:

$$\kappa = \frac{\vec{q}}{dT/dx}, \quad (13)$$

Finite size effects arise due to the separation of the hot and cold regions being shorter than the phonon mean free path [609, 610, 611]. In NEMD, the true bulk thermal conductivity, $k_\infty$, must be extrapolated from the following relationship $\kappa^{-1} = AL_x^{-1} + k_\infty^{-1}$, where $\kappa$ is the thermal conductivity of the system with supercell length $L_x$.

The Green-Kubo method is carried out on a system in its equilibrium state, whereby the system is held at constant temperature. The thermal conductivity is determined from the ensemble average of the heat-current autocorrelation function, $\langle J(0) \cdot J(t) \rangle$, such that:

$$\kappa = \frac{V}{3k_B T^2} \int_0^\infty \langle J(0) \cdot J(t) \rangle \, dt, \quad (14)$$

where $V$ is the volume of the system, $T$ is the temperature, and $k_B$ is the Boltzmann constant. For long simulation times, $J(t)$ becomes uncorrelated with respect to $J(0)$, such that $\langle J(0) \cdot J(t) \rangle$ tends to zero, enabling convergence of $\kappa$. A benefit of the Green-Kubo method is that finite size effects appear to be less severe than for the NEMD method; however, long simulation times are required and the force on an atom due to each of its neighbors must be precisely defined, which for many-body empirical potentials is not trivial [607, 611].

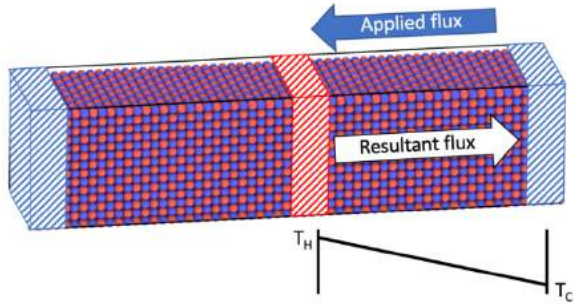

**Figure 30.** A schematic of the NEMD method, whereby a UO$_2$ supercell with a large aspect ratio is used. A heat flux is applied, which causes a thermal gradient. This, in turn, causes a resultant heat flux. Under steady-state, the applied heat flux and the resultant heat flux are equal and opposite.

Inherent to both methods are uncertainties regarding the choice of empirical potential and its ability to describe thermal transport in the system being studied. Commonly, pairwise empirical potential forms have been employed for oxides, whereby the interaction energy of pairs of U-U, U-O, and O-O atoms are independent of the positions of other atoms in the system. The total system energy is given by a sum over all pair interactions. Typically, a pair potential consists of long-range Coulomb and short-range interactions. Often partial ionic charges are used, meaning that the degree of ionicity can be used as a fitting parameter. However, the majority of fit parameters are still contained within the short-range interactions. Most UO$_2$ potentials use a Buckingham form [612] to describe the repulsive force due to the Pauli-exclusion principle and intermediate-range dispersive forces due to Van der Waals interactions. Additionally, a Morse potential [613] can be included to capture a degree of covalency in UO$_2$; for example, in the Basak [614] and Yakub [615] potentials. Several UO$_2$ potentials, such as the Morelon [616] and Jackson [617] potentials, utilize the 4-range formulation of the Buckingham potential, which retains the exponential description of forces at short ranges and the dispersive term at longer distances, with splines used near the potential minima. This approach avoids the unphysical energies at short ranges due to the asymptotic nature of the r$^{-6}$ dispersion term. A number of potentials account for ionic polarizability using the shell model, representing a deviation from a strictly pairwise description that enables the dielectric constant and the Cauchy violation in the elastic constants to be reproduced [618]. More recently, Cooper, Rusthon and Grimes (CRG) extended a Coulombic-Buckingham-Morse pair potential form to include many-body interactions through the embedded atom method (EAM), allowing the temperature dependence of bulk modulus to be captured, which was also later demonstrated by Soulie et al. [619] for the Sattonay potential [620]. The Sattonnay and Li [621] potentials moved beyond a rigid-ion model to include charge transfer between atoms. A number of detailed studies [602, 618, 622, 623, 624, 625] have compared the properties of contemporary potentials to expertimental or DFT data, such as thermal expansion, mechanical properties, and defect energies. The ability of a potential to capture such properties is the primary metric used to evaluate the quality of the potential.

However, frequently, empirical potentials are not fit to the phonon properties of a material and the ability of a potential to accurately describe thermal transport must be rigorously validated. Notably, classical MD struggles to reproduce the low thermal conductivity of UO$_2$, particulary at low temperatures [602, 626, 627, 628, 629], whereas MD potentials tend to perform better for ThO$_2$ [154, 630, 631]. It has been suggested that this is due to spin-phonon resonance associated with low lying $f$-electron states in UO$_2$ that cannot be included directly in classical MD but when accounted for in the Klemens-Callaway model, greatly improve the comparison between MD and experiment [25, 26, 180]. Chernatynskiy et al. applied a non-MD lattice dynamic approach, finding that several potentials performed reasonably well for UO$_2$ and suggested an anharmonic correction to improve results [632].

Recently, the CRG [923] potential was rigorously accessed for accurate predictions of phonon dispersion, lifetime, and branch specific conductivity for UO$_2$ and ThO$_2$ [152]. It was found that the CRG potential captures the dispersion of the acoustic branches well, but exhibits significant discrepancies for the optical branches, leading to an overestimate of phonon lifetime and conductivity for both materials. Similary, it has been found that several UO$_2$ potentials accurately capture the dispersion of the acoustic branches, while inaccuracies in the optical branches result in a poor description of thermal properties [602]. These findings suggest that the empirical potential needs to be further optimized for robust prediction of thermal conductivity both in perfect crystals and in the presence of complex defects. Moreover, the methods discussed here are only suitable for modeling classical thermal transport due to phonons and do not address polaron contributions, which can be significant at high temperatures in semi-conductors like UO$_2$ [633], or spin-phonon resonance processes.

### 4.4.2 Point defect scattering using molecular dynamics

The thermal transport of defective crystals can be addressed using the same methods outlined above. First, the defect-free lattice thermal conductivity should be determined as a baseline. Subsequently, defects can be introduced to the system by adding new atoms at interstitial sites, removing atoms to create vacancies, or substituting one atom for another to create impurity defects [25, 627, 634, 635, 636]. The now-defective system should then be allowed to equilibrate before repeating the thermal conductivity assessment of the system, taking care to treat finite size effects. This process can be repeated with varying defect concentrations. The phonon thermal conductivity calculated as a function of concentration can be fitted with either the simple model represented by Equation 4 or the Klemens-Callaway model to determine the effective phonon scattering cross-section for each defect or defect grouping [25, 28].

Broadly speaking, impurity defects have a lesser impact on thermal conductivity than interstitials and vacancies, and of the fission products, Xe has the most significant impact on the thermal conductivity of $UO_2$ [25, 26, 555]. An additional complexity for ionic materials is the fact that scattering due to charge compensating defects must also be considered. For example, Figure 31 shows the significant impact that a particular charge compensating mechanism has on the thermal conductivity of $UO_2$ doped with 5 wt% Gd [635]. If Gd impurities are compensated by $U^{5+}$ substituted at a $U^{4+}$ site the thermal conductivity is significantly higher than if oxygen vacancies ($V_O$) are used. This is despite the lower concentration of $V_O$ needed to charge compensate 5 wt% Gd. By reducing the number of phonon-scattering centers, binding increases the thermal conductivity with respect to a random distribution of defects. This demonstrates the complexity that must be considered for all defects in oxide nuclear fuel, be they formed by fission product accommodation, irradiation damage, or by varying off-stoichiometry. Understanding the scattering due to individual point defects allows the coupling of phenomena, such as fission gas behavior or radiation damage, to thermal conductivity models [26, 28]. Lastly, phonon thermal conductivity calculated using either the NEMD or the Green-Kubo method can be fitted with either the simple model represented by Equation 4 or the Klemens-Callaway model to determine the effective phonon scattering cross-section for each case [25, 28].

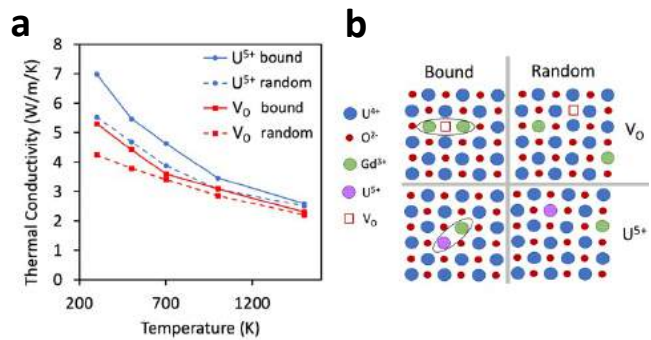

**Figure 31.** (a) The thermal conductivity of $UO_2$ containing 5 wt% Gd calculated using NEMD by Qin et al. Each data set corresponds to a different charge compensating mechanism, as shown by (b) the schematic of atomic configurations. Adapted with permission from [635]. Copyright 2020 Elsevier B.V.

### 4.4.3 Extended defects

MD simulations can be employed to simulate the impact that extended defects (such as dislocations, grain boundaries, voids, and bubbles) have on phonon-mediated thermal conductivity. For example, the reduction in thermal conductivity of $UO_2$ for several concentrations of edge dislocations has been predicted to develop an analytical form suitable for use in a fuel performance code [637]. The impact of grain boundaries on $UO_2$ thermal conductivity has also been estimated and extrapolated from MD simulations, revealing only a small impact for micrometer-sized grains (as are typically present in fresh fuel) [638, 639, 640, 641]. However, because grain boundaries serve as sinks for point defects, it is expected that the impediment to thermal transport caused by grain boundaries will increase under irradiation [407, 552]. MD simulations have been used to investigate the degradation of thermal conductivity due to porosity [553] and the accumulation of fission gas (Xe) [642] or decay gas (He) [643] into bubbles. As expected, an increased porosity reduces the thermal conductivity; however, Chen et al. [642] and Lee et al. [643] both showed a decrease in thermal conductivity as the voids were filled with Xe and He, respectively. For He this was attributed to the resolution of He into the lattice, whereas for Xe resolution did not occur and it was attributed to $UO_2$ atom distortion at the bubble-$UO_2$ interface. Nonetheless, Tonks et al. demonstrated that the thermal conductivity degradation due to dispersed gas in the lattice greatly outweighs that of gas in the bubbles [26].

### 4.5 Impact of grain boundaries and porosity

While the impact of atomistic-level defects is best captured using RTA, the effect of multiple grain boundaries on phonon thermal conductivity is better captured using effective medium approximations. The impact of grain boundaries on thermal conductivity may be represented using a series resistance model:

$$\frac{1}{\kappa_e} = \frac{1}{\kappa_0} + \frac{R_{GB}}{d}, \quad (15)$$

where the effective thermal conductivity, $\kappa_e$, is a function of the single crystal conductivity, $\kappa_0$, the average grain size, $d$, and the grain boundary interfacial thermal, or Kapitza, resistance $R_{GB}$. At low temperatures, where the Debye model serves as a good approximation for the density of states, the acoustic mismatch model and the diffusive mismatch model have been developed to explain the temperature drop across boundaries [644]. Unfortunately, neither model provides good predictive capability across a wide range of grain boundary types. These failures result from interface scattering processes which depend sensitively on atomic structure and are difficult to treat in analytical models. Recent studies have used molecular dynamics [552, 645] to understand the impact of atomic structure on thermal transport across grain boundaries. One of the first efforts involving $UO_2$ was performed by Watanabe and coworkers [552]. Their results demonstrated that the calculated value of the grain boundary resistance depends strongly on the choice of interatomic potential. A more recent molecular dynamic study [645] that considered larger grain sizes reported Kaptiza resistances that were 50% smaller than reported by Watanabe. It was suggested that this difference is due to the polycrystalline averaging scheme employed by Watanabe.

Corrections to thermal conductivity resulting from volumetric porosity have also been accounted for by including a correction factor:

$$\kappa_p = \kappa_e f(p, T), \qquad (16)$$

using the Loeb expression or Maxwell-Eucken formula, which account for fractional volume of porosity, shape factor, and temperature [530]. This expression does not account for pore size and breaks down for nanometer-sized pores that have dimensions on the scale of the phonon mean free path. Clearly, more accurately accounting for nanovoids requires models that consider the atomic structure of the defects. In parallel with grain boundaries, recent molecular dynamics modeling work has aimed to understand the role of atomic structure at pore/bubble boundaries [642, 646]. Zhu et al. [646] used non-equilibrium molecular dynamics to study the impact of Xe bubbles on thermal transport in $UO_2$. They found that many defects are produced at the bubble interface, leading to a substantial increase in the interfacial thermal resistance and an overall decrease in the thermal conductivity.

## 4.6 Comparison between experiment and modeling

In this section, we review efforts to correlate microstructural defects produced by ion irradiation to thermal conductivity using the Boltzmann transport framework and MD. Very little work had been done in this area until recently due the difficulty in accurately characterizing the range of defects produced by ion irradiation and accurately measuring thermal conductivity of thin damage layers. As mentioned above, the most recent work in this area has been the experimental measurement of low-temperature conductivity in defect-bearing crystals. At reduced temperatures, the relative contribution of defects to the total scattering time is the largest, allowing the most information to be collected on fundamental mechanisms of phonon-defect interactions.

### 4.6.1 Impact of point defects in low temperature, low dose irradiated $UO_2$

One of the first efforts along these lines was the work of Weisensee et al. [570]. They investigated the effect of room temperature Ar ion irradiation on thermal conductivity of thin film of $UO_2$ using TDTR. While the lattice strain associated with the lattice mismatch between substrate and film prevented a detailed study of baseline microstructure, it was suggested that the $UO_2$ was slightly hyperstoichiometric. XRD data revealed a broadening of the diffraction peaks and a shift to lower angles, an indication that the $U_4O_9$ phase was not formed under irradiation. A simple model of phonon thermal conductivity was used to extract an effective non-dimensional scattering strength for point defects in $UO_2$.

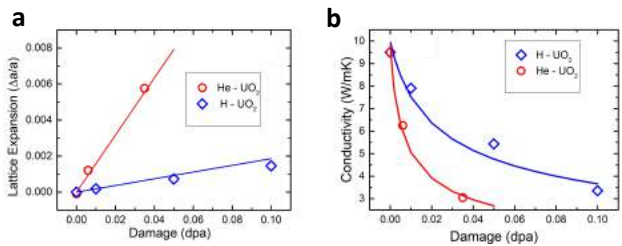

**Figure 32.** Analysis of thermal conductivity degradation in $H^+$ and $He^{++}$ ion irradiated $UO_2$. Adapted with permission from [28]. Copyright 2020 Acta Materialia Inc.

In another study targeting the impact of point defects, polycrystalline $UO_2$ was irradiated using 2.6 MeV protons and 3.9 MeV helium ions at 300 and 200°C, respectively [28]. After irradiation, the sample thermal conductivity and lattice constant were measured using MTR and XRD respectively (Figure 32). The reduction in conductivity and lattice expansion did not follow a similar trend with dose for these two different irradiation conditions. The results were analyzed using simple models for lattice expansion and thermal conductivity reduction, informed by atomistic simulations rom the literature. It was concluded that microstructure evolution under H and He ion irradiation produced different types of defects owing to differences in defect clustering and electronic-ionization-induced mobility of the defects.

### 4.6.2 Impact of dislocation loops in high temperature proton irradiated $CeO_2$

A study targeting the impact of dislocation loops on conductivity by Kafizov et al. [315] involved measuring the conductivity reduction caused by proton irradiation of $CeO_2$ at 700°C. As revealed by TEM analysis, the microstructure was dominated by dislocation loops. A rate theory model parametrized by analyzing two sets of samples irradiated at 600 and 700°C revealed that the conductivity reduction in 700°C irradiated sample is due to dislocation loops, whereas at 600°C the reduction is primarily attributed to point defects (Figure 33) [297]. One interesting observation that follows from these studies is that thermal conductivity reduction due to loops is unusually high and is attributed to long range strain fields [315].

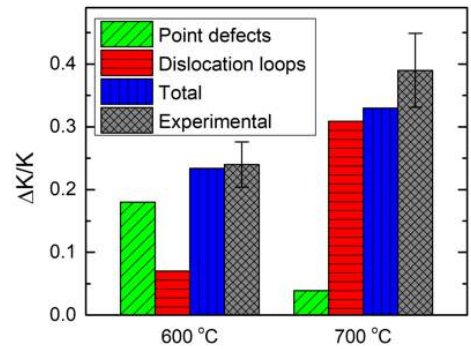

**Figure 33.** Analysis of thermal conductivity degradation due to dislocation loops in proton irradiated $CeO_2$. Adapted with permission from [297]. Copyright 2021 Acta Materialia Inc.

### 4.6.3 Mixed impact of point defects and loops in $ThO_2$

More recently, efforts have been directed at drawing a closer connection between microstructure and thermal conductivity using more comprehensive defect characterization techniques, discussed in section 3 of this review, and a more complete accounting of the physical mechanisms of phonon transport.

Dennett and coworkers experimentally studied the impact of 2 MeV proton irradiation on thermal transport in single crystal thorium dioxide grown using the hydrothermal synthesis method using MTR [29]. $ThO_2$ was irradiated at 25°C from 0.016 to 0.16 dpa and at 600°C from 0.47 to 0.79 dpa. At 25°C, the thermal conductivity in the irradiated layer dropped to 20% of the pristine value for all doses investigated, while at 600°C, 40% of the pristine conductivity was retained at both doses. For both temperatures investigated, the thermal conductivity was found to be consistent between all doses, suggesting a saturation in the defect population imparted via irra-diation at these levels. Top-down Raman spectra collected following irradiation suggested the presence of defect clusters in both cases, with a higher density or larger clusters expected at 600°C due to the higher mobility of defects. The retention of higher thermal conductivity at 600°C is attributed to the concentration of defects into larger clusters and loops, reducing the total number of phonon scattering sites retained in the lattice [520, 554, 583].

Further measurements of the thermal conductivity in defected single crystal $ThO_2$ were conducted on a subset of the crystals irradiated at 600°C over a temperature range of 77 to 300 K using MTR (see Figure 34) [132]. These data provide a low-temperature comparison for BTE calculations of defect-affected thermal transport (similar to the 300K and higher BTE calculation of transport in U-doped $ThO_2$ shown in Figure 29). To compute the thermal conductivity of defect-bearing $ThO_2$, dislocation loop characteristics were measured directly using transmission electron microscopy (TEM) in order to compute scattering rates from stacking fault type defects as shown Table 2 (as these loops are observed to be faulted Frank loops). In addition, point defect contributions to scattering were estimated through the use of a rate theory model for defect generation and evolution [297] and used to seed the calculation of the scattering rates based on Equation 12. Using fitting parameters only to account for native impurities in the as-synthesized crystals, good agreement was found between MTR-measured conductivity and that computed using the BTE in the SMRT approximation as shown in Figure 34 for $ThO_2$ irradiated to 0.16 and 0.47 dpa.

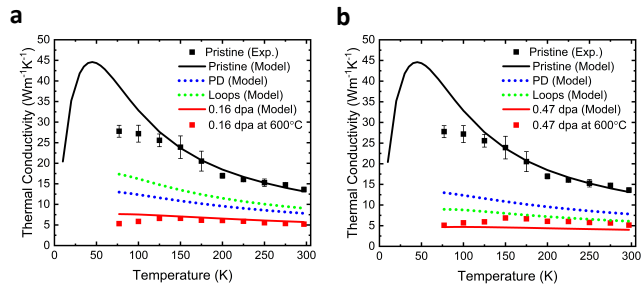

**Figure 34.** Comparison of MTR-measured thermal conductivity and BTE computed thermal conductivity for single crystal $ThO_2$ irradiated with 2 MeV protons at (a) 0.16 dpa at 600°C and (b) 0.47 dpa at 600°C. Computed conductivities considering only one type of extrinsic defect (dislocation loops or point defects) are shown in the dotted lines, while the total conductivity considering both defect populations is shown in the solid red line in each. Adapted with permission from [132]. Copyright 2021 Dennett et al. (Published by Elsevier Ltd.).

#### 4.6.4 Impact of grain boundaries

The segregation of defects at grain boundaries is thought to have a significant impact on thermal conductivity in the HBS. Generally, the HBS corresponds to a rim in the nuclear fuel where the fuel remains coldest and has largest damage owing to a higher fission rate. In this region, extensive damage is accommodated by grain subdivision and subsequent defect segregation. The net beneficial effect on intrinsic conductivity suggested by Rondinella and Wiss [6] has motivated recent work that has looked at understanding the role that interfaces have on phonon transport in fuel surrogates.

Hua et al. [647] reported the Kapitza resistance for grain boundaries in large-grained $CeO_2$ using an investigative approach first developed for Si bicrystals [648, 649]. This study provided a correlation between the measured Kapitza resistance and misorientation angle between grain boundaries. To form this correlation, it was assumed that the plane of the grain boundaries coincided with the surface normal. In another study involving $UO_2$, experimental measurements of thermal conductivity were performed on a series of polycrystalline samples having grain sizes ranging from ~100 nm to several micrometerss [650]. Using equation 15, it was found that the extracted Kapitza resistance was considerably larger than calculated using molecular dynamics [552, 645]. It was suggested that the difference between measurement and model may be due to local changes in oxygen stoichiometry near the boundary.

Another study by Khafizov et al. [407], of more direct connection to thermal transport behavior in the HBS, involved measuring conductivity in highly non-stoichiometric nanocrystalline ceria [407]. In this study, it was found that the measured conductivity was much larger than that predicted assuming that the oxygen defects were uniformly distributed. Electron energy loss spectroscopy revealed that oxygen defects were segregating at grain boundaries leaving the grain interior relatively defect free (Figure 35). The authors speculated that the increase in the grain boundary resistance due to segregation was more than offset by a reduction in phonon scattering by intragranular oxygen defects. This supposition has since been supported by MD and multiscale modeling studies of $UO_2$ [512, 552]. Watanabe et al. found that the grain boundary thermal resistance that was two orders of magnitude larger than that expected from pristine interfaces calculated using molecular dynamics simulation [552], while Bai et al. found that grain boundary segregation leads to higher intragranular thermal conductivity [512].

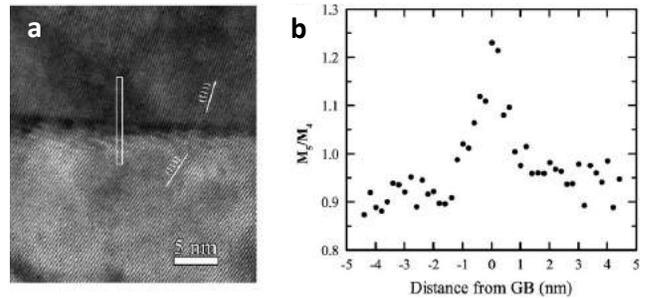

**Figure 35.** Impact of defect segregation on interface thermal resistance of grain boundaries. Adapted with permission from [407]. Copyright 2019 The American Ceramic Society.

### 4.7 Future Outlook

While the works described above have shown some success in connecting models of varying complexity to experimental measurements of thermal transport in defected actinide oxides, gaps still exist in the ability to capture the unique influence of the entire spectrum of radiation-induced defects on phonon scattering. Models beyond the RTA are needed to accurately treat conductivity at sufficiently low temperatures where normal processes dominate over Umklapp processes. Furthermore, the RTA can be problematic in certain classes of materials, such as epitaxial films, that introduce symmetries not found in the bulk. MD models may help address the shortcomings of the RTA but require interatomic potentials that accurately capture phonon transport properties. First principles descriptions of the details of the phonon band structure can provide important benchmarks for developing new, more accurate interatomic potentials. While the computational tools needed to perform this task are available, this will no doubt be a challenge for strongly electron correlated materials.

Currently, the Green's function T-matrix method is the most fundamental approach to capture the impact of defects on phonon scattering rates [651, 652]. Work using this method has demonstrated the shortcomings of using perturbation theory to treat the impact of vacancies in diamond, silicon carbide, and boron arsenide. These methods are currently finding applications in predicting scattering rates of phonons with extended defects such as clusters and dislocation loops [653, 654]. However, extending these approaches to materials that exhibit strong electron correlation remains a difficult task for first principles calculations.

# 5 Concluding Remarks

This review focused primarily on phonon thermal transport in two actinide oxides: one currently in use in commercial nuclear reactors, $UO_2$, and one advanced fuel candidate material, $ThO_2$. A comprehensive understanding of phonon transport in these actinide oxides under irradiation remains a grand science challenge. This area of research contains rich physics including the $5f$ electron challenge, defect generation and evolution, and the development of experimentally validated mesoscale models of thermal conductivity in the presence of defects. New computational tools such a DMFT-DFT and new experimental tools enabled by new synthesis capability are ushering in a new era, enabling scientists to study, in exquisite detail, the fundamental mechanisms that control phonon transport.

Meeting this challenge will provide new opportunities for the nuclear fuels community. For example, accurate calculations of electronic structure in $5f$ systems will provide the scientific community with key information for constructing accurate interatomic potentials required for high-throughput compositional screening of advanced fuels. The development of a comprehensive understanding of thermal energy transport, while intimately connected to advanced nuclear fuels, can provide opportunities for energy-related technologies beyond nuclear energy. Examples include tailoring electron scattering in textured nanocrystals to enhance thermal conductivity of metal interconnects, simultaneously controlling electron and phonon transport across interfaces in thermoelectric devices, specifying the interaction of electrons with dislocations in topological insulators, and harnessing strong electron correlation to realize new paradigms for quantum materials.

**AUTHOR INFORMATION**

**Corresponding Author** - david.hurley@inl.gov

<u>Author Contributions</u>

The manuscript was written through contributions of all authors. All authors have given approval to the final version of the manuscript.

**Biographies**

*David H. Hurley*
Dr. David H. Hurley is a Laboratory Fellow and a group lead in the Materials Science and Engineering Department at Idaho National Laboratory. He received his Ph.D. in Materials Science and Engineering from Johns Hopkins University in 1999. He was a Japanese Society for the Promotion of Science Postdoctoral Fellow in the Applied Solid State Physics Laboratory at Hokkaido University. He is the recipient of the INL Outstanding Innovation Award for development of a hot-cell compatible instrument for nanoscale thermal transport measurements. Currently, he is the founding director of the Center for Thermal Transport under Irradiation, an Energy Frontier Research Center funded by the US Department of Energy, Office of Basic Energy Sciences.

*Anter El-Azab*
Dr. Anter El-Azab is a Professor of Materials Science and Engineering at Purdue University. He is Science Question Lead for the Center for Thermal Energy Transport under Irradiation, funded by the U.S. DOE's Energy Frontier Research Centers program. Prior to coming to Purdue, he was a Professor of Computational Science and Materials Science at Florida State University and a research scientist at Pacific Northwest National Laboratory. He obtained his doctorate degree at the University of California, Los Angeles. His research and teaching focus on microstructure complexity in materials.

*Matthew S. Bryan*
Dr. Matthew S. Bryan is a postdoctoral researcher in the Materials Science and Technology Division at Oak Ridge National Laboratory. He obtained his PhD from Indiana University in physics in 2019. His work focuses on using inelastic neutron and X-ray scattering to study phonon dispersion, line widths, and density of states and how these measurements relate to thermal transport in actinide bearing materials.

*Michael W.D. Cooper*
Dr. Michael Cooper is currently a Scientist in the Materials Science and Technology Division of Los Alamos National Laboratory, where he was also a postdoc from 2015 to 2018. He obtained a PhD in Materials Science from Imperial College London in 2015. His research interests focus on the application of classical molecular dynamics and density functional theory to understanding the atomic scale processes that govern nuclear fuel behavior. In particular, he derives parameters for defects, fission products, and thermophysical properties relevant to ceramic fuels of various forms, such as oxides, nitrides, and silicides.

*Cody A. Dennett*
Dr. Cody A. Dennett is currently an Applied Physicist in the Materials Science and Engineering Department at the Idaho National Laboratory. He obtained his PhD from the Massachusetts Institute of Technology in 2019. He was the recipient of the INL Russell L. Heath Distinguished Postdoctoral Fellowship and prior to INL was a US Department of Energy National Nuclear Security Administration Stewardship Science Doctoral Fellow. His work focuses on laser-based measurements of thermophysical properties in dynamic or extreme environments to understand defect kinetics and transport phenomena in metals and ceramics.

*Krzysztof Gofryk*
Dr. Krzysztof Gofryk is a distinguished staff scientist and a group lead for actinide science in the Irradiated Fuels and Materials Department at the Idaho National Laboratory (INL). He obtained his Ph.D. from the Institute of Low Temperature and Structure Research (Polish Academy of Sciences) in 2006. Before joining INL, he worked at Los Alamos and Oak Ridge National Laboratories. He is a recipient of the DOE Early Career Award, the Presidential Early Career Award for Scientists and Engineers, and INL's exceptional achievement award. Dr. Gofryk's work focuses on strongly correlated electron systems at low temperatures and strong magnetic fields including transport and magnetism in 5f-electron materials, quantum criticality, heavy fermion physics, and spin-orbit coupled systems.

*Lingfeng He*
Dr. Lingfeng He a distinguished staff scientist and High-Resolution Materials Characterization Group leader in Advanced Characterization Department at Idaho National Laboratory (INL). He is the deputy director of the Center for Thermal Energy Transport under Irradiation, an Energy Frontier Research Center funded by the US Department of Energy, Office of Basic Energy Sciences. He received his Ph.D. in Materials Science at the Chinese Academy of Sciences in 2009. He worked as a post-doctoral research associate and assistant scientist at Nagaoka University of Technology in Japan and the University of Wisconsin-Madison before joining INL in 2014. He is the recipient of INL Laboratory Director's 2020 Exceptional Scientific Achievement Award.

*Marat Khafizov*
Dr. Marat Khafizov is an associate professor in Mechanical and Aerospace Engineering at The Ohio State University. He received his Ph.D. degree in Physics from University of Rochester in 2008. He was a postdoctoral research associate and R&D scientist at Idaho National Laboratory (2010-2014), before joining OSU in 2014. His research focus is on understanding the impact of defects and radiation damage on physical properties of materials. He is a Phonon Transport Thrust Lead for the Center for Thermal Energy Transport under Irradiation.

*Gerard H. Lander*
Dr. Gerry Lander obtained his Ph.D. at Cambridge (UK) in 1967. He worked for 19 years at Argonne. For the last 5 years, he was the Director of Argonne's Pulsed Neutron Source. In 1986 he returned to Europe to work at the EU Transuranium Institute in Karlsruhe, where he was Director from 2002-2006, when he retired. He is a Fellow of the APS, was a John Wheatley Fellow at LANL, a consultant to the Japanese Atomic Energy Commission, was awarded an Honorary Professorship by the Polish Academy of Science, and received the Walter Halg Prize from the European Neutron Scattering Association in 2011. He is now a Visiting Professor at Bristol University, UK.

*Michael E. Manley*
Dr. Michael E. Manley is a Senior Researcher at Oak Ridge National Laboratory in the Materials Science and Technology Division. He received his PhD in Materials Science from the California Institute of Technology in 2001 and was awarded the 18th Louis Rosen Prize for his thesis. He was a Director's Postdoctoral Fellow at Los Alamos National Laboratory before being promoted to Technical Staff Member. He then moved to Lawrence Livermore National Laboratory in 2006 and then to his current position at Oak Ridge in 2012. He won the 2021 TMS FMD Distinguished Scientist Award for his research on the thermophysical properties of materials and the discovery of intrinsic localized modes in anharmonic materials. He is Science Question Lead for the Center for Thermal Energy Transport under Irradiation, funded by the U.S. DOE's Energy Frontier Research Centers program.

*J. Matthew Mann*
Dr. Matthew Mann is a Senior Research Chemist for the Sensors Directorate with the Air Force Research Laboratory. He received his Ph.D. in Inorganic Chemistry from Clemson University in 2009. He is the recipient of the 2019 Dr. Brian Kent Award for his innovative research in the area of neutron detection. His current research interests include crystal growth and characterization of radioactive materials, neutron detectors, novel laser hosts, and next generation high powered electronics.

*Chris A. Marianetti*
Dr. Chris Marianetti is an Associate Professor of Materials Science and Applied Physics and Applied Mathematics at Columbia University. Marianetti's research focuses on computing the behavior of crystalline materials from the first-principles of quantum mechanics. Applications span the periodic table, with particular emphasis on materials with strong electronic correlations. Dr. Marianetti received a Ph.D. in Materials Science and Engineering from MIT in 2004, and B.S./M.S. from The Ohio State University. He held postdoctoral positions in the Department of Physics at Rutgers University and at Lawrence Livermore National Laboratory. Dr. Marianetti was the recipient of the NSF CAREER and DARPA Young Faculty awards.

*Karl Rickert*
Dr. Karl Rickert is currently a Senior Chemist at KBR, contracted to the Sensors Directorate at the Air Force Research Laboratory. Prior to this, he was an Oak Ridge Institute for Science and Education Postdoctoral Fellow at the Air Force Institute of Technology after receiving his PhD from Northwestern University in 2016. His work focuses on the growth and characterization of actinide oxides and their implementation in detection devices.

*Farida A. Selim*
Dr. Farida Selim is an associate Professor of Physics and a member of the Center for Photochemical Science at Bowling Green State University. She obtained her PhD in a joint program between Harvard and Alexandria University in Egypt. She is the lead of Thrust 1 "Point Defects" for the Center for Fundamental Understanding of Transport under Reactor Extreme, an Energy Frontier Research Center funded by the US Department of Energy, Office of Basic Energy Sciences. She is known for inventing a new positron annihilation spectroscopy technique (Gamma Induced Positron Spectroscopy). Her research spans a wide range of areas in defect studies, exploring novel electronic phenomena in wide band gap oxides and developing new instrumentation for defect and photoemissions studies.

*Michael R. Tonks*
Dr. Michael R. Tonks is a Professor and Alumni Professor of Materials Science and Engineering and Nuclear Engineering at the University of Florida. He received his Ph.D. in Mechanical Engineering from the University of Illinois at Urbana-Champaign in 2008. Before coming to UF he was a staff scientist at Idaho National Laboratory for six years and an Assistant Professor at Penn State for two years. He is a recipient of the Presidential Early Career Award for Scientists and Engineers and the INL Early Career Exceptional Achievement Award. His work uses multiscale modeling and simulation to investigate the co-evolution of microstructure and properties in materials in harsh environments.

*Janelle P. Wharry*
Dr. Janelle P. Wharry is an Associate Professor in the School of Materials Engineering at Purdue University. Her research aims to understand structure-property-functionality relationships in irradiated materials, with an emphasis on deformation mechanisms and mechanical behavior at the nano/microscale. She is a recipient of the US Department of Energy Early Career Award and National Science Foundation CAREER Award. She is an Editor of Materials Today Communications. She received her Ph.D. from the University of Michigan in 2012.

**ACKNOWLEDGMENT**
D.H.H., A.E., M.S.B., C.A.D., K.G., L.H., M.K., M.E.M., C.A.M., J. M. M., K. R. and J.P.W. acknowledge support from TETI (Thermal Energy Transport under Irradiation), an Energy Frontier Research Center funded by the US Department of Energy, Office of Science, Office of Basic Energy Sciences. F.A.S. acknowledges support from FUTURE (Fundamental Understanding of Transport under Reactor Extremes), an Energy Frontier Research Center funded by the U.S. Department of Energy, Office of Science, Office of Basic Energy Sciences. M.R.T. and M.W.D.C. acknowledge support from the Nuclear Energy Advanced Modeling and Simulation (NEAMS) program funded by the US Department of Energy, Office of Nuclear Energy.

## References

[1] BP (2020) Statistical Review of World Energy, https://www.bp.com/en/global/corporate/energy-economics/statistical-review-of-world-energy.html (Accessed March 1, 2021).

[2] Energy Density Calculations of Nuclear Fuel, https://whatisnuclear.com/energy-density.html (Accessed March 1, 2021).

[3] Lamarsh, J. R.; Baratta, A. J. *Introduction to Nuclear Engineering*; Prentice Hall: New Jersey, **2001**; p 117.

[4] Ronchi, C.; Sheindlin, M.; Musella, M.; Hyland, G. J. Thermal Conductivity of Uranium Dioxide up to 2900 K from Simultaneous Measurement of the Heat Capacity and Thermal Diffusivity, *J. Appl. Phys.* **1999**, 85, 776-789.

[5] Ronchi, C.; Sheindlin, M.; Staicu, D.; Kinoshita, M. Effect of Burn-Up on the Thermal Conductivity of Uranium Dioxide up to 100 MWdt$^{-1}$. *J. Nucl. Mater.* **2004**, 327, 58-76.

[6] Rondinella, V. V.; Wiss, T. The High Burn-Up Structure in Nuclear Fuel, *Materials Today.* **2010**, 13, 24–32.

[7] Lassmann, K.; Walker, C. T.; Van de Laar, J.; Lindström, F. Modelling the High Burnup $UO_2$ Structure in LWR Fuel. *J. Nucl. Mater.* **1995**, 226, 1-8.

[8] Walker, C. T.; Staicu, D.; Sheindilin, M.; Papaioannou, D.; Goll, W.; Sontheimer, F. On the Thermal Conductivity of $UO_2$ Nuclear Fuel at a High Burn-Up of Around 100 MWd/kgHM. *J. Nucl. Mater.* **2006**, 350, 19-39.

[9] Roostaii, B.; Kazeminejad, H.; Khakshournia, S. Including High Burnup Structure Effect in the $UO_2$ Fuel Thermal Conductivity Model. *J. Nucl. Mater.* **2021**, 131, 103561.

[10] Sawbridge, P. T.; Baker, C.; Cornell, R. M.; Jones, K. W.; Reed, D.; Ainscough, J. B. The Irradiation Performance of Magnesia-Doped $UO_2$ fuel. *J. Nucl. Mater.* **1980**, 95, 119-128.

[11] Arborelius, J.; Backman, K.; Hallstadius, L.; Limbäck, M.; Nilsson, J.; Rebensdorff, B.; Zhou, G.; Kitano, K.; Löfstrom R.; Rönnberg, G. Advanced Doped $UO_2$ Pellets in LWR Applications. *J. Nucl. Sci. Technol.* **2006**, 43, 967-976.

[12] Cooper, M. W. D.; Stanek, C. R.; Andersson, D. A. The Role of Dopant Charge State on Defect Chemistry and Grain Growth of Doped $UO_2$. *Acta Mater.* **2018**, 150, 403-413.

[13] Cooper, M. W. D.; Pastore, G.; Che, Y.; Matthews, C.; Forslund, A.; Stanek, C. R.; Shivan, K.; Tverberg, T.; Gamble, K. A.; Mays, B.; et al. Fission Gas Diffusion and Release for $Cr_2O_3$-Doped $UO_2$: From the Atomic to the Engineering Scale. *J. Nucl. Mater.* **2021**, 545, 152590.

[14] Jovani-Abril, R.; Gibilaro, M.; Janssen, A.; Eloirdi, R.; Somers, J.; Spino, J.; Malmbeck, R. Synthesis of Nc-$UO_2$ by Controlled Precipitation in Aqueous Phase. *J. Nucl. Mater.* **2016**, 477, 298-304.

[15] Spino, J.; Santa-Cruz, H.; Jovani-Abril, R.; Birtcher, R.; Rerriro, C. Bulk-Nanochrystalline Oxide Nuclear Fuels – An Innovative Material Option for Increasing Fission Gas Retention, Plasticity and Radiation-Tolerance. *J. Nucl. Mater.* **2012**, 422, 27-44.

[16] Zhou, W; Zhou, W. Enhanced Thermal Conductivity Accident Tolerant Fuels for Improved Reactor Safety – A Comprehensive Review. *Ann. of Nucl. Energy* **2018**, 119, 66–86.

[17] Gong, B.; Yao, T.; Lei, P.; Cai, L.; Metzger, K. E.; Lahoda, E. J.; Boylan, F. A.; Mohamad, A., Harp, J.; Nelson, A. T.; et al. $U_3Si_2$ and $UO_2$ Composites Densified by Spark Plasma Sintering for Accident-Tolerant Fuels. *J. Nucl. Mater.* **2020**, 534, 152147.

[18] S.C. Finkeldei, J.O. Kiggans, R.D. Hunt, A.T. Nelson, K.A. Terrani, Fabrication of $UO_2$-Mo Composite Fuel with Enhanced Thermal Conductivity from Sol-Gel Feedstock. *J. Nucl. Mater.* **2019**, 520, 56–64.

[19] Li, B.; Yang, Z.; Jia, J.; Zhang, P.; Gao, R.; Liu, T.; Li, R.; Huang, H.; Sun, M.; Mazhao, D. High Temperature Thermal Physical Performance of SiC/$UO_2$ composites up to 1600 °C. *Ceram. Int.* **2018**, 44, 10069–10077.

[20] Yao, T.; Xin, G.; Scott, S. M.; Gong, B.; Lian, J. Thermally-Conductive and Mechanically-Robust Graphene Nanoplatelet Reinforced $UO_2$ Composite Nuclear Fuels. *Sci. Rep.* **2018**, 8, 2987.

[21] Li, B.; Yang, Z.; Jia, J.; Zhong, Y.; Liu, X.; Zhang, P.; Gao, R.; Liu, T.; Li, R.; Huang, H.; et al. High Temperature Thermal Physical Performance of BeO/$UO_2$ Composites Prepared by Spark Plasma Sintering (SPS). *Scripta Mater.* **2018**, 142, 70–73.

[22] Yang, J. H.; Kim, D.-J.; Kim, K. S.; Koo, Y.-H. $UO_2$–UN Composites with Enhanced Uranium Density and Thermal Conductivity. *J. Nucl. Mater.* **2015**, 465, 509–515.

[23] Yeo, S.; Baney, R.; Subhash, G.; Tulenko, J. The Influence of SiC Particle Size and Volume Fraction on the Thermal Conductivity of Spark Plasma Sintered $UO_2$–SiC Composites. *J. Nucl. Mater.* **2013**, 442, 245–252.

[24] Ishimoto, S.; Hirai, M.; Ito, K.; Korei, Y. Thermal Conductivity of $UO_2$-BeO Pellet. *J. Nucl. Sci. Technol.* **1996**, 33, 134–140.

[25] Liu, X. Y.; Cooper, M. W. D.; McClellan, K.J.; Lashley, J. C.; Byler, D.D.; Bell, B. D. C.; Grimes, R. W.; Stanek, C. R.; Andersson, D. A. Molecular Dynamics Simulation of Thermal Transport in $UO_2$ Containing Uranium, Oxygen, and Fission-product Defects. *Phys. Rev. Appl.* **2016**, 6, 044015.

[26] Tonks, M. R.; Liu, X. Y.; Andersson, D.; Perez, D.; Chernatynskiy, A.; Pastore, G.; Stanek, C. R.; Williamson, R. Development of a Multiscale Thermal Conductivity Model for Fission Gas in $UO_2$. *J. Nucl. Mater.* **2016**, 469, 89-98.

[27] Pang, J. W.; Buyers, W. J.; Chernatynskiy, A.; Lumsden, M. D.; Larson, B. C.; Phillpot, S. R. Phonon Lifetime Investigation of Anharmonicity and Thermal Conductivity of $UO_2$ by Neutron Scattering and Theory. *Phys. Rev. Lett.* **2013**, 110, 157401.

[28] Khafizov, M.; Riyad, M. F.; Wang, Y. Z.; Pakarinen, J.; He, L. F.; Yao, T. K.; El-Azab, A.; Hurley, D., Combining Mesoscale Thermal Transport and X-ray Diffraction Measurements to Characterize Early-Stage Evolution of Irradiation-Induced Defects in Ceramics. *Acta Mater.* **2020,** 193, 61-70.

[29] Dennett, C. A.; Hua, Z. L.; Khanolkar, A.; Yao, T. K.; Morgan, P. K.; Prusnick, T. A.; Poudel, N.; French, A.; Gofryk, K.; He, L. F.; et al. The Influence of Lattice Defects, Recombination, and Clustering on Thermal Transport in Single Crystal Thorium Dioxide. *APL Mater.* **2020,** 8, 111103.

[30] Klemens, P. G. The Scattering of Low-Frequency Lattice Waves by Static Imperfections. *Proc. Phys. Soc.* **1955**, A68, 1113-1128.

[31] Callaway, J.; von Baeyer, H. C. Effect of Point Imperfections on Lattice Thermal Conductivity. *Phys. Rev.* **1960**, 120, 1149-1154.

[32] Slack, G. A. Thermal Conductivity of Pure and Impure Silicon, Silicon Carbide, and Diamond. *J. Appl. Phys.* **1964**, 35, 3460-3466.

[33] Berman, R.; Foster, E. L.; Schneidmesser, B.; Tirmizi, S. M. A. Effects of Irradiation on the Thermal Conductivity of Synthetic Sapphire. *J. Appl. Phys.* **1960**, 31, 2156-2159.

[34] Broido, D. A.; Malorny, M.; Birner, G.; Mingo, N.; Stewart, D. A. Intrinsic Lattice Thermal Conductivity of Semiconductors from First Principles. *Appl. Phys. Lett.* **2007**, 91, 231922.

[35] Ward, A.; Broido, D. A.; Stewart, D. A.; Deinzer, G. Ab Initio Theory of the Lattice Thermal Conductivity in Diamond. *Phys. Rev. B* **2009**, 80, 125203.

[36] Esfarjani, K.; Chen, G.; Stokes, H. T. Heat Transport in Silicon from First-Principles Calculations. *Phys. Rev. B* **2011**, 84, 085204.

[37] Mingo, N.; Yang, L. Phonon Transport in Nanowires Coated with an Amorphous Material: An Atomistic Green's Function Approach. *Phys. Rev. B* **2003**, 68, 245406.

[38] Pernot, G.; Stoffel, M.; Savic, I.; Pezzoli, F.; Chen, P.; Savelli, G.; Jacquot, A.; Schumann, J.; Denker, U.; Mönch I.; et al. Precise Control of Thermal Conductivity at the Nanoscale Through Individual Phonon-Scattering Barriers. *Nat. Mater.* **2010**, 9, 491–495.

[39] Protik, N. H.; Carrete, j.; Katcho, N. A.; Mingo, N.; Broido, D. Ab Initio Study of the Effect of Vacancies on the Thermal Conductivity of Boron Arsenide. *Phys. Rev. B* **2016**, 94, 045207.

[40] Delaire, O.; Ma, J.; Marty, K.; May, A. F.; McGuire, M. A.; Du, M. -H.; Singh, D. J.; Podlesnyak, A.; Ehlers, G.; Lumsden, M. D.; et al. Giant Anharmonic Phonon Scattering in PbTe. *Nat. Mater.* **2011**, 10, 614–619.

[41] Burkel, E. Phonon Spectroscopy by Inelastic X-ray Scattering. *Rep. Prog. Phys.* **2000**, 63, 171-232.

[42] Cahill, D. G.; Ford, W. K.; Goodson, K. E.; Mahan, G. D.; Majumdar, A.; Maris, H. J.; Merlin, R.; Phillpot, S. R. Nanoscale Thermal Transport. *J. Appl. Phys.* **2003**, 93, 793-818.

[43] Broido, D. A.; Ward, A.; Mingo, N. Lattice Thermal Conductivity of Silicon from Empirical Interatomic Potentials. *Phys. Rev. B* **2005**, 72, 014308.

[44] Omini M.; Sparavigna, A. Beyond the Isotropic-Model Approximation in the Theory of Thermal Conductivity. *Phys. Rev. B* **1996**, 53, 9064-9073.

[45] Turney, J. E.; Landry, E. S.; McGaughey A. J. H.; Amon, C. H. Predicting Phonon Properties and Thermal Conductivity from Anharmonic Lattice Dynamics Calculations and Molecular Dynamics Simulations. *Phys. Rev. B.* **2009**, 79, 064301.

[46] Squires, G. L. *Introduction to the Theory of Thermal Neutron Scattering;* Cambridge Univ. Press: Cambridge, **2012;** p 25.

[47] Kittel, C. *Introduction to Solid State Physics*; Wiley: New York, **1976;** p 89.

[48] Fultz, B. Vibrational Thermodynamics of Materials. *Prog. in Mater. Sci.* **2010**, 55, 247-352.

[49] Dolling, G.; Cowley, R. A.; Woods, A. D. B. The Crystal Dynamics of Uranium Dioxide. *Can. J. Phys.* **1965**, 43, 1397-1413.

[50] Clausen, K.; Hayes, W.; Macdonald, J. E.; Osborn, R.; Schnabel, P. G.; Hutchings, M. T.; Magerl, A. Inelastic Neutron Scattering Investigation of the Lattice Dynamics of $ThO_2$ and $CeO_2$. *J. Chem. Soc., Faraday Trans. 2: Mol. Chem. Phys.* **1987**, 83, 1109–1112.

[51] Brockhouse, B. N.; Stewart, A. T. Scattering of Neutrons by Phonons in an Aluminum Single Crystal. *Phys. Rev.* **1955**, 100, 756-757.

[52] Woods, A. D. B.; Cochran, W.; Brockhouse, B. N. Lattice Dynamics of Alkali Halide Crystals. *Phys. Rev.* **1960**, 119, 980-999.

[53] Verbeni, R.; Sette, F.; Krisch, M. H.; Bergmann, U.; Gorges, B.; Halcoussis, C.; Martel, K.; Masciovecchio, C.; Ribois, J. F.; Ruocco, G.; et al. X-ray Monochromator with $2 \times 10^8$ Energy Resolution. *J. Synchrotron Radiat.* **1996**, 3, 62-64.

**[54]** Sinn, H.; Alp, E.E.; Barraza, J.; Bortel, G.; Burkel, E.; Shu, D.; Sturhahn, W.; Sutter, J.P.; Toellner, T.S.; et al. An Inelastic X-ray Spectrometer with 2.2 meV Energy Resolution. *Nucl. Instrum. Meth. Phys. Res. A* **2001**, 467-468, 1545–1548.

[55] Alatas, A.; Leu, B.M.; Zhao, J.; Yavas, H.; Toellner, T.S.; Alp, E. E. Improved Focusing Capability for Inelastic X-ray Spectrometer at 3-ID of the APS: A Combination of Toroidal and Kirkpatrick-Baez (KB) Mirrors. *Nucl. Instrum. Methods Phys. Res. Sect. A* **2011**, 649, 166–168.

[56] Toellner, T. S.; Alatas, A.; Said, A. H. Six-Reflection meV-Monochromator for Synchrotron Radiation. *J. Synchrotron Radiat.* **2011**, 18, 605–611.

[57] Manley, M. E.; Lander, G. H.; Sinn, H.; Alatas, A.; Hults, W. L.; McQueeney, R. J.; Smith, J. L.; Willit, J. Phonon Dispersion in Uranium Measured using Inelastic X-ray Scattering. *Phys. Rev. B* **2003**, 67, 052302.

[58] Wong, J.; Krisch, M.; Farber, D. L.; Occelli, F.; Schwartz, A. J.; Chiang, T.-C., Wall, M.; Boro, C., Xu, R. Phonon Dispersion of FCC d-Plutonium-Gallium by Inelastic X-ray Scattering. *Science* **2003**, 301, 1078-1080.

[59] Manley, M. E.; Jeffries, J. R.; Said, A. H.; Marianetti, C. A.; Cynn, H.; Leu, B. M.; Wall, M. A. Measurement of the Phonon Density of States of $PuO_2$(+2% Ga): A Critical Test of Theory. *Phys. Rev. B* **2012**, 85, 132301.

[60] Maldonada, P.; Paolasini, L.; Oppeneer, P. M.; Forrest, T. R.; Prodi, A.; Magnani, N.; Bosak, A.; Lander, G. H.; Caciuffo, R. Crystal Dynamics and Thermal Properties of Neptunium Dioxide. *Phys. Rev. B* **2016**, B93, 144301.

[61] Stone, M. B.; Niedziela, J. L.; Abernathy, D. L.; DeBeer-Schmitt, L.; Ehlers, G.; Garlea, O.; Granroth, G. E.; Graves-Brook, M.; Kolesnikov, A. I.; Podlesnyak, A.; et al. A Comparison of Four Direct Geometry Time-of-Flight Spectrometers at the Spallation Neutron Source. *Rev. Sci. Instrum.* **2014**, 85, 045113.

[62] Janoschek, M.; Das, P.; Chakrabarti, B.; Abernathy, D. L.; Lumsden, M. D.; Lawrence, J. M.; Thompson, J. D.; Lander, G. H.; Mitchell, J. N.; Richmond, S.; et al. The Valence-Fluctuating Ground State of Plutonium. *Sci. Adv.* **2015**, 1, e1500188.

[63] Pang, J. W. L.; Chernatynskiy, A.; Larson, B, C.; Buyers, W. J. L.; Abernathy, D. L.; McClellan, K. J.; Phillpot, S. R. Phonon Density of States and Anharmonicity of $UO_2$. *Phys. Rev. B* **2014**, 89, 115132. 38

[64] Yin, Q.; Savrasov, S. Y. Origin of the Low Thermal Conductivity in Nuclear Fuels. *Phys. Rev. Lett.* **2008**, 100, 225504.

[65] Rennie, S.; Lawrence Bright, E.; Darnbrough, J. E.; Paolasini, L; Bosak, A; Smith, A. D.; Mason, N.; Lander, G. H.; Springell, R. Study of Phonons in Irradiated Epitaxial Thin Films of $UO_2$. *Phys. Rev. B* **2018**, 97, 224303.

[66] Weisensee, P. B.; Feser, J. P.; Cahill, D. G. Effect of Ion Irradiation on the Thermal Conductivity of $UO_2$ and $U_3O_8$ Epitaxial Layers. *J. Nucl. Mater.* **2013**, 443, 212-217.

[67] Bryan, M. S.; Fu, L.; Rickert, K.; Turner, D.; Prusnick, T. A.; Mann, J. M.; Abernathy, D. L.; Marianetti, C. A., Manley, M. E. Nonlinear Propagating Modes Beyond the Phonons in Fluorite-Structured Crystals. *Comm. Phys.* **2020**, 3, 217.

[68] Manley, M. E.; A. J. Sievers; J. W. Lynn; S. A. Kiselev; N. I. Agladze; Y. Chen; A. Llobet; A. Alatas. Intrinsic Localized Modes Observed in the High-Temperature Vibrational Spectrum of NaI. *Phys. Rev. B* **2009**, 79, 134304.

[69] Hohenberg, P.; Kohn, W. Inhomogeneous Electron Gas. *Phys. Rev.* **1964**, 136, 864-871.

[70] Kohn, W.; Sham, L. J. Self-Consistent Equations Including Exchange and Correlation Effects. *Phys. Rev.* **1965**, 140, 1133-1138.

[71] Kohn, W. Nobel Lecture: Electronic Structure of Matter-Wave Functions and Density Functionals. *Rev. Mod. Phys.* **1999**, 71, 1253-1266.

[72] Jones, R. O.; Gunnarsson, O. The Density Functional Formalism, Its Applications and Prospects. *Rev. Mod. Phys.* **1989**, 61, 689-746.

[73] Jones, R. O. Density Functional Theory: Its Origins, Rise to Prominence, and Future. *Rev. Mod. Phys.* **2015**, 87, 897-923.

[74] Kummel, S.; Kronik, L. Orbital-Dependent Density Functionals: Theory and Applications. *Rev. Mod. Phys.* **2008**, 80, 3-60.

[75] Ernzerhof, M.; Scuseria, G. E. Assessment of the Perdew-Burke-Ernzerhof Exchange-Correlation Functional. *J. Chem. Phys.* **1999**, 110, 5029-5036.

[76] Adamo, C.; Barone, V. Toward Reliable Density Functional Methods Without Adjustable Parameters: The PBE0 model. *J. Chem. Phys.* **1999**, 110, 6158-6170.

[77] Oba, F.; Togo, A.; Tanaka, I.; Paier, J.; Kresse, G. Defect Energetics in ZnO: A Hybrid Hartree-Fock Density Functional Study. *Phys. Rev. B* **2008**, 77, 245202.

[78] Franchini, C.; Bayer, V.; Podloucky, R.; Paier, J.; Kresse, G. Density Functional Theory Study of MnO by a Hybrid Functional Approach. *Phys. Rev. B* **2005**, 72, 045132.

[79] Tran, F.; Blaha, P.; Schwarz, K.; Novak, P. Hybrid Exchange-Correlation Energy Functionals for Strongly Correlated Electrons: Applications to Transition-Metal Monoxides. *Phys. Rev. B* **2006**, 74, 155108.

[80] Franchini, C.; Podloucky, R.; Paier, J.; Marsman, M.; Kresse, G. Ground-State Properties of Multivalent Manganese Oxides: Density Functional and Hybrid Density Functional Calculations. *Phys. Rev. B* **2007**, 75, 195128.

[81] Rodl, C.; Fuchs, F.; Furthmuller, J.; Bechstedt, F. Quasiparticle Band Structures of the Antiferromagnetic Transition-Metal Oxides MnO, FeO, CoO, and NiO. *Phys. Rev. B* **2009**, 79, 235114.

[82] Paier, J.; Marsman, M.; Kresse, G. Why Does the b3lyp Hybrid Functional Fail for Metals? *J. Chem. Phys.* **2007**, 127, 024103.

[83] Heyd, J.; Scuseria, G. E.; Ernzerhof, M. Hybrid Functionals Based on a Screened Coulomb Potential. *J. Chem. Phys.* **2003**, 118, 8207-8215.

[84] Eyert, V. The Metal-Insulator Transitions of $VO_2$: A Band Theoretical Approach. *Annalen Der Physik* **2002**, 11, 650-702.

[85] Grau-crespo, R.; Wang, H.; Schwingenschloegl, U. Why the Heyd-Scuseria-Ernzerhof hybrid Functional Description of $VO_2$ Phases is Not Correct. *Phys. Rev. B* **2012**, 86, 081101.

[86] Kosuge, K. The Phase Transition in $VO_2$. *J. Phys. Soc. Japan* **1967**, 22, 551-557.

[87] Qazilbash, M. M.; Burch, K. S.; Whisler, D.; Shrekenhamer, D.; Chae, B. G.; Kim, H. T.; Basov, D. N. Correlated Metallic State of Vanadium Dioxide. *Phys. Rev. B* **2006**, 74, 205118.

[88] Paier, J.; Hirschl, R.; Marsman, M.; Kresse, G. The Perdew-Burke-Ernzerhof Exchange-Correlation Functional Applied to the g2-1 Test Set using a Plane-Wave Basis Set. *J. Chem. Phys.* **2005**, 122, 234102.

[89] Anisimov, V. I.; Aryasetiawan, F.; Lichtenstein, A. I. First-Principles Calculations of the Electronic Structure and Spectra of Strongly Correlated Systems: The LDA+U Method. *J. Phys. Condens. Matter* **1997**, 9, 767-808.

[90] Ivady, V.; Armiento, R.; Szasz, K.; Janzen, E.; Gali, A.; Abrikosov, I. A. Theoretical Unification of Hybrid-DFT and DFT Plus U Methods for the Treatment of Localized Orbitals. *Phys. Rev. B* **2014**, 90, 035146.

[91] Agapito, L. A.; Curtarolo, S.; Nardelli, M. B. Reformulation of DFT Plus U as a Pseudohybrid Hubbard Density Functional for Accelerated Materials Discovery. *Phys. Rev. X* **2015**, 5, 011006

[92] Luttinger, J. M.; Ward, J. C. Ground-State Energy of a Many-Fermion System. II. *Phys. Rev.* **1960**, 118, 1417-1427.

[93] Baym, G.; Kadanoff, L. P. Conservation Laws and Correlation Functions. *Phys. Rev.* **1961**, 124, 287-299.

[94] Dedominicis, C.; Martin, P. C. Stationary Entropy Principle + Renormalization in Normal + Superfluid Systems. I. Algebraic Formulation. *J. Math. Phys.* **1964**, 5, 14-30.

[95] Chitra, R.; Kotliar, G. Effective-Action Approach to Strongly Correlated Fermion Systems. *Phys. Rev. B* **2001**, 63, 115110.

[96] Gutzwiller, M. C. Effect of Correlation on Ferromagnetism of Transition Metals. *Phys. Rev. Lett.* **1963**, 10, 159-162.

[97] Kanamori, J. Electron Correlation and Ferromagnetism of Transition Metals. *Prog. Theor. Phys.* **1963**, 30, 275-289.

[98] Hubbard, J. Electron Correlations in Narrow Energy Bands. *Proc. R. Soc. A* **1963**, 276, 238-257.

[99] Lieb, E. H.; Wu, F. Y. Absence of Mott Transition in an Exact Solution of Short-Range 1-Band Model in 1 Dimension. *Phys. Rev. Lett.* **1968**, 20, 1445-1448.

[100] Lieb, E. H.; Wu, F. Y. The One-Dimensional Hubbard Model: A Reminiscence. *Physica A* **2003**, 321, 1-27.

[101] Georges, A.; Kotliar, G.; Krauth, W.; Rozenberg, M. J. Dynamical Mean-Field Theory of Strongly Correlated Fermion Systems and the Limit of Infinite Dimensions. *Rev. Mod. Phys.* **1996**, 68, 13-125.

[102] Kotliar, G.; Savrasov, S. Y.; Haule, K.; Oudovenko, V. S.; Parcollet, O.; Marianetti, C. A. Electronic Structure Calculations with Dynamical Mean-Field Theory. *Rev. Mod. Phys.* **2006**, 78, 865-951.

[103] Gull, E.; Millis, A. J.; Lichtenstein, A. I.; Rubtsov, A. N.; Troyer, M.; Werner, P. Continuous-Time Monte-Carlo Methods for Quantum Impurity Models. *Rev. Mod. Phys.* **2011**, 83, 349.

[104] Haule, K. Quantum Monte Carlo Impurity Solver for Cluster Dynamical Mean-Field Theory and Electronic Structure Calculations with Adjustable Cluster Base. *Phys. Rev. B* **2007**, 75, 155113.

[105] Troyer, M.; Wiese, U. J. Computational Complexity and Fundamental Limitations to Fermionic Quantum Monte Carlo Simulations. *Phys. Rev. Lett.* **2005**, 94, 170201.

[106] Park, H.; Millis, A. J.; Marianetti, C. A. Total Energy Calculations Using DFT Plus DMFT: Computing the Pressure Phase Diagram of the Rare Earth Nickelates. *Phys. Rev. B* **2014**, 89, 245133.

[107] Park, H.; Millis, A. J.; Marianetti, C. A. Computing Total Energies in Complex Materials Using Charge Self-Consistent DFT+DMFT. *Phys. Rev. B* **2014**, 90, 235103.

(108) Haule, K.; Birol, T. Free Energy from Stationary Implementation of the DFT+DMFT Functional. *Phys. Rev. Lett.* **2015**, 115, 256402.

(109) Delange, P.; Ayral, T.; Simak, S. I.; Ferrero, M.; Parcollet, O.; Biermann, S.; Pourovskii, L. Large Effects of Subtle Electronic Correlations on the Energetics of Cacancies in Alpha-Fe. *Phys. Rev. B* **2016**, 94, 100102.

(110) Leonov, I.; Anisimov, V.I.; Vollhardt, D. First-Principles Calculation of Atomic Forces and Structural Distortions in Strongly Correlated Materials. *Phys. Rev. Lett.* **2014**, 112, 146401.

(111) Haule, K.; Pascut, G. L. Mott Transition and Magnetism in Rare Earth Nickelates and Its Fingerprint on the X-ray Scattering. *Sci. Rep.* **2017**, 7, 10375.

(112) Anisimov, V. I.; Zaanen, J.; Andersen, O. K. Band Theory and Mott Insulators - Hubbard-U Instead of Stoner-I. *Phys. Rev. B* **1991**, 44, 943-954.

(113) Fu, L.; Kornbluth, M.; Cheng, Z.; Marianetti, C. A. Group Theoretical Approach to Computing Phonons and Their Interactions. *Phys. Rev. B* **2019**, 100, 014303.

(114) Martin, R. M. *Electronic Structure: Basic Theory and Practical Methods;* Cambridge University Press: New York, **2008**; p 52.

(115) Baroni, S.; deGironcoli, S.; DalCorso, A.; Giannozzi, P. Phonons and Related Crystal Properties from Density-Functional Perturbation Theory. *Rev. Mod. Phys.* **2001**, 73, 515-562.

(116) Kunc, K.; Martin, R. M. Ab Initio Force Constants of GaAs: A New Approach to Calculation of Phonons and Dielectric Properties. *Phys. Rev. Lett.* **1982**, 48, 406-409.

(117) Parlinski, K.; Li, Z.; Kawazoe, Y. First-Principles Determination of the Soft Mode in Cubic $ZrO_2$. *Phys. Rev. Lett.* **1997**, 78, 4063-4066.

(118) Ziman, J. M. *Electrons and Phonons: the Theory of Transport Phenomena in Solids;* Oxford University Press: London, **1960**; p 264.

(119) Fugallo, G.; Lazzeri, M.; Paulatto, L.; Mauri, F. Ab Initio Variational Approach for Evaluating Lattice Thermal Conductivity. *Phys. Rev. B* **2013**, 88, 045430.

(120) Fugallo, G.; Colombo, L. Calculating Lattice Thermal Conductivity: A Synopsis. *Physica Scripta* **2018**, 93, 043002.

(121) Lindsay, L.; Katre, A.; Cepellotti, A.; Mingo, N. Perspective on Ab Initio Phonon Thermal Transport. *J. Appl. Phys.* **2019**, 126, 050902.

(122) Feng, T. L.; Ruan, X. L. Quantum Mechanical Prediction of Four-Phonon Scattering Rates and Reduced Thermal Conductivity of Solids. *Phys. Rev. B* **2016**, 93, 045202.

(123) Feng, T. L.; Lindsay, L.; Ruan, X. L. Four-Phonon Scattering Significantly Reduces Intrinsic Thermal Conductivity of Solids. *Phys. Rev. B* **2017**, 96, 161201.

(124) Ravichandran, N. K.; Broido, D. Unified First-Principles Theory of Thermal Properties of Insulators. *Phys. Rev. B* **2018**, 98, 085205.

(125) Ravichandran, N. K.; Broido, D. Phonon-Phonon Interactions in Strongly Bonded Solids: Selection Rules and Higher-Order Processes. *Phys. Rev. X* **2020**, 10, 021063.

(126) Carbogno, C.; Ramprasad, R.; Scheffler, M. Ab Initio Green-Kubo Approach for the Thermal Conductivity of Solids. *Phys. Rev. Lett.* **2017**, 118, 175901.

(127) Kang, J.; Wang, L. W. First-Principles Green-Kubo Method for Thermal Conductivity Calculations. *Phys. Rev. B* **2017**, 96, 020302.

(128) Lu, Y.; Yang, Y.; Zhang, P. Thermodynamic Properties and Structural Stability of Thorium Dioxide. *J. Phys. Condens. Matter* **2012**, 24, 225801.

(129) Szpunar, B.; Szpunar, J. A. Theoretical Investigation of Structural and Thermo-Mechanical Properties of Thoria up to 3300 K Temperature. *Solid State Sci.* **2014**, 36, 35-40.

(130) Szpunar, B.; Szpunar, J. A.; Sim, K. S. Theoretical Investigation of Structural and Thermo-Mechanical Properties of Thoria. *J. Phys. Chem. Solids* **2016**, 90, 114-120.

(131) Malakkal, L.; Szpunar, B.; Zuniga, J. C.; Siripurapu, R. K.; Szpunar, J. A. First Principles Calculation of Thermo-Mechanical Properties of Thoria Using Quantum Espresso. *Int. J. Comput. Mater. Sci. Eng.* **2016**, 5, 1650008.

(132) Dennett, C. A.; Deskins, W. R.; Khafizov, M.; Hua, Z.; Khanolkar, A.; Bawane, K.; Fu, L.; Mann, J. M.; Marianetti, C. A.; He, L.; et al. An Integrated Experimental and Computational Investigation of Defect and Microstructural Effects on Thermal Transport in Thorium Dioxide. *Acta Mater.* **2021**, 213, 116934.

(133) Toropova, A.; Marianetti, C. A.; Haule, K.; Kotliar, G. One-Electron Physics of the Actinides. *Phys. Rev. B* **2007**, 76, 155126.

(134) Wen, X.; Martin, R. L.; Henderson, T. M.; Scuseria, G. E. Density Functional Theory Studies of the Electronic Structure of Solid State Actinide Oxides. *Chem. Rev.* **2013**, 113, 1063-1096.

(135) Boettger, J. C.; Ray, A. K. All-Electron LCGTO Calculations for Uranium Dioxide. *Int. J. Quantum Chem.* **2000**, 80, 824-830.

(136) Prodan, I. D.; Scuseria, G. E.; Martin, R. L. Assessment of Metageneralized Gradient Approximation and Screened Coulomb Hybrid Density Functionals on Bulk Actinide Oxides. *Phys. Rev. B* **2006**, 73, 045104.

(137) Prodan, I. D.; Scuseria, G. E.; Martin, R. L. Covalency in the Actinide Dioxides: Systematic Study of the Electronic Properties Using Screened Hybrid Density Functional Theory. *Phys. Rev. B* **2007**, 76, 033101.

(138) Laskowski, Robert; Madsen, Georg K.; Blaha, Peter; Schwarz, Karlheinz Magnetic Structure and Electric-Field Gradients of Uranium Dioxide: An Ab Initio Study. *Phys. Rev. B* **2004**, 69, 140408.

(139) Zhou, F.; Ozolins, V. Crystal Field and Magnetic Structure of $UO_2$. *Phys. Rev. B* **2011**, 83, 085106.

(140) Suzuki, M. T.; Magnani, N.; Oppeneer, P. M. Microscopic Theory of the Insulating Electronic Ground States of the Actinide Dioxides $AnO_2$ (An= U, Np, Pu, Am, and Cm). *Phys. Rev. B* **2013**, 88, 195146.

(141) Dudarev, S. L.; Liu, P.; Andersson, D. A.; Stanek, C. R.; Ozaki, T.; Franchini, C. Parametrization of LSDA Plus U for Noncollinear Magnetic Configurations: Multipolar Magnetism in $UO_2$. *Phys. Rev. Mater.* **2019**, 3, 083802.

(142) Pi, S. T.; Nanguneri, R.; Savrasov, S. Calculation of Multipolar Exchange Interactions in Spin-Orbital Coupled Systems. *Phys. Rev. Lett.* **2014**, 112, 077203.

(143) Sanati, M.; Albers, R. C.; Lookman, T.; Saxena, A. Elastic Constants, Phonon Density of States, and Thermal Properties of $UO_2$. *Phys. Rev. B* **2011**, 84, 014116.

(144) Wang, B. T.; Zhang, P.; Lizarraga, R.; Di Marco, I.; Eriksson, O. Phonon Spectrum, Thermodynamic Properties, and Pressure-Temperature Phase Diagram of Uranium Dioxide. *Phys. Rev. B* **2013**, 88, 104107.

(145) Kolorenc, J.; Shick, A. B.; Lichtenstein, A. I. Electronic Structure and Core-Level Spectra of Light Actinide Dioxides in the Dynamical Mean-Field Theory. *Phys. Rev. B* **2015**, 92, 085125.

(146) Pourovskii, L. V.; Khmelevskyi, S. Quadrupolar Superexchange Interactions, Multipolar Order, and Magnetic Phase Transition in $UO_2$. *Phys. Rev. B* **2019**, 99, 094439.

[147] Lanata, N.; Yao, Y. X.; Deng, X. Y.; Dobrosavljevic, V.; Kotliar, G. Slave Boson Theory of Orbital Differentiation with Crystal Field Effects: Application to $UO_2$. *Phys. Rev. Lett.* **2017**, 118, 126401.

[148] Slack, G. A. Nonmetallic Crystals with High Thermal-Conductivity. *J. Phys. Chem. Solids* **1973**, 34, 321-335.

[149] Lu, Y.; Yang, Y.; Zhang, P. Thermodynamic Properties and Structural Stability of Thorium Dioxide. *J. Phys. Condens. Matter* **2012**, 24, 225801.

[150] Kaur, G.; Panigrahi, P.; Valsakumar, M. C. Thermal Properties of $UO_2$ with a Non-Local Exchange-Correlation Pressure Correction: A Systematic First Principles DFT Plus U Study. *Model. Sim. Mater. Sci. Eng.* **2013**, 21, 065014.

[151] Nakamura, H.; Machida, M. First-Principles Calculation Study on Phonon Thermal Conductivity of Thorium and Plutonium Dioxides: Intrinsic Anharmonic Phonon-Phonon and Extrinsic Grain-Boundary-Phonon Scattering Effects. *J. Nucl. Mater.* **2019**, 519, 45-51.

[152] Jin, M.; Khafizov, M.; Jiang, C.; Zhou, S.; Marianetti, C.; Bryan, M.; Manley, M. E.; Hurley, D. H. Assessment of Empirical Interatomic Potential to Predict Thermal Conductivity in $ThO_2$ and $UO_2$. *J. Phys. Conden. Matter* **2021**, 33, 275402.

[153] Liu, J. Y.; Dai, Z. H.; Yang, X. X.; Zhao, Y. C.; Meng, S. Lattice Thermodynamic Behavior in Nuclear Fuel $ThO_2$ from First Principles. *J. Nucl. Mater.* **2018**, 511, 11-17.

[154] Malakkal, L.; Prasad, A.; Jossou, E.; Ranasinghe, J.; Szpunar, B.; Bichler, L.; Szpunar, J. Thermal Conductivity of Bulk and Porous $ThO_2$: Atomistic and Experimental Study. *J. Alloys Compd.* **2019**, 798, 507-516.

[155] Jones, W. M.; Gordon, J.; Long, E. A. The Heat Capacities of Uranium, Uranium Trioxide, and Uranium Dioxide from 15 K to 300 K. *J. Chem. Phys.* **1952**, 20, 695-699.

[156] Osborne D. W.; Westrum, E. F. Specific Heat of $NpO_2$ at Low Temperature. *J. Chem. Phys.* **1953** 21, 1884-1887.

[157] Frazer, B. C.; Shirane, G.; Cox, D. E.; Olsen, C. E. Neutron Diffraction Study of Antiferromagnetism in $UO_2$. *Phys. Rev.* **1965,** 140, A1448-A1452.

[158] Willis, B. T. M.; Taylor, R. J. Neutron Diffraction Study of Antiferromagnetism in $UO_2$. *Phys. Lett.* **1965**, 17, 188-190.

[159] Burlet, P; Rossat-Mignod, J; Quezel, S.; Vogt, O.; Spirlet, J-C.; Rebizant, J. Neutron Diffraction on Actinides. *J. Less-Common. Met*. **1986**, 121, 121-139.

[160] Blackburn, E.; Caciuffo, R.; Magnani, N.; Santini P.; Brown P J.; Enderle, M.; Lander, G. H. Spherical Neutron Spin Polarimetry of Anisotropic Magnetic Fluctuations in $UO_2$. *Phys. Rev. B* **2005**, 72, 184411.

[161] Paixão, J. A.; Detlefs. C.; Longfield, M. J.; Caciuffo, R.; Santini, P.; Bernhoeft, N.; Rebizant, J.; Lander, G. H. Triple-q Octupolar Ordering in $NpO_2$. *Phys. Rev. Lett.* **2002**, 89, 187202.

[162] Wilkins, S. B.; Caciuffo, R.; Detlefs, C.; Rebizant, J.; Colineau, E.; Wastin, F.; Lander, G. H. Direct Observation of Electric-Quadrupolar Order in $UO_2$. *Phys. Rev.* **2006**, B73, 060406.

[163] Santini, P.; Carretta, S.; Amoretti, G.; Caciuffo, R.; Magnani, N.; Lander, G. H. Multipolar Interactions in *f*-Electron Systems: The Paradigm of Actinide Dioxides. *Rev. Mod. Phys.* **2009**, 81, 807-863.

[164] Caciuffo, R.; Santini, P.; Carretta, S.; Amoretti, G.; Hiess, A.; Magnani, N.; Regnault, L-P.; Lander, G. H. Multipolar, Magnetic, and Vibrational Lattice Dynamics in the Low-Temperature Phase of Uranium Dioxide. *Phys. Rev. B* **2011**, 84, 104409.

[165] Carretta, S.; Santini, P.; Caciuffo, R.; Amoretti, G. Quadrupolar Waves in Uranium Dioxide, *Phys. Rev. Lett.* **2010**, 105, 167201.

[166] Lander, G. H.; Caciuffo, R. The Fifty Years it Has Taken to Understand the Dynamics of $UO_2$ in Its Ordered State. *J. Phys. Cond. Matter* **2020**, 32, 374001.

[167] Faber, J.; Lander, G. H. Neutron Diffraction Study of $UO_2$: Antiferromagnetic State. *Phys. Rev. B* **1976**, 14, 1151-1164.

[168] Bryan, M. S.; Pang, J. W. L.; Larson, B. C.; Chernatynskiy, A.; Abernathy, D. L.; Gofryk, K.; Manley, M. E. Impact of Anharmonicity on the Vibrational Entropy and Specific Heat of $UO_2$. *Phys. Rev. Mater.* **2019**, 3, 065405.

[169] Maldonado, P.; Paolasini, L.; Oppeneer, P. M.; Prodi, A.; Magnani, N.; Bosak, A.; Lander, G. H.; Caciuffo, R. Crystal Dynamics and Thermal Properties of Neptunium Dioxide. *Phys. Rev. B* **2016**, 93, 144301.

[170] Flotow, H.E.; Osborne, D.W.; Fried, S.M.; Malm, J.G. Heat Capacity of $^{242}PuO_2$ from 12 to 3500 K and of $^{244}PuO_2$ from 4 to 25 K. Entropy, Enthalpy, and Gibbs Energy of Formation of $PuO_2$ at 298.15 K. *J. Chem. Phys*. **1976**, 65, 1124-1129.

[171] Sandenaw, Th. A. Heat Capacity of Plutonium Dioxide Below 325 K. *J. Nucl. Mater.* **1963**, 10, 165-172.

[172] Martel, L.; Hen, A.; Tokunaga, Y.; Kinnart, F.; Magnani, N.; Colineau, E.; Griveau, J-C.; Caciuffo, R. Magnetization, Specific Heat, $^{17}O$ NMR, and $^{237}Np$ Mössbauer study of $U_{0.15}Np_{0.85}O_2$. *Phys. Rev. B* **2018**, 98, 014410.

[173] Willis, B. T. M.; Hazell, R. G. Re-analysis of Single-Crystal Neutron-Diffraction Data on $UO_2$ Using Third Cumulants. *Acta Cryst.* **1980**, A36, 582-584.

[174] Santini, P.; Carretta, S.; Magnani, N.; Amoretti, G.; Caciuffo, R. Hidden Order and Low-Energy Excitations in $NpO_2$. *Phys. Rev. Lett.* **2006**, 97, 207203.

[175] Suzuki, M.-T.; Magnani, M.; Oppeneer, P. M. First-Principles Theory of Multipolar Order in Neptunium Dioxide. *Phys. Rev. B* **2010**, 82, 241103(R).

[176] Walstedt, R. E.; Tokunaga, Y.; Kambe, S.; Sakai, H. NMR Evidence for *f*-Electron Excitations in the Multipolar Ground State of $NpO_2$. *Phys. Rev. B* **2018**, 98, 144403.

[177] Vălu, O.S.; Beneš, O.; Konings, R.J.M.; Griveau, J.-C.; Colineau, E. The Low-Temperature Heat Capacity of the $(Th,Pu)O_2$ Solid Solution. *J. Phys. Chem. Solids* **2015,** 86, 194-206.

[178] Magnani, N.; Santini, P.; Amoretti, G.; Caciuffo, R. Perturbative Approach to J Mixing in *f*-Electron Systems: Application to Actinide Dioxide. *Phys. Rev. B* **2005**, 71, 054405.

[179] Mann, M.; Thompson, D.; Serivalsatit, K.; Tritt, T. M.; Ballato, J.; Kolis, J. Hydrothermal Growth and Thermal Property Characterization of $ThO_2$ Single Crystals. *Cryst. Growth Des.* **2010**, 10, 2146-2151.

[180] Gofryk, K.; Du, S.; Stanek, C. R.; Lashley, J. C.; Liu, X. Y.; Schulze, R. K.; Smith, J. L.; Safarik, D. J.; Byler, D. D.; McClellan, K. J.; et al. Anisotropic Thermal Conductivity in Uranium Dioxide. *Nat. Commun.* **2014**, 5, 4551.

[181] Gevers, R.; De Batist, R.; De Coninck, R.; Denayer, M.; Nagels, P.; Penninck, R.; Van den Bosch, A.; Van Lierde, W. Physical Properties of $UO_2$ single crystals. *European Atomic Energy Community - EURATOM, EURAEC Report No. 1776*; **1967**.

[182] Slack, G. Thermal Conductivity of $CaF_2$, $MnF_2$, $CoF_2$, and $ZnF_2$ crystals. *Phys. Rev.* **1961**, 122, 1451-1464.

[183] Moore, J.; McElroy, D. Thermal Conductivity of Nearly Stoichiometric Single Crystal and Polycrystalline $UO_2$. *J. Am. Ceram. Soc.* **1971**, 54, 40-46.

[184] Hofmann, M.; Lorenz, T.; Uhrig, G. S.; Kierspel, H.; Zabara, O.; Freimuth, A.; Kageyama, H.; Ueda. Y. Strong Damping of Phononic Heat Current by Magnetic Excitations in SrCu$_2$(BO$_3$)$_2$. *Phys. Rev. Lett*. **2001**, 87, 047202.

[185] Paolasini, L.; Chaney, D.; Bosak. A.; Lander, G. H.; Caciuffo, R. Ansiotropy in Cubic UO$_2$ Caused by Electronic-Lattice Interactions. *Phys. Rev. B* **2021**, 104, 024305.

[186] Macedo, P. M.; Capps, W.; Wachtman, J. O. Elastic Constants of Single Crystal ThO$_2$ at 25˚C. *J. Am. Ceram. Soc.* **1964**, 47, 651-651.

[187] Wachtman, J. B.; Wheat, M. L.; Anderson, H. J.; Bates, J. L. Elastic Constants of Single Crystal UO$_2$ at 25˚C. *J. Nucl. Mater.* **1965**, 16, 39-41.

[188] Brandt, O. G. and Walker, C. T. Temperature Dependence of Elastic Constants and Thermal Expansion for UO$_2$. *Phys. Rev. Lett.* **1967**, 18, 11-13.

[189] Brandt, O. G.; Walker, C. G. Ultrasonic Attenuation and Elastic Constants for Uranium Dioxide. *Phys. Rev.* **1968** 170, 528-541.

[190] White, G. K.; Sheard, F. W. The Thermal Expansion at Low Temperatures of UO$_2$ and UO$_2$/ThO$_2$. *J. Low Temp. Phys.* **1974**, 14, 445-457.

[191] Antonio, D. J.; Weiss, J. T.; Shanks, K. S.; Ruff, J. P. C.; Jaime, M.; Saul, A.; Swinburne, T.; Salamon, M.; Shrestha, K.; Lavina, B.; et al. Piezomagnetic Switching and Complex Phase Equilibria in Uranium Dioxide. *Commun. Mater.* **2021**, 2, 17.

[192] Sasaki, K.; Obata, Y. Studies of the Dynamical Jahn-Teller Effect on Static Magnetic Susceptibility. *J. Phys. Soc. Japan* **1970**, 28, 1157-1167.

[193] Caciuffo, R.; Amoretti, G.; Santini, P.; Lander, G. H; Kulda, J.; Du Plessis, P. V. Magnetic Excitations and Dynamical Jahn-Teller Distortions in UO$_2$. *Phys. Rev. B* **1999**, 59, 13892-13900.

[194] Dolling, G.; Cowley, R. A. Observation of Magnon-Phonon Interaction at Short Wavelengths. *Phys. Rev. Lett.* **1966**, 16, 683-685.

[195] Cowley, R. A.; Dolling, G. Magnetic Excitations in Uranium Dioxide. *Phys. Rev.* **1968**, 167, 464-477.

[196] Jaime, M.; Saul, A.; Salamon, M.; Zapf, V. S.; Harrison, N.; Durakiewicz, T.; Lashley, J. C.; Andersson, D. A.; Stanek, C. R.; Smith, J. L.; et al. Piezomagnetism and Magnetoelastic Memory in Uranium Dioxide. *Nat. Commun.* **2017**, 8, 99.

[197] Bar'yakhtar, V. G.; Vitebskii, I. M.; Yablonskii, D. A. Magnetoelastic Effects in Noncollinear Antiferromagnets. *Zh. Eksp. Teor. Fiz.* **1985**, *89*, 189–202.

[198] Ronchi, C.; Hiernaut, J. P. Experimental Measurement of Pre-Melting and Melting of Thorium Dioxide. *J. Alloy and Compd.* **1996**, 240, 179-185.

[199] Manara, D.; Ronchi, C.; Sheindlin, M.; Lewsi, M.; Brykin, M. Melting of Stoichiometric and Hyperstoichiometric Uranium Dioxide. *J. Nucl. Mat.* **2005**, 342, 148-163.

[200] De Bruycker, F.; Boboridis, K.; Manara, D.; Poml, P.; Rini, M.; Konings, R. J. M. Reassessing the Melting Temperature of PuO$_2$. *Mat. Today.* **2010**, 13, 52-55.

[201] Capriotti, L.; Quaini, A.; Bohler, R.; Boboridis, K.; Luzzi, L.; Manara, D. A Laser Heating Study of the CeO$_2$ Solid/Liquid Transition: Challenges Related to a Refractory Compound with a Very High Oxygen Pressure. *High Temp-High Press.* **2014**, 44, 69-82.

[202] Neck, V.; Altmaier, M.; Muller, R.; Bauer, A.; Fanghanel, Th.; Kim, J. I. Solubility of Crystalline Thorium Dioxide. *Radiochim. Acta.* **2003**, 91, 253-262.

[203] Neck, V.; Kim, J. I. Solubility and Hydrolysis of Tetravalent Actinides. *Radiochim. Acta.* **2001**, 89, 1-16.

[204] Hayes, S. A.; Yu, P.; O'Keefe, T. J.; O'Keefe, M. J.; Stoffer, J. O. The Phase Stability of Cerium Species in Aqueous Systems - I. E-pH Diagram for the Ce-HClO$_4$-H$_2$O System. *J. Electrochem. Soc.* **2002**, 149, C623-C630.

[205] Yu, P.; Hayes, S. A.; O'Keefe, T. J.; O'Keefe, M. J.; Stoffer, J. O. The Phase Stability of Cerium Species in Aqueous Systems - II. The Ce(III/IV)-H$_2$O-H$_2$O$_2$/O$_2$ Systems. Equilibrium Considerations and Pourbaix Diagram Calculations. *J. Electrochem. Soc.* **2006**, 153, C74-C79.

[206] Byrappa, K.; Yoshimura, M. *Handbook of Hydrothermal Technology: a technology for crystal growth and materials processing*; Noyes Publications: Park Ridge, NJ, **2001**; p 1.

[207] Elwell, D.; Scheel, H. J. *Crystal Growth from High Temperature Solutions*; Academic Press: London-New York-San Francisco, **1975;** p 1.

[208] Suresh, D.; Rohatgi, P. K.; Coutures, J. P. Use of Solar Furnaces – I Materials Research. *Solar Energy.* **1981**, 26, 377-390.

[209] Capper P. Bulk Crystal Growth: Methods and Materials. In, *Springer Handbook of Electronic and Photonic Materials;* Springer Handbooks: Cham, **2017**; p 1.

[210] Rao, C. N. R.; Biswas, K. Arc and Skull Methods. In, *Essentials of Inorganic Materials Synthesis;* John Wiley & Sons: Hoboken, **2015**; p 37.

[211] Scott, B. L.; Joyce, J. J.; Durakiewicz, T. D.; Martin, R. L.; McCleskey, T. M.; Bauer, E.; Luo, H.; Jia, Q. High Quality Epitaxial Thin Films of Actinide Oxides, Carbides, and Nitrides: Advancing Understanding of Electronic Structure of f-Element Materials. *Coord. Chem. Rev.* **2014**, 266-267, 137-154.

[212] Herrick, C. C.; Behrens, R. G. Growth of Large Uraninite And Thorianite Single Crystal from the Melt Using A Cold-Crucible Technique. *J. Cryst. Growth.* **1981,** 51, 183-189.

[213] Rasmussen, J. J.; Nelson, R. P. Growth of UO$_2$ Single Crystals by Closed-Capsule Zone Melting. *J. Nucl. Mater.* **1970**, 36, 237-240.

[214] Chapman, A. T.; Clark, G. W. Growth of UO$_2$ Single Crystals Using the Floating-Zone Technique. *J. Am. Ceram. Soc.* **1965**, 48, 494-495.

[215] Yust, C. S.; McHargue, C. J. Dislocation Substructures in Deformed Uranium Dioxide Single Crystals. *J. Nucl. Mat.* **1969**, 31, 121-137.

[216] Scott, J. J.; Patchett, J. E. Preparation Methods and Quality Control of Fused Urania. *Meet. on Charact. of Uranium Dioxide.* **1961**, TID-7637, 481.

[217] Bard, R. J. *The Preparation of Uranium Dioxide Crystals*; Rep. LA-2076; Los Alamos Sci. Lab.: Los Alamos, NM, **1957**.

[218] Brite, D. W.; Anderson, H. J. High Density UO$_2$ for Nuclear Fuel Applications. *Meet. on Charact. of Uranium Dioxide.* **1961**, TID-7637, 408.

[219] Childs, B. G.; Harvey, P. J.; Hallett, J. B. Color Centers and Point Defects in Irradiated Thoria. *J. Am. Ceram. Soc*. **1970**, 53, 431-435.

[220] Griffiths, T. R.; Dixon, J. Optical Absorption Spectroscopy of Thorium Dioxide. *J. Chem. Soc. Faraday Trans.* **1992**, 88, 1149-1160.

[221] Harvey, P. J.; Childs, B.G.; Moerman, J. Optical Absorbance and Fluorescence in Pure and Doped ThO$_2$. *J. Am. Ceram. Soc.* **1973**, 56, 134-136.

[222] Harvey, P. J.; Hallett, J. B. Factors Influencing the Photoluminescence Behavior of Thoria. *J. Electrochem. Soc.: Solid-State Sci. and Tech.* **1976**, 123, 398-403.

[223] Rodine, E. T.; Land, P. L. Electronic and Defect Structure of Single-Crystal ThO2 by Thermoluminescence. *Phys. Rev. B.* **1971**, 4, 2701-2724.

[224] Sakurai, T.; Kamada, O.; Ishigame, M. Uranium Dioxide Crystals Grown by A Solar Furnace. *J. Cryst. Growth.* **1968**, 2, 326-327.

[225] Laszlo, T. S.; Sheehan, P. J.; Gannon, R. E. Thoria Single Crystals Grown by Vapor Deposition In A Solar Furnace. *J. Phys. Chem. Solids.* **1967**, 28, 313-316.

[226] Hedley, W. H.; Roehrs, R. J. *High Temperature Preparation of $UO_2$.* Uranium Dioxide: Properties and Nuclear Applications; US Government Printing Office: Washington D.C., **1961**.

[227] Chapman, A. T.; Clark, G. W.; Hendrix, D. E.; Yust, C. S.; Cavin, O. B. Substructure and Perfection in $UO_2$ Single Crystals. *J. Am. Ceram. Soc.* **1970**, 53, 46-51.

[228] Naito, K.; Kamegashira, N.; Nomura, Y. Single Crystals of Uranium Oxides by Chemical Transport Reactions. *J. Cryst. Growth.* **1971**, 8, 219-220.

[229] Singh, R. N.; Coble, R. L. Growth Of Uranium Dioxide Single Crystals By Chemical Vapor Deposition. *J. Cryst. Growth.* **1974**, 21, 261-266.

[230] Nomura, Y.; Kamegashira, N.; Naito, K. Synthesis of Single Crystals of Nonstoichiometric Uranium Oxides by Chemical Transport Reactions. *J. Cryst. Growth.* **1981**, 52, 279-284.

[231] Faile, S. P. Growth of Uranium Dioxide Crystals Using Tellurium Tetrachloride and Argon. *J. Cryst. Growth.* **1978**, 43, 133-134.

[232] Kamegashira, N.; Ohta, K.; Naito, K. Growth of Single Crystals of $U_{1-x}Th_xO_2$ Solid Solutions by Chemical Transport Reactions. *J. Cryst. Growth.* **1978**, 44, 1-4.

[233] Kolberg, D.; Wastin, F.; Rebizant, J.; Boulet, P.; Lander, G. H.; Schoenes, J. Magnetic Susceptibility and Spin-Lattice Interactions in $U_{1-x}Pu_xO_2$ single crystals. *Phys. Rev. B.* **2002**, 66, 214418.

[234] Prohaska, J.; Tromel, M.; Rager, H. EPR Study of $Fe^{3+}$ and $Mn^{2+}$ in $CeO_2$ and $ThO_2$. *Appl. Magn. Reson.* **1993**, 5, 387-398.

[235] Wanklyn, B. M.; Garrard, B. J. The Flux Growth of Large Thoria and Ceria Crystals. *J. Cryst. Growth.* **1984**, 66, 346-350.

[236] Linares, R. C. Growth and Properties of $CeO_2$ and $ThO_2$ Single Crystals. *J. Phys. Chem. Solids.* **1967**, 28, 1285-1286.

[237] Harari, M. A.; Thery, M. J., Collongues, M. R. Sur La Preparation De Monocristaux D'Oxydes $MO_2$. *Rev. Int. Hautes Temper. et Refract.* **1967**, 4, 207-209.

[238] Grodkiewicz, W. H.; Nitti, D. J. Oxide Crystal Growth by Flux Evaporation. *J. Am. Ceram. Soc.* **1966**, 49, 576-576.

[239] Finch, C. B.; Clark, G. W. Single-Crystal Growth of Cerium Dioxide from Lithium Ditungstate Solvent. *J. Appl. Phys.* **1966**, 37, 3910-3910.

[240] Chaminade, J. P.; Olazcuaga, R.; Le Polles, G.; Le Flem, G.; Hagenmuller, P. Crystal Growth and Characterization of $Ce_{1-x}Pr_xO_2$ (x=0.05) Single Crystals. *J. Cryst. Growth.* **1988**, 87, 463-465.

[241] Baker, J. M.; Copland, G. M.; Wanklyn, B. M. Nuclear Moments and Hyperfine Structure Anomaly for Gadolinium. *J. Phys. C: Solid State Phys.* **1969**, 2, 862-869.

[242] Rizzoli, C.; Spirlet, J. C. Single Crystal Growth of Actinide Dioxides. *Inorg. Chim. Acta.* **1984**, 94, 112-113.

[243] Finch, C. B.; Clark, G. W. Single-Crystal Growth of Thorium Dioxide from Lithium Ditungstate Solvent. *J. Appl. Phys.* **1965**, 36, 2143-2145.

[244] Kolopus, J. L.; Finch, C. B.; Abraham, M. M. ESR of $Pb^{3+}$ Centers in $ThO_2$. *Phys. Rev. B.* **1970**, 2, 2040-2045.

[245] Chase, A. B.; Osmer, J. A. Localized Cooling in Flux Crystal Growth. *J. Am. Ceram. Soc.* **1967**, 50, 325-328.

[246] Garton, G.; Smith, S. H.; Wanklyn, B. M. Crystal Growth from The Flux Systems $PbO-V_2O_5$ and $Bi_2O_3-V_2O_5$. *J. Cryst. Growth.* **1972**, 13-14, 588-592.

[247] Amoretti, G.; Giori, D. C.; Ori, O.; Schianchi G., Calestani, G., Rizzoli, C.; Spirlet, J. C. Actinide and Rare Earth Ions in Single Crystals of $ThO_2$: Preparation, EPR, Studies and Related Problems. *J. Less Common Met.* **1986**, 122, 35-45.

[248] Amelinckx, S.; Bressers, J.; Nevelsteen, K.; Smets, E.; Van Lierde, W. Growth of $UO_2$ Single Crystals. *Eur. At. Energy Community - EURATOM [Rep.].* **1963**, EUR 345.e.

[249] Hillebrand, W. F. A Further Example of The Isomorphism of Thoria and Uranium Dioxide. *Bull. U. S. Geol. Soc.* **1893**, 113, 41-43.

[250] Phipps, K. D.; Sullenger, D. B. Plutonium Dioxide: Preparation of Single Crystals. *Science.* **1964**, 145, 1048-1049.

[251] Finch, C. B.; Clark, G. W. High-Temperature Solution Growth of Single-Crystal Plutonium Dioxide. *J. Cryst. Growth.* **1972**, 12, 181-182.

[252] Modin, A.; Yun, Y.; Suzuki, M-T.; Vegelius, J.; Werme, L.; Nordgren, J.; Oppeneer, P. M.; Butorin, S. M. Indication of Single-Crystal $PuO_2$ Oxidation from O 1s X-ray Absorption Spectra. *Phys. Rev. B* **2011**, 83, 075113.

[253] Modin, A.; Suzuki, M-T.; Vegelius, J.; Yun, Y.; Shuh, D. K.; Werme, L.; Nordgren, J.; Oppeneer, P. M.; Butorin, S. M. 5f-Shell Correlation Effects in Dioxides of Light Actinides Studied by O 1s X-ray Absorption and Emission Spectroscopies and First-Principles Calculations. *J. Phys. Condens. Matter.* **2015**, 27, 315503.

[254] Tani, E.; Yoshimura, M.; Somiya, S. Crystallization and Crystal Growth of $CeO_2$ Under Hydrothermal Conditions. *J. Mater. Sci. Lett.* **1982**, 461-462.

[255] Sen, A.; Bachhav, M.; Vurpillot, F.; Mann, J. M.; Morgan, P. K., Prusnick, T. A.; Wharry, J. P. Influence on Field Conditions on Quantitative Analysis of Single Crystal Thorium Dioxide by Atom Probe Tomography. *Ultramicroscopy.* **2021**, 220, 113167.

[256] Knight, S.; Korlacki, R.; Dugan, C.; Petrosky, J. C.; Mock, A.; Dowben, P. A.; Mann, J. M.; Kimani, M. M.; Schubert, M. Infrared-Active Phonon Modes in Single-Crystal Thorium and Uranium Dioxide. *J. Appl. Phys.* **2020**, 127, 125103.

[257] Mock, A.; Dugan, C.; Knight, S.; Korlacki, R.; Mann, J. M.; Kimani, M. M.; Petrosky, J. C.; Dowben, P. A.; Schubert, M. Band-to-Band Transitions and Critical Points in the Near-Infrared to Vacuum Ultraviolet Dielectric Function of Single Crystal Urania and Thoria. *Appl. Phys. Lett.* **2019**, 114, 211901.

[258] Rickert, K.; Prusnick, T. A.; Kimani, M. M.; Moore, Merriman, C. A.; Mann, J. M. Assessing $UO_2$ Sample Quality with μ-Raman Spectroscopy. *J. Nucl. Mat.* **2019**, 514, 1-11.

[259] Kelly, T. D.; Petrosky, J. C.; McClory, J. W.; Mann, J. M.; Kolis, J. W. Analysis of Oxygen Shell Splitting in Hydrothermally Grown Single Crystal $ThO_2$ (200). *Phys. Status Solidi RRL.* **2015**, 9, 668-672.

[260] Kelly, T. D.; Petrosky, J. C.; Turner, D.; McClory, J. W.; Mann, J. M.; Kolis, J. W.; Zhang, X.; Dowben, P. A. The Unoccupied Electronic Structure Characterization of Hydrothermally Grown ThO$_2$ Single Crystals. *Phys. Status Solidi RRL.* **2014**, 8, 283-286.

[261] Kelly, T. D.; Petrosky, J. C.; McClory, J. W.; Zens, T.; Turner, D.; Mann, J. M.; Kolis, J. W.; Santana, J. A. C.; Dowben, P. A. The Debye Temperature for Hydrothermally Growth ThO$_2$ Single Crystals. *Mater. Res. Soc. Symp. Proc.* **2013**, 1576, 996.

[262] Castilow, J.; Zens, T. W.; Mann, J. M.; Kolis, J. W.; McMillen, C. D.; Petrosky, J. Hydrothermal Synthesis and Characterization of ThO$_2$, U$_x$Th$_{1-x}$O$_2$, and UO$_x$. *Mater. Res. Soc. Symp. Proc.* **2013**, 1576, 973.

[263] Turner, D. B.; Kelly, T. D.; Peterson, G. R.; Reding, J. D.; Hengehold, R. L.; Mann, J. M.; Kolis, J. W.; Zhang, X.; Dowben, P. A.; Petrosky, J. C. Electronic Structure of the Hydrothermally Synthesized Single Crystal U$_{0.22}$Th$_{0.78}$O$_2$. *Phys. Status Solidi B.* **2016**, 253, 1970-1976.

[264] Rickert, K.; Prusnick, T. A.; Hunt, E.; Kimani, M. M.; Chastang, S.; Brooks, D. L.; Moore, E. A.; Petrosky, J. C.; Mann, J. M. Inhibiting Laser Oxidation of UO$_2$ via Th Substitution. *J. Nucl. Mater.* **2019**, 517, 254-262.

[265] Dugan, C. L.; Peterson, G. G.; Mock, A.; Young, C.; Mann, J. M.; Nastasi, M.; Schubert, M.; Wang, L.; Mei, W-N.; Tanabe, I.; et al. Electrical and Materials Properties of Hydrothermally Grown Single Crystal (111) UO$_2$. *Eur. Phys. J. B.* **2018**, 91, 67.

[266] Young, C.; Petrosky, J.; Mann, J. M.; Hunt, E. M.; Turner, D.; Dowben, P. A. The Lattice Stiffening Transition in UO$_2$ Single Crystals. *J. Phys. Condens. Matter.* **2017**, 29, 035005.

[267] Young, C.; Petrosky, J.; Mann, J. M.; Hunt, E. M.; Turner, D.; Kelly, T. The Work Function of Hydrothermally Synthesized UO$_2$ and the Implications for Semiconductor Device Fabrication. *Phys. Status Solidi RRL.* **2016**, 10, 687-690.

[268] Meredith, N. A.; Wang, S.; Diwu, J.; Albrecht-Schmitt, T. E. Two Convenient Low-Temperature Routes to Single Crystals of Plutonium Dioxide. *J. Nucl. Mater.* **2014**, 454, 164-167.

[269] Kawabuchi, K.; Magari, S. Growth Morphology of UO$_2$, ZrO$_2$, HfO$_2$, and ThO$_2$ Crystallites Grown from Oxide Plasmas. *J. Cryst. Growth.* **1980**, 49, 81-84.

[270] Van Lierde, W.; Strumane, R.; Smets, E.; Amelinckx, S. The Preparation of Uranium Oxide Single Crystals By Sublimation. *J. Nucl. Mater.* **1962**, 5, 250-253.

[271] Schlechter, M. The Preparation Of Crystalline UO$_2$ Crystals By Thermal Decomposition of UOCl$_2$ Dissolved In Molten UCl$_4$. *J. Nucl. Mater.* **1968**, 28, 228-229.

[272] Schlechter, M. The Preparation Of PuO$_2$ Crystals By Thermal Decomposition of Pu(SO$_4$)$_2$ Dissolved In Chloride Melts. *J. Nucl. Mater.* **1970**, 37, 82-88.

[273] Robins, R. G. Uranium Dioxide Single Crystals by Electrodeposition. *J. Nucl. Mater.* **1961**, 3, 294-301.

[274] Scott, F. A.; Mudge, L. K. The Electrolytic Preparation of Single Crystals of Uranium Dioxide. *J. Nucl. Mater.* **1963**, 9, 245-251.

[275] Schlechter, M. The Preparation Of UO$_2$ By Electrolysis of UO$_2$Cl$_2$ Dissolved In Molten NaCl-KCl-MgCl$_2$ Eutectic. *J. Nucl. Mater.* **1963**, 10, 145-146.

[276] Kordyukevich, V. O.; Kuznetsov, V. I.; Otstavnov, Y. D.; Smirnov, N. N. Electrochemical Preparation of Single Crystals of $^{235}$UO$_2$. *At. Energ.* **1977**, 42, 145-147.

[277] Graham, J. T.; Zhang, Y. W.; Weber, W. J. Irradiation-Induced Defect Formation and Damage Accumulation in Single Crystal CeO$_2$. *J. Nucl. Mater.* **2018**, 498, 400-408.

[278] Evans, W. R.; Allred, D. D. Determining Indices of Refraction for ThO$_2$ Thin Films Sputter Under Different Bias Voltages from 1.2 to 6.5 eV by Spectroscopic Ellipsometry. *Thin Solid Film.* **2006**, 515, 847-853.

[279] Navinsek, B. Epitaxial Growth of UO$_2$ Thin Films Produced by Cathode Sputtering. *J. Nucl. Mater.* **1971**, 40, 338-340.

[280] Strehle, M. M.; Heuser, B. J.; Elbakhshwan, M. S.; Han, X.; Gennardo, D. J.; Pappas, H. K.; Ju, H. Characterization of Single Crystal Uranium-Oxide Thin Films Grown via Reactive-Gas Magnetron Sputtering on Yttria-Stabilized Zirconia and Sapphire. *Thin Solid Films* **2012**, 520, 5616-5626.

[281] Z. Bao, Z.; Springell, R.; H. Walker, H. C.; Leiste, H.; Kuebel, K.; Prang, R.; Nisbet, G.; Langridge, S; Ward, R. C. C.; Gouder, T.; et al. Antiferromagnetism in UO$_2$ Thin Epitaxial Films. *Phys. Rev. B* **2013**, 88, 134426.

[282] Rennie, S.; Bright, E. L; Sutcliffe, J. E.; Darnbrough, J. E.; Burrows, R.; Rawle, J.; Nicklin, C.; Lander, G.H.; Springell, R. The Role of Crystal Orientation in the Dissolution of UO$_2$ Thin Films. *Corros. Sci.* **2018,** 145, 162-169.

[283] Elbakhshwan, M. S.; Miao, Y.; Stubbins, J. F.; Heuser, B. J. Mechanical Properties of UO$_2$ Thin Films Under Heavy Ion Irradiation Using Nanoindentation and Finite Element Modeling. *J. Nucl. Mater.* **2016**, 479, 548-558.

[284] Darnbrough, J. E.; Harker, R. M.; Griffiths, I.; Wermeille, D.; Lander, G. H.; Springell, R. Interaction Between U/UO$_2$ Bilayers and Hydrogen Studied by In-Situ X-ray Diffraction. *J. Nucl. Mater.* **2018**, 502, 9-19

[285] Whapham, A. D.; Sheldon, B. E. Radiation Damage in Uranium Dioxide, *Philos. Mag.* **1965**, 12, 1179-1102.

[286] Weber, W. J. Alpha-Irradiation Damage in CeO$_2$, UO$_2$ and PuO$_2$. *Radiat. Eff. Defects Solids* **1984**, 83, 145-156.

[287] Weber, W. J.. Thermal Recovery of Lattice-Defects in Alpha-Irradiated UO$_2$ Crystals. *J. Nucl. Mater.* **1983**, 114, 213-221.

[288] Weber, W. J. Ingrowth of Lattice-Defects in Alpha Irradiated UO$_2$ Single-Crystals. *J. Nucl. Mater.* **1981**, 98, 206-215.

[289] Matzke, Hj.; Blank, H.; Coquerelle, M.; Lassmann, K.; Ray, I. L. F.; Ronchi, C.; Walker, C.T. Oxide Fuel Transients, *J. Nucl. Mater.* **1989**, 166, 165-178.

[290] Evans, J. H.; Vanveen. A.; Westerduin, K. T. A TEM and TDS Study of Gas-Release from Bubbles in Krypton-Implanted Uranium-Dioxide. *J. Nucl. Mater.* **1992**, 195, 250-59.

[291] Whapham, A. D. Electron Microscope Observation of Fission-Gas Bubble Distribution in UO$_2$. *Nucl. Applicat.* **1966**, 2, 123-130.

[292] Kashibe, S.; Une, K.; Nogita, K. Formation and Growth of Intragranular Fission-Gas Bubbles in UO$_2$ Fuels with Burnup of 6-83 GWd/T. *J. Nucl. Mater.* **1993**, 206, 22-34.

[293] Une, K.; Nogita, K.; Kashibe, S.; Imamura, M. Microstructural Change and Its Influence on Fission-Gas Release in High Burnup UO$_2$ Fuel. *J. Nucl. Mater.* **1992**, 188, 65-72.

[294] Rest, J.; Cooper, M. W. D.; Spino, J.; Turnbull, J. A.; Van Uffelen, P.; Walker, C. T. Fission Gas Release from UO$_2$ Nuclear Fuel: A Review. *J. Nucl. Mater.* **2019**, 513, 310-45.

[295] Ye, B.; Kirk, M. A.; Chen, W. Y.; Oaks, A.; Rest, J.; Yacout, A.; Stubbins, J. F. TEM Investigation of Irradiation Damage in Single Crystal CeO$_2$. *J. Nucl. Mater.* **2011**, 414, 251-256.

[296] Ye, B.; Oaks, A.; Kirk, M.; Yun, D.; Chen, W. Y.; Holtzman, B.; Stubbins, J. F. Irradiation Effects in UO$_2$ and CeO$_2$. *J. Nucl. Mater.* **2013**, 441, 525-529.

297 Chauhan, V. S.; Pakarinen, J.; Yao, T.; He, L.; Hurley, D. H.; Khafizov, M. Indirect Characterization of Point Defects in Proton Irradiated Ceria. *Materialia* **2021**, 15, 101019.

298 Miao, Y. B.; Aidhy, D.; Chen, W. Y.; Mo, K.; Oaks, A.; Wolf, D.; Stubbins, J. F. The Evolution Mechanism of the Dislocation Loops in Irradiated Lanthanum Doped Cerium Oxide. *J. Nucl. Mater.* **2014**, 445, 209-217.

299 Matzke, H., Xenon Migration and Trapping in Doped $ThO_2$. *J. Nucl. Mater*. **1967**, 21, 190-198.

300 Ogawa, T.; Verrall, R. A.; Matzke, H.; Lucuta, P. G. Release of Ion-Implanted Kr from $(Th,U)O_2$ - Effect of Matrix Oxidation. *Solid State Ionics* **1991**, 49, 211-216.

301 Jegadeesan, P.; Amirthapandian, S.; Panigrahi, B. K.; Kaur, G.; Kumar, D. S.; Magudapathy, P.; Ananthasivan, K. Aggregation and Ordering of Helium in Thoria. *J. Alloy. Compd*. **2018**, 743, 196-202.

302 Jin, M.; Jiang, C.; Gan, J.; Hurley, D. H. Systematic Analysis on the Primary Radiation Damage in $Th_{1-x}U_xO_2$ Fluorite Systems. *J. Nucl. Mater.* **2020**, 536, 152144

303 Bawane, K.: Liu, X.; Yao, T.; Khafizov, M.; French, A.; Mann, J. M.; Shao, L.; Gan, J.; Hurley, D. H.; He, L. TEM Characterization of Dislocation Loops in Proton Irradiated Single Crystal $ThO_2$. *J. Nucl. Mater.* **2021**, 552, 152998

304 Was, G. S.; Allen, T. R., Radiation Damage from Different Particle Types. In *Radiation Effects in Solids*, Springer Netherlands: 2007; Vol. 235, pp 65-98.

305 Cureton, W. F.; Palomares, R. I.; Tracy, C. L.; O'Quinn, E. C.; Walters, J.; Zdorovets, M.; Ewing, R. C.; Toulemonde, M.; Lang, M. Effects of Irradiation Temperature on the Response of $CeO_2$, $ThO_2$, and $UO_2$ to Highly Ionizing Radiation. *J. Nucl. Mater.* **2019**, 525, 83-91.

306 Cureton, W. F.; Palomares, R. I.; Walters, J.; Tracy, C. L.; Chen, C.-H.; Ewing, R. C.; Baldinozzi, G.; Lian, J.; Trautmann, C.; Lang, M. Grain Size Effects on Irradiated $CeO_2$, $ThO_2$, and $UO_2$. *Acta Mater*. **2018**, 160, 47-56.

307 Garcia, P.; Gilabert, E.; Martin, G.; Carlot, G.; Sabathier, C.; Sauvage, T.; Desgardin, P.; Barthe, M.-F. Helium Behaviour in $UO_2$ Through Low Fluence Ion Implantation Studies. *Nucl. Instrum. Meth. Phys. Res. B* **2014**, 327, 113-116.

308 Garcia, P.; Martin, G.; Sabathier, C.; Carlot, G.; Michel, A.; Martin, P.; Dorado, B.; Freyss, M.; Bertolus, M.; Skorek, R.; et al. Nucleation and Growth of Intragranular Defect and Insoluble Atom Clusters in Nuclear Oxide Fuels. *Nucl. Instrum. Meth. Phys. Res. B* **2012**, 277, 98-108.

309 Haddad, Y.; Delauche, L.; Gentils, A.; Garrido, F. In Situ Characterization of Irradiation-Induced Microstructural Evolution in Urania Single Crystals at 773 K. *Nucl. Instrum. Meth. Phys. Res. B* **2018**, 435, 25-30.

310 He, L.; Bai, X. M.; Pakarinen, J.; Jaques, B. J.; Gan, J.; Nelson, A. T.; El-Azab, A.; Allen, T. R. Bubble Evolution in Kr-Irradiated $UO_2$ During Annealing. *J. Nucl. Mater.* **2017**, 496, 242-250.

311 He, L.; Valderrama, B.; Hassan, A.-R.; Yu, J.; Gupta, M.; Pakarinen, J.; Henderson, H. B.; Gan, J.; Kirk, M. A.; Nelson, A. T.; et al. Bubble Formation and Kr Distribution in Kr-Irradiated $UO_2$. *J. Nucl. Mater.* **2015**, 456, 125-132.

312 He, L. F.; Gupta, M.; Kirk, M. A.; Pakarinen, J.; Gan, J.; Allen, T. R. In Situ TEM Observation of Dislocation Evolution in Polycrystalline $UO_2$. *JOM* **2014**, 66, 2553-2561.

313 He, L. F.; Gupta, M.; Yablinsky, C. A.; Gan, J.; Kirk, M. A.; Bai, X.-M.; Pakarinen, J.; Allen, T. R. In Situ TEM Observation of Dislocation Evolution in Kr-Irradiated $UO_2$ Single Crystal. *J. Nucl. Mater*. **2013**, 443, 71-77.

314 He, L. F.; Pararinen, J.; Kirk, M. A.; Gan, J.; Nelson, A. T.; Bai, X.-M.; El-Azab, A.; Allen, T. R. Microstructure Evolution in Xe-Irradiated $UO_2$ at Room Temperature. *Nucl. Instrum. Meth. Phys. Res*. B **2014**, 330, 55-60.

315 Khafizov, M.; Pakarinen, J.; He, L.; Hurley, D. H. Impact of Irradiation Induced Dislocation Loops on Thermal Conductivity in Ceramics. *J. Am. Ceram. Soc*. **2019**, 102, 7533-7542.

316 Marchand, B.; Moncoffre, N.; Pipon, Y.; Bererd, N.; Garnier, C.; Raimbault, L.; Sainsot, P.; Epicier, T.; Delafoy, C.; Fraczkiewicz, M.; et al. Xenon Migration in $UO_2$ Under Irradiation Studied by SIMS Profilometry. *J. Nucl. Mater*. **2013**, 440, 562-567.

317 Onofri, C.; Sabathier, C.; Baumier, C.; Bachelet, C.; Palancher, H.; Warot-Fonrose, B.; Legros, M. Influence of Exogenous Xenon Atoms on the Evolution Kinetics of Extended Defects in Polycrystalline $UO_2$ Using In Situ TEM. *J. Nucl. Mater.* **2018**, 512, 297-306.

318 Onofri, C.; Sabathier, C.; Palancher, H.; Carlot, G.; Miro, S.; Serruys, Y.; Desgranges, L.; Legros, M. Evolution of Extended Defects in Polycrystalline $UO_2$ Under Heavy Ion Irradiation: Combined TEM, XRD and Raman Study. *Nucl. Instrum. Meth. Phys. Res. B* **2016**, 374, 51-57.

319 Pakarinen, J.; He, L.; Gupta, M.; Gan, J.; Nelson, A.; El-Azab, A.; Allen, T. R. 2.6 MeV Proton Irradiation Effects on the Surface Integrity of Depleted $UO_2$. *Nucl. Instrum. Meth. Phys. Res*. B **2014**, 319, 100-106.

320 Pakarinen, J.; Khafizov, M.; He, L.; Wetteland, C.; Gan, J.; Nelson, A. T.; Hurley, D. H.; El-Azab, A.; Allen, T. R. Microstructure Changes and Thermal Conductivity Reduction in $UO_2$ Following 3.9 MeV $He^{2+}$ Ion Irradiation. *J. Nucl. Mater.* **2014**, 454, 283-289.

321 Palomares, R. I.; Shamblin, J.; Tracy, C. L.; Neuefeind, J.; Eqing, R. C.; Trautmann, C.; Lang, M. Defect Accumulation in Swift Heavy Ion-Irradiated $CeO_2$ and $ThO_2$. *J. Mater. Chem. A* **2017**, 5, 12193-12201.

322 Tracy, C. L.; Lang, M.; Pray, J. M.; Zhang, F.; Popov, D.; Park, C.; Trautmann, C.; Bender, M.; Severin, D.; Skuratov, V. A.; et al. Redox Response of Actinide Materials to Highly Ionizing Radiation. *Nat. Commun.* **2015**, 6, 6133.

323 Tracy, C. L.; McLain Pray, J.; Lang, M.; Popov, D.; Park, C.; Trautmann, C.; Ewing, R. C. Defect Accumulation in $ThO_2$ Irradiated With Swift Heavy Ions. *Nucl. Instrum. Meth. Phys. Res. B* **2014**, 326, 169-173.

324 Onofri, C.; Legros, M.; Lechelle, J.; Palancher, H.; Baumier, C.; Bachelet, C.; Sabathier, C. Full Characterization of Dislocations in Ion-Irradiated Polycrystalline $UO_2$. *J. Nucl. Mater.* **2017**, 494, 252-259.

325 Onofri, C.; Sabathier, C.; Baumier, C.; Bachelet, C.; Drouan, D.; Gerardin, M.; Legros, M. Extended Defect Change in $UO_2$ During In Situ TEM Annealing. *Acta Mater*. **2020**, 196, 240-251.

326 Onofri, C.; Sabathier, C.; Baumier, C.; Bachelet, C.; Palancher, H.; Legros, M. Evolution of Extended Defects in Polycrystalline Au-Irradiated $UO_2$ Using In Situ TEM: Temperature and Fluence Effects. *J. Nucl. Mater.* **2016**, 482, 105-113.

327 Onofri, C.; Sabathier, C.; Carlot, G.; Drouan, D.; Bachelet, C.; Baumier, C.; Gerardin, M.; Bricout, M. Changes in Voids Induced by Ion Irradiations in $UO_2$: In Situ TEM Studies. *Nucl. Instrum. Meth. Phys. Res. B* **2020**, 463, 76-85.

328 He, L.; Khafizov, M.; Jiang, C.; Tyburska-Puschel, B.; Jaques, B. Xiu, P.; Xu, P.; Meyer, M. K.; Sridharan, K.; Butt, D. P.; et al. Phase and Defect Evolution in Uranium-Nitrogen-Oxygen System Under Irradiation. *Acta Mater.* **2021**, 208, 116778.

329 Wiss, T. Radiation Effects in $UO_2$. In, *Comprehensive Nuclear Materials*; Elsevier: The Netherlands, **2012**; p 465.

[330] De Bona, E; Benedetti, A.; Dieste, O.; Staicu, D.; Wiss, T.; Konings, R. J. M. Radiation Effects in Alpha-Doped $UO_2$. *Nucl. Instrum. Meth. Phys. Res. B* **2020**, 468, 54-59.

[331] Soullard, J. High-Voltage Electron-Microscope Observations of $UO_2$. *J. Nucl. Mater.* **1985**, 135, 190-196.

[332] Leinders, G.; Bes, R.; Pakarinen, J.; Kvashnina, K.; Verwerft, M. Evolution of the Uranium Chemical State in Mixed-Valence Oxides. *Inorg. Chem.* **2017**, 56, 6784-6787.

[333] Schmitt, R.; Nenning, A.; Kraynis, O.; Korobko, R.; Frenkel, A. I.; Lubomirsky, I.; Haile, S. M.; Rupp, J. L. M. A Review of Defect Structure and Chemistry in Ceria and Its Solid Solutions. *Chem. Soc. Rev.* **2020**, 49, 554-592.

[334] Amidani, L.; Vaughan, G. B. M.; Plakhova, T. V.; Romanchuk, A. Y.; Gerber, E.; Svetogorov, R.; Weiss, S.; Joly, Y.; Kalmykov, S. N.; Kvashnina, K. O. The Application of HEX and HERFD XANES for Accurate Structural Characterisation of Actinide Nanomaterials: The Case of $ThO_2$, *Chem. Eur. J.* **2020**, 27, 252-263.

[335] Allen, G. C.; Tempest, P. A.; Garner, C. D.; Ross, I.; Jones, D. J. EXAFS - a New Approach to the Structure of Uranium-Oxides. *J. Phys. Chem.* **1985,** 89, 1334-1339.

[336] Jones, D. J.; Roziere, J.; Allen, G. C.; Tempest, P. A. The Structural Determination of Fluorite-Type Oxygen Excess Uranium-Oxides Using EXAFS Spectroscopy. *J. Chem. Phys.* **1986,** 84, 6075-6082.

[337] Conradson, S. D.; Manara, D.; Wastin, F.; Clark, D. L.; Lander, G. H.; Morales, L. A.; Rebizant, J.; Rondinella, V. V. Local Structure and Charge Distribution in the $UO_2$-$U_4O_9$ System. *Inorg. Chem.* **2004,** 43, 6922-6935.

[338] Conradson, S. D.; Begg, B. D.; Clark, D. L.; den Auwer, C.; Ding, M.; Dorhout, P. K.; Espinosa-Faller, F. J.; Gordon, P. L.; Haire, R. G.; Hess, N. J.; et al. Charge Distribution and Local Structure and Speciation in the $UO_{2+x}$ and $PuO_{2+x}$ Binary Oxides for x <= 0.25. *J. Solid. State. Chem.* **2005,** 178, 521-535.

[339] Martin, P.; Ripert, M.; Petit, T.; Reich, T.; Hennig, C.; D'Acapito, F.; Hazemann, J. L.; Proux, O. A XAS Study of the Local Environments of Cations in $(U,Ce)O_2$. *J. Nucl. Mater.* **2003,** 312, 103-110.

[340] Martin, P.; Grandjean, S.; Valot, C.; Carlot, G.; Ripert, M.; Blanc, P.; Hennig, C. XAS Study of $(U_{1-y}Pu_y)O_{x-2}$ Solid Solutions. *J. Alloy Compd.* **2007,** 444, 410-414.

[341] Kvashnina, K. O.; Butorin, S. M.; Martin, P.; Glatzel, P. Chemical State of Complex Uranium Oxides. *Phys. Rev. Lett.* **2013,** 111, 253002.

[342] Shi, W. Q.; Yuan, L. Y.; Wang, C. Z.; Wang, L.; Mei, L.; Xiao, C. L.; Zhang, L.; Li, Z. J.; Zhao, Y. L.; Chai, Z. F. Exploring Actinide Materials Through Synchrotron Radiation Techniques. *Adv. Mater.* **2014,** 26, 7807-7848.

[343] Ramanantoanina, H.; Kuri, G.; Martin, M.; Bertsch, J. Study of Electronic Structure in the L-Edge Spectroscopy of Actinide Materials: $UO_2$ as an Example. *Phys. Chem. Chem. Phys.* **2019,** 21, 7789-7801.

[344] Degueldre, C.; Bertsch, J.; Kuri, G.; Martin, M. Nuclear Fuel in Generation II and III Reactors: Research Issues Related to High Burn-Up. *Energ. Environ. Sci.* **2011,** 4, 1651-1661.

[345] Lang, M.; Tracy, C. L.; Palomares, R. I.; Zhang, F.; Severin, D.; Bender, M.; Trautmann, C.; Park, C.; Prakapenka, V. B.; Skuratov, V. A.; et al. Characterization of Ion-Induced Radiation Effects in Nuclear Materials Using Synchrotron X-ray Techniques. *J. Mater. Res.* **2015**, 30, 1366-1379.

[346] Selim, F. A. Positron Annihilation Spectroscopy of Defects in Nuclear and Irradiated Materials - A Review. *Mater. Charact.* **2021**, 174, 110952.

[347] Schultz, P. J.; Lynn, K. G. Interaction of Positron Beams with Surfaces, Thin Films, and Interfaces. *Rev. Mod. Phys.* **1988**, 60, 701-779.

[348] Mills Jr. A. P.; Platzman, P. M.; Brown, B. L. Slow-Positron Emission from Metal Surfaces. *Phys. Rev. Lett.* **1978**, 41 1076-1079.

[349] Hugenschmidt, C. Positrons in Surface Physics. *Surf. Sci. Rep.* **2016**, 71, 547-594.

[350] Auguste, R.; Liedke, M. O.; Selim, F. A.; Uberuaga, B. P.; Wagner, A.; Hosemann, P. Measurement and Simulation of Vacancy Formation in 2-MeV Self-irradiated Pure Fe. *JOM* **2020**, 72, 2436-2444.

[351] Pells, G. P. Radiation-Damage Effects in Alumina. *J. Am. Ceram. Soc.* **1994**, 77, 368-377.

[352] Popov, A. I.; Kotomin, E. A.; Maier, J. Basic Properties of the F-Type Centers in Halides, Oxides and Perovskites. *Nucl. Instrum. Meth. Phys. Res. B* **2010**, 268, 3084-3089.

[353] Lushchik, A.; Lushchik, C.; Schwartz, K.; Vasil'chenko, E.; Kaerner, T.; Kudryavtseva, I.; Isakhanyan, V.; Shugai, A. Stabilization and Annealing of Interstitials Formed by Radiation in Binary Metal Oxides and Fluorides. *Nucl. Instrum. Meth. Phys. Res. B* **2008**, 266, 2868-2871.

[354] Griffiths, T. R.; Dixon, J. Fast-Neutron Radiation on Single-Crystal Thorium-Dioxide – An Optical-Absorption Spectroscopy Study of the F(O) Color Center. *J. Chem. Soc. Faraday Trans.* **1992**, 88, 3475-3482.

[355] Xiao, H. Y.; Weber, W. J. Oxygen Vacancy Formation and Migration in $Ce_xTh_{1-x}O_2$ Solid Solution. *J. Phys. Chem. B* **2011**, 115, 6524-6533.

[356] Marabelli, F.; Wachter, P. Covalent Insulator $CeO_2$ – Optical Reflectivity Measurements. *Phys. Rev. B* **1987**, 36, 1238-1243.

[357] Patsalas, P.; Logothetidis, S.; Sygellou, L.; Kennou, S. Structure-Dependent Electronic Properties of Nanocrystalline Cerium Oxide Films. *Phys. Rev. B* **2003**, 68, 035104.

[358] Schoenes, J. Electronic-Transitions, Crystal-Field Effects and Phonons in $UO_2$. *Phys. Rep.* **1980**, 63, 301-336.

[359] Costantini, J. M.; Lelong, G.; Guillaumet, M.; Gourier, D.; Takaki, S.; Ishikawa, N.; Watanabe, H.; Yasuda, K. Optical Reflectivity of Ion-Irradiated Cerium Dioxide Sinters. *J. Appl. Phys.* **2019**, 126, 175902.

[360] Costantini, J.-M.; Lelong, G.; Guillaumet, M.; Weber, W. J.; Takaki, S.; Yasuda, K. Color Center Production and Recovery in Electron-Irradiated Magnesium Aluminate Spinel and Ceria. *J. Phys. Condens. Matter.* **2016**, 28, 325901.

[361] Schoenes, J. Optical-Properties and Electronic-Structure of $UO_2$. *J. Appl. Phys.* **1978**, 49, 1463-1465.

[362] Wang, J. W.; Ewing, R. C.; Becker, U. Electronic Structure and Stability of Hyperstoichiometric $UO_{2+x}$ Under Pressure. *Phys. Rev. B* **2013**, 88, 024109.

[363] Anderson, D. A.; Uberuaga, B. P.; Nerikar, P. V.; Unal, C.; Stanek, C. R. U and Xe Transport in $UO_{2\pm x}$: Density Functional Theory Calculations. *Phys. Rev. B* **2011**, 84, 054105.

[364] Conradson, S. D.; Durakiewicz, T.; Espinosa-Faller, F. J.; An, Y. Q.; Andersson, D. A.; Bishop, A. R.; Boland, K. S.; Bradley, J. A.; Byler, D. D.; Clark, D. L.; et al. Possible Bose-Condensate Behavior in a Quantum Phase Originating in a Collective Excitation in the Chemically and Optically Doped Mott-Hubbard System $UO_{2+x}$. *Phys. Rev. B* **2013**, 88, 115135.

[365] Marlow, P. G.; Russel, J. P.; Hardy, J. R. Raman Scattering in Uranium Dioxide. *Philos. Mag.* **1966**, 14, 409-410.

[366] Keramidas, V. G.; White, W. B. Raman Spectra of Oxides with the Fluorite Structure. *J. Chem. Phys.* **1973**, 59, 1561–1562.

[367] Rao, R.; Bhagat, R. K.; Salke, N. P.; Kumar, A. Raman Spectroscopic Investigation of Thorium Dioxide-Uranium Dioxide (ThO$_2$-UO$_2$) Fuel Materials. *Appl. Spectrosc.* **2014**, 68, 44-48.

[368] Mohun, R.; Desgranges, L.; Lechelle, J.; Simon, P.; Guimbretiere, G.; Canizares, A.; Duval, F.; Jegou, C.; Magnin, M.; Clavier, N.; et al. Charged Defects During Alpha-Irradiation of Actinide Oxides as Revealed by Raman and Luminescence Spectroscopy. *Nucl. Instrum. Meth. Phys. Res. B* **2016**, 374, 67-70.

[369] Weber, W. H.; Hass, K. C.; McBride, J. R. Raman-Study of CeO$_2$ – 2$^{nd}$-Order Scattering, Lattice-Dynamics, and Particle-Size Effects. *Phys. Rev. B* **1993**, 48, 178-185.

[370] Nakajima, A.; Yoshihara, A.; Ishigame, M. Defect-Induced Raman-Spectra in Doped CeO$_2$. *Phys. Rev. B* **1994**, 50, 13297-13307.

[371] McBride, J. R.; Hass, K. C.; Poindexter, B. D.; Weber, W. H. Raman and X-ray Studies of Ce$_{1-x}$Re$_x$O$_{2-y}$, Where Re=La, Pr, Nd, Ee, Gd, and Tb. *J. Appl. Phys.* **1994**, 76, 2435-2441.

[372] Lacina, W. B.; Pershan, P. S. Phonon Optical Properties of Ca$_{1-x}$Sr$_x$F$_2$. *Phys. Rev. B* **1970**, 1, 1765-1786.

[373] Haridasan, T. M.; Govindarajan, J.; Nerenberg, M. A.; Jacobs, P. W. M. Impurity Modes Due to Interstitials in CaF$_2$. *Phys. Rev. B* **1979**, 20, 3462-3473.

[374] Haridasan, T. M.; Govindarajan, J.; Nerenberg, M. A.; Jacobs, P. W. M. Phonon Resonances Associated with a Vacancy in CaF$_2$. *Phys. Rev. B* **1979**, 20, 3474-3480.

[375] Li, L.; Chen, F.; Lu, J. Q.; Luo, M. F. Study of Defect Sites in Ce$_{1-x}$M$_x$O$_{2-\delta}$ (x=0.2) Solid Solutions Using Raman Spectroscopy. *J. Phys. Chem. A* **2011**, 115, 7972-7977.

[376] Desgranges, L.; Baldinozzi, G.; Simon, P.; Guimbretiere, G.; Canizares, A. Raman Spectrum of U$_4$O$_9$: A New Interpretation of Damage Lines in UO$_2$. *J. Raman Spectrosc.* **2012**, 43, 455-458.

[377] Talip, Z.; Wiss, T.; Raison, P. E.; Paillier, J.; Manara, D.; Somers, J.; Konings, R. J. M. Raman and X-ray Studies of Uranium-Lanthanum-Mixed Oxides Before and After Air Oxidation. *J. Am. Ceram. Soc.* **2015**, 98, 2278-2285.

[378] Mohun, R.; Desgranges, L.; Jégou, C.; Boizot, B.; Cavani, O.; Canizarès, A.; Duval, F.; He, C.; Desgardin, P.; Barthe, M.F.; et al. Quantification of Irradiation-Induced Defects in UO$_2$ Using Raman and Positron Annihilation Spectroscopies. *Acta Mater.* **2019,** 164, 512-519.

[379] Colmenares, C.; Howell, R.; McCreary, T. Oxidation of Uranium Studied by Gravimetric and Positron Annihilation Techniques. *Technical Report (UCRL-85549) USA*, **1981**.

[380] Upadhyaya, D.; Muraleedharan, R.; Sharma, B. Study of Positron Lifetime Spectra in UO$_2$ Powders. *J. Nucl. Mater.* **1982**, 105, 219-222.

[381] Upadhyaya, D. D.; Muraleedharan, R. V.; Sharma, B. D.; Prasad, K. G. Positron Lifetime Studies on Thorium Oxide Powders, *Philos. Mag.* **1981**, 45, 509-518.

[382] Troev, T.; Penev, I.; Protochristov, H. Positron Anihilation in UO$_2$. *Phys. Lett. A* **1984**, 100A, 221-224.

[383] Chollet, M.; Krsjak, V.; Cozzo, C.; Bertsch, J. Positron Annihilation Spectroscopy Study of Lattice Defects in Non-Irradiated Doped and Un-Doped Fuels *EPJ Nuclear Sciences & Technologies* **2017,** 3, 3.

[384] Mogensen, M.; Sammes, N. M.; Tompsett, G. A. Physical, Chemical and Electrochemical Properties of Pure and Doped Ceria. *Solid State Ionics* **2000**, 129, 63-94.

[385] Ohmichi, T.; Fukushima, S.; Maeda, A.; Watanabe, H. On the Relation Between Lattice-Parameter and O/M Ratio for Uranium-Dioxide Trivalent Rare-Earth-Oxide Solid-Solution. *J. Nucl. Mater.* **1981**, 102, 40-46.

[386] Shannon, R. D., Revised Effective Ionic-Radii and Systematics Studies of Interatomic Distances in Halides and Chalcogenides. *Acta Crystallogr. A* **1976,** 32, 751-767.

[387] Grieshammer, S.; Zacherle, T.; Martin, M. Entropies of Defect Formation in Ceria From First Principles. *Phys. Chem. Chem. Phys.* **2013**, 15, 15935-15942.

[388] Bruneval, F.; Freyss, M.; Crocombette, J. P. Lattice Constant in Nonstoichiometric Uranium Dioxide From First Principles. *Phys. Rev. Mater.* **2018,** 2, 023801.

[389] Manara, D.; Renker, B. Raman Spectra of Stoichiometric and Hyperstoichiometric Uranium Dioxide. *J. Nucl. Mater.* **2003**, 321, 233–237.

[390] Elorrieta, J. M.; Bonales, L. J.; Rodríguez-Villagra, N.; Baonza, V. G.; Cobos, J. A Detailed Raman and X-ray Study of UO$_{2+X}$ Oxides and Related Structure Transitions. *Phys. Chem. Chem. Phys.* **2016,** 18, 28209-28216.

[391] Guéneau, C.; Baichi, M.; Labroche, D.; Chatillon, C.; Sundman, B. Thermodynamic Assessment of the Uranium–Oxygen System. *J. Nucl. Mater.* **2002**, 304, 161-175.

[392] Agarwal, R.; Parida, S. C. Phase Diagrams and Thermodynamic Properties of Thoria, Thoria–Urania, and Thoria–Plutonia. In, *Thoria-based Nuclear Fuels.* Springer, London **2013**, 71-105.

[393] Benz, R. Thorium-Thorium Dioxide Phase Equilibria. *J. Nucl. Mater.* **1969**, 29, 43-49.

[394] Darnell, A.; McCollum, W. A. *High Temperature Reactions of Thorium and Thoria and the Vapor Pressure of Thoria*; Report NAA-SR-6498; Office of Technical Services, Dept. Comm.: Washington, D.C., **1960.**

[395] Ackerman, R.; Rauh, E.; Thorn, R. A Thermodynamic Study of the Thorium-Oxygen System at High Temperatures. *J. Phys. Chem.* **1963**, 67, 762-769.

[396] Murphy, S.; Cooper, M. W. D.; Grimes, R. W. Point Defects and Non-Stoichiometry in Thoria. *Solid State Ionics* **2014**, 267, 80-87.

[397] Bergeron, A.; Manara, D.; Beneš, O.; Eloirdi, R.; Piro, M.; Corcoran, E. Point Defects and Non-Stoichiometry in Thoria: Description of the Th-U-Pu-O Quaternary System. *J. Nucl. Mater.* **2018**, 512, 324-348.

[398] Schmalzried, H. Chemical Kinetics of Solids; VCH Publishers: New York, **1995**; p 19.

[399] Piro, M.; Bergeron, A.; Corcoran, E. Thermodynamic Modelling of Thoria-Urania and Thoria-Plutonia Fuels. In, *Conference: Materials Modelling and Simulations for Nuclear Fuels,* **2017**; p 1-1.

[400] Wachsman, E. Solid-State Ionic Devices. *The Electrochemical Society Interface* **2007,** 27-28.

[401] Hassan, A.-R.; El-Azab, A.; Yablinsky, C.; Allen, T. R. Defect Disorder in UO$_2$. *J. Solid State Chem.* **2013**, 204, 136-145.

[402] Cooper, M. W. D.; Murphy, S.; Andersson. D. A. The Defect Chemistry of UO$_{2\pm x}$ From Atomistic Simulations. *J. Nucl. Mater.* **2018**, 504, 251-260.

[403] Soulié, A.; Bruneval, F.; Marinica, M.-C.; Murphy, S.; Crocombette, J.-P. Influence of Vibrational Entropy on the Concentrations of Oxygen Interstitial Clusters and Uranium Vacancies in Non-Stoichiometric UO$_2$. *Phy. Rev. Mater.* **2018**, 2, 083607.

404 Matzke, Hj. Radiation Damage in Crystalline Insulators, Oxides, and Ceramic Nuclear Fuels. *Radiat. Eff.* **1982**, 64, 3-33.

405 Wiss, T; Hiernaut, J.-P.; Roudil, D.; Colle, J.-Y.; Maugeri, E.; Talip, Z.; Janssen, A.; Rondinella, V.; Konings, R.; Matzke, H.-J.; et al. Evolution of Spent Nuclear Fuel in Dry Storage Conditions for Millennia and Beyond. *J. Nucl. Mater.* **2014**, 451, 198-206.

406 Cahill, D. G. Thermal-Conductivity Measurement by Time-Domain Thermoreflectance. *MRS Bull.* **2018**, 43, 782-788.

407 Khafizov, M.; Park, I.-W.; Chernatynskiy, A.; He, L.; Lin, J.; Moore, J. J.; Swank, D.; Lillo, T.; Phillpot, S. R.; El-Azab, A.; et al. Thermal Conductivity in Nanocrystalline Ceria Thin Films. *J. Am. Ceram. Soc.* **2014**, 97, 562-569.

408 Was, G. S.; Busby, J. T.; Allen, T.; Kenik, E. A.; Jensson, A.; Bruemmer, S. M.; Gan, J.; Edwards, A. D.; Scott, P. M.; Andreson, P. L. Emulation of Neutron Irradiation Effects with Protons: Validation of Principle. *J. Nucl. Mater.* **2002**, 300, 198-216.

409 Popel, A. J.; Spurgeon, S. R.; Matthews, B.; Olszta, M. J.; Thye Tan, B.; Gouder, T.; Eloirdi, R.; Buck, E. C.; Farnan, I. An Atomic-Scale Understanding of $UO_2$ Surface Evolution During Anoxic Dissolution. *ACS Appl. Mater. Interfaces* **2020**, 12, 39781-39786.

410 Clark, R. A.; Conroy, M. A.; Lach, T. G.; Buck, E. C.; Pellegrini, K. L.; McNamara, B. K.; Schwantes, J. M. Distribution of Metallic Fission-Product Particles in the Cladding Liner of Spent Nuclear Fuel. *npj Mater. Degrad.* **2020**, 4, 4.

411 Feng, B.; Sugiyama, I.; Hojo, H.; Ohta, H.; Shibata, N.; Ikuhara, Y. Atomic Structures and Oxygen Dynamics of $CeO2$ Grain Boundaries. *Sci. Rep.* **2016**, 6, 20288.

412 Fortner, J. A.; Buck, E. C. The Chemistry of the Light Rare-Earth Elements as Determined by Electron Energy Loss Spectroscopy. *Appl. Phys. Lett.* **1996**, 68, 3817-3819.

413 Garvie, L. A. J.; Buseck, P. R. Determination of $Ce^{4+}/Ce^{3+}$ in Electron-Beam-Damaged $CeO_2$ by Electron Energy-Loss Spectroscopy. *J. Phys. Chem. Solids* **1999**, 60, 1943-1947.

414 Pakarinen, J.; He, L. F.; Hassan, A. R.; Wang, Y. Q.; Gupta, M.; El-Azab, A.; Allen, T. R. Annealing-Induced Lattice Recovery in Room-Temperature Xenon Irradiated $CeO_2$: X-ray Diffraction and Electron Energy Loss Spectroscopy Experiments. *J. Mater. Res.* **2015**, 30, 1555-1562.

415 Colella, M.; Lumpkin, G. R.; Zhang, Z.; Buck, E. C.; Smith, K. L. Determination of the Uranium Valence State in the Brannerite Structure Using EELS, XPS, and EDX. *Phys. Chem. Miner.* **2005**, 32, 52-64.

416 Fortner, J. A.; Buck, E. C.; Ellison, A. J. G.; Bates, J. K. EELS Analysis of Redox in Glasses for Plutonium Immobilization. *Ultramicroscopy* **1997**, 67, 77-81.

417 Spurgeon, S. R.; Sassi, M.; Ophus, C.; Stubbs, J. E.; Ilton, E. S.; Buck, E. C. Nanoscale Oxygen Defect Gradients in $UO_{2+x}$ Surfaces. *Proc. Natl. Acad. Sci.* **2019**, 116, 17181-17186.

418 Kellogg, G. L.; Tsong, T. T. Pulsed-Laser Atom-Probe Field-Ion Microscopy. *J. Appl. Phys.* **1980**, 51, 1184-1193.

419 Vella, A. On the Interaction of an Ultra-Fast Laser with a Nanometric Tip by Laser Assisted Atom Probe Tomography: A Review. *Ultramicroscopy* **2013**, 132, 5–18.

420 Valderrama, B.; Henderson, H. B.; Gan, J.; Manuel, M. V. Influence of Instrument Conditions on the Evaporation Behavior of Uranium Dioxide with UV Laser-Assisted Atom Probe Tomography. *J. Nucl. Mater.* **2015**, 459, 37-43.

421 Devaraj, A.; Colby, R.; Hess, W. P.; Perea, D. E.; Thevuthasan, S. Role of Photoexcitation and Field Ionization in the Measurement of Accurate Oxide Stoichiometry by Laser-Assisted Atom Probe Tomography. *J. Phys. Chem. Lett.* **2013**, 4, 993–998.

422 Kirchhofer, R.; Teague, M. C.; Gorman, B. P. Thermal Effects on Mass and Spatial Resolution During Laser Pulse Atom Probe Tomography of Cerium Oxide. *J. Nucl. Mater.* **2013**, 436, 23–28.

423 Vella, A.; Mazumder, B.; Da Costa, G.; Deconihout, B. Field Evaporation Mechanism of Bulk Oxides Under Ultra Fast Laser Illumination. *J. Appl. Phys.* **2011**, 110, 044321.

424 Verberne, R.; Saxey, D. W.; Reddy, S. M.; Rickard, W. D. A.; Fougerouse, D.; Clark, C. Analysis of Natural Rutile ($TiO_2$) by Laser-Assisted Atom Probe Tomography. *Microsc. Microanal.* **2019**, 25, 539–546.

425 Marquis, E. A.; Yahya, N. A.; Larson, D. J.; Miller, M. K.; Todd. R. I. Probing the Improbable: Imaging C Atoms in Alumina. *Materials Today* **2010**, 13, 34–36.

426 Valderrama, B.; Henderson, H. B.; Yablinsky, C. A.; Gan, J.; Allen, T. R.; Manuel, M. V. Investigation of Material Property Influenced Stoichiometric Deviations as Evidenced During UV Laser-Assisted Atom Probe Tomography in Fluorite Oxides. *Nucl. Instrum. Meth. Phys. Res. B* **2015**, 359, 107–114.

427 Silaeva, E. P.; Karahka, M.; Kreuzer, H. J. Atom Probe Tomography and Field Evaporation of Insulators and Semiconductors: Theoretical Issues. *Curr. Opin. Solid State Mater. Sci.* **2013**, 17, 211–216.

428 Herbig, M. Spatially Correlated Electron Microscopy and Atom Probe Tomography: Current Possibilities and Future Perspectives. *Scripta Mater.* **2018**, 148, 98-105.

429 Makineni, S. K.; Lenz, M.; Kontis, P.; Li, Z.; Kumar, A.; Felfer, P. J.; Neumeier, S.; Herbig, M.; Spiecker, E.; Raabe, D.; et al. Correlative Microscopy-Novel Methods and Their Applications to Explore 3D Chemistry and Structure of Nanoscale Lattice Defects: A Case Study in Superalloys. *JOM* **2018**, 70, 1736-1743.

430 He, L.; Bachhav, M.; Keiser, D. D.; Madden, J. W.; Perez, E.; Miller, B. D.; Gan, J.; Van Renterghem, W.; Leenaers, A.; Van den Berghe, S. STEM-EDS/EELS and APT Characterization of ZrN Coatings on UMo Fuel Kernels. *J. Nucl. Mater.* **2018**, 511, 174-182.

431 Lach, T. G.; Olszta, M. J.; Taylor, S. D.; Yano, K. H.; Edwards, D. J.; Byun, T. S.; Chou, P. H.; Schreiber, D. K. Correlative STEM-APT Characterization of Radiation-Induced Segregation and Precipitation of In-Service BWR 304 Stainless Steel. *J. Nucl. Mater.* **2021**, 549, 152894.

432 Wang, X.; Hatzoglou, C.; Sneed, B.; Fan, Z.; Guo, W.; Jin, K.; Chen, D.; Bei, H. B.; Wang, Y. Q.; Weber, W. J.; et al. Interpreting Nanovoids in Atom Probe Tomography Data for Accurate Local Compositional Measurements. *Nat. Commun.* **2020**, 11, 1022.

433 Oberdorfer, C.; Eich, S. M.; Lutkemeyer, M.; Schmitz, G. Applications of a Versatile Modelling Approach to 3D Atom Probe Simulations. *Ultramicroscopy* **2015**, 159, 184-194.

434 Hofer, F.; Golob, P. Quantification of Electron Energy-Loss Spectra with K-Shell and L-Shell Ionization Cross-Sections. *Micron. Microsc. Acta* **1988**, 19, 73-86.

435 Amirifar, N.; Larde, R.; Talbot, E.; Pareige, P.; Rigutti, L.; Mancini, L.; Houard, J.; Castro, C.; Sallet, V.; Zehani, E.; et al. Quantitative Analysis of Doped/Undoped ZnO Nanomaterials Using Laser Assisted Atom Probe Tomography: Influence of the Analysis Parameters. *J. Appl. Phys.* **2015**, 118, 215703.

[436] Karahka, M.; Xia, Y.; Kreuzer, H. J. The Mystery of Missing Species in Atom Probe Tomography of Composite Materials. *Appl. Phys. Lett.* **2015**, 107, 062105.

[437] Graves, P. R. Raman Microprobe Spectroscopy of Uranium Dioxide Single Crystals and Ion Implanted Polycrystals. *Appl. Spectrosc.* **1990**, 44, 1665-1667.

[438] Guimbretière, G.; Desgranges, L.; Canizarès, A.; Carlot, G.; Caraballo, R.; Jégou, C.; Duval, F.; Raimboux, N.; Ammar, M. R.; Simon, P. Determination of In-Depth Damaged Profile by Raman Line Scan in a Pre-Cut $He^{2+}$ Irradiated $UO_2$. *Appl. Phys. Lett.* **2012**, 100, 251914.

[439] Desgranges, L.; Guimbretière, G.; Simon, P.; C. Jegou, C.; Caraballo, R. A Possible New Mechanism for Defect Formation in Irradiated $UO_2$. *Nucl. Instrum. Meth. Phys. Res. B* **2013**, 315, 169–172.

[440] Nakae, N.; Iwata, Y.; Kirihara, T. Thermal Recovery of Defects in Neutron-Irradiated $UO_2$. *J. Nucl. Mater.* **1979**, 80, 314-322.

[441] Roudil, D.; Barthe MF.; Jégou, C.; Gavazzi, A.; Vella, F. Investigation of Defects in Actinide-Doped $UO_2$ by Positron Annihilation Spectroscopy. *J. Nucl. Mater.* **2012**, 420, 63-68.

[442] Yamamoto, Y.; Kishino, T.; shiyama, T.; Iwase, A.; Hori, F. Study on Lattice Defects in $CeO_2$ by Means of Positron Annihilation Measurements. *J. Phys. Conf. Ser.* **2016**, 674, 012015.

[443] Wiktor, J.; Barthe, MF.; Jomard, G.; Torrent, M.; Freyss, M.; Bertolus, M. Coupled Experimental and DFT+ U Investigation of Positron Lifetimes in $UO_2$. *Phys. Rev. B* **2014**, 90, 184101.

[444] Djourelov, N.; Marchand, B.; Marinov, H.; Moncoffre, N.; Pipon, Y.; Nédélec, P.; Toulhoat, N.; Sillou, D. Variable Energy Positron Beam Study of Xe-Implanted Uranium Oxide. *J. Nucl. Mater.* **2013**, 432, 287-293.

[445] Evans, H.; Evans, J.; Rice-Evans, P.; Smith, D.; Smith, C. A Slow Positron Study Into the Recovery Behaviour of As-Polished and Krypton-Implanted Uranium Dioxide. *J. Nucl. Mater.* **1992**, 199, 79-83.

[446] Barthe, M.; Labrim, H.; Gentils, A.; Desgardin, P.; Corbel, C.; Esnouf, S.; Piron, J. P. Positron Annihilation Characteristics in $UO_2$: For Lattice and Vacancy Defects Induced by Electron Irradiation. *Phys. Status Solidi C* **2007**, 4, 3627-3632.

[447] Labrim, H.; Barthe, M.; Desgardin, P.; Sauvage, T.; Blondiaux, G.; Corbel, C.; Piron, J. P. Vacancy Defects Induced in Sintered Polished $UO_2$ Disks by Helium Implantation. *Appl. Surf. Sci.* **2006**, 252, 3256-3261.

[448] Labrim, H.; Barthe, M.; Desgardin, P.; Sauvage, T.; Corbel, C.; Blondiaux, G.; Piron, J. P. Thermal Evolution of the Vacancy Defects Distribution in 1MeV Helium Implanted Sintered $UO_2$. *Nucl. Instrum. Meth. Phys. Res. B* **2007**, 261, 883-887.

[449] Djourevol, N.; Marchand, B.; Marinov, H.; Moncoffre, N.; Pipon, Y.; Bererd, N.; Nedelec, P.; Raimbault, L.; Epicier, T. Study of the Temperature and Radiation Induced Microstructural Changes in Xe-Implanted $UO_2$ by TEM, STEM, SIMS and Positron Spectroscopy *J. Nucl. Mater.* **2013**, 443, 562-569.

[450] Nogita, K.; Une, K. High Resolution TEM Observation and Density Estimation of Xe Bubbles in High Burnup $UO_2$ Fuels. *Nucl. Instrum. Meth. Phys. Res. B* **1998**, 141, 481-486.

[451] Sabathier, C.; Vincent, L.; Garcia, P.; Garrido, F.; Carlot, G.; Thome, L.; Martin, P.; Valot, C. In Situ TEM of Temperature-Induced Fission Product Precipitation in $UO_2$. *Nucl. Instrum. Meth. Phys. Res. B* **2008**, 266, 3027-3032.

[452] Michel, A.; Sabathier, C.; Carlot, G.; Kaitasov, O.; Bouffard, S.; Garcia, P.; Valot, C. An In Situ TEM Study of the Evolution of Xe Bubble Populations in $UO_2$. *Nucl. Instrum. Meth. Phys. Res. B* **2012**, 272, 218-221.

[453] Michel, A.; Sabathier, C.; Carlot, G.; Cabie, M.; Bouffard, S.; Garcia, P. A TEM Study of Bubbles Growth with Temperature in Xenon and Krypton Implanted Uranium Dioxide. *Diffusion in Materials* **2012**, 323-325, 191-196.

[454] Lamontagne, J.; Desgranges, L.; Valot, C.; Noirot, J.; Blay, T.; Roure, I.; B. Pasquet, B. Fission Gas Bubble Characterisation in Irradiated $UO_2$ Fuel by SEM, EPMA and SIMS. *Microchim. Acta* **2006**, 155, 183-187.

[455] Martin, P. M.; Vathonne, E.; Carlot, G.; Delorme, R.; Sabathier, C.; Freyss, M.; Garcia, P.; Bertolus, M.; Glatzel, P.; Proux, O. Behavior of Fission Gases in Nuclear Fuel: XAS Characterization of Kr in $UO_2$. *J. Nucl. Mater.* **2015**, 466, 379-392.

[456] Garcia, P.; Martin, P.; Carlot, G.; Castelier, E.; Ripert, M.; Sabatheir, C.; Valot, C.; D'Acapito, F.; Hazemann, J.-L.; Proux, O.; et al. A Study of Xenon Aggregates in Uranium Dioxide Using X-ray Absorption Spectroscopy. *J. Nucl. Mater.* **2006**, 352, 136-143.

[457] Martin, P.; Garcia, P.; Carlot, G.; Sabathier, C.; Valot, C.; Nassif, V.; Proux, O.; Hazemann, J. L. XAS Characterisation of Xenon Bubbles in Uranium Dioxide. *Nucl. Instrum. Meth. Phys. Res. B* **2008**, 266, 2887-2891.

[458] Sabathier, C.; Martin, G.; Michel, A.; Carlot, G.; Maillard, S.; Bachelet, C.; Fortuna, F.; Kaitasov, O.; Oliviero, E.; Garcia, P. In-Situ TEM Observation of Nano-Void Formation in $UO_2$ Under Irradiation. *Nucl. Instrum. Meth. Phys. Res. B* **2014**, 326, 247-250.

[459] Perrin-Pellegrino, C.; Dumont, M.; Fadel Keita, M.; Neisius, T.; Mikaelian, G.; Mangelinck, D.; Carlot, G.; Maugis, P. Characterization by APT and TEM of Xe Nano-Bubbles in $CeO_2$. *Nucl. Instrum. Meth. Phys. Res. B* **2020**, 469, 24-27.

[460] Asoka-Kumar, P.; Alatalo, M.; Ghosh, V. J.; Kruseman, A. C.; Nielsen, B.; Lynn, K. G. Increased Elemental Specificity of Positron Annihilation Spectr, *Phys. Rev. Let.* **1996**, 77, 2097-2100.

[461] Nagai, Y.; Hasegawa, M.; Tang, T.; Hempel, A.; Yubuta, K.; Shimamura, T.; Kawazoe, Y.; Kawai, A.; Kano, F. Positron Confinement in Ultrafine Embedded Particles: Quantum-Dot-Like State in an Fe-Cu Alloy *Phys. Rev. B* **2000**, 61, 6574-6578.

[462] Agarwal, S.; Liedke, M. O.; Jones, A. C. L.; Reed, E.; Kohnert, A. A.; Uberuaga, B. P.; Wang, Y. Q.; Cooper, J., Kaomi, D. A New Mechanism for Void-Cascade Interaction From Nondestructive Depth-Resolved Atomic-Scale Measurements of Ion Irradiation–Induced Defects in Fe. *Science Advances* **2020**, 6, eaba8437.

[463] Jiang, J.; Wu, Y.C.; Liu, X. B.; Wang, R. S.; Nagai, Y.; Inoue, K.; Shimizu, Y.; Toyama, T. Microstructural Evolution of RPV Steels Under Proton and Ion Irradiation Studied by Positron Annihilation Spectroscopy. *J. Nucl. Mater.* **2015**, 458, 326-334.

[464] Jiang, W.; Conroy, M. A.; Kruska, K.; Overman, N. R.; Droubay, T. C.; Gigx, J.; Shao, L.; Devanathan R. Nanoparticle Precipitation in Irradiated and Annealed Ceria Doped with Metals for Emulation of Spent Fuel. *J. Phys. Chem.* **2017**, 121, 22465-22477.

[465] Kuri, G.; Miexczynsk, C.; Martin, M.; Betrsch, J.; Borca, C. N.; Delafoy, Ch. Local Atomic Structure of Chromium Bearing Precipitates in Chromia Doped Uranium Dioxide Investigated by Combined Micro-Beam X-ray Diffraction and Absorption Spectroscopy *J. Nucl. Mater.* **2014**, 449, 158-167.

[466] Yun, D.; Oaks, A. J.; Chen, W.-y.; Kirk, M. A.; Rest, J.; Insopov, Z. Z.; Yacout, A. M.; Stubbins, J. F. Kr and Xe Irradiations in Lanthanum (La) Doped Ceria: Study at the High Dose Regime. *J. Nucl. Mater.* **2011**, 418, 80-86.

[467] Bachhav, M.; Gan, J.; Keiser, D.; Giglio, J.; Jadernas, D.; Leenaers, A.; Van den Berghe, S. A Novel Approach to Determine the Local Burnup in Irradiated Fuels using Atom Probe Tomography (APT). *J. Nucl. Mater.* **2020**, 528, 151853.

[468] Bachhav, M.; He, L.; Kane, J.; Liu, X.; Gan, J.; Vurpiliot, F. Atom Probe Tomography for Burnup and Fission Product Analysis for Nuclear Fuels. *Microsc. Microanal.* **2020**, 26, 3086-3088.

[469] Ryazanov, A. I.; Kinoshita, C. Growth Kinetics of Dislocation Loops in Irradiated Ceramic Materials. *Nucl. Instrum. Meth. Phys. Res. B* **2002**, 191, 65-73.

[470] Kiritani, M.; Yoshida, N.; Takata, H.; Maehara, Y. Growth of Interstitial Type Dislocation Loops and Vacancy Mobility in Electron-Irradiated Metals. *J. Phys. Soc. Japan* **1975**, 38, 1677-1686.

[471] Kinoshita, C.; Hayashi, K.; Kitajima, S. Kinetics of Point-Defects in Electron-Irradiated MgO. *Nucl. Instrum. Meth. Phys. Res. B* **1984**, *1*, 209-218.

[472] Mansur, L. K. Theory and Experimental Background on Dimensional Changes in Irradiated Alloys. *J. Nucl. Mater.* **1994**, 216, 97-123.

[473] Ghoniem, N. M.; Cho, D. D. The Simultaneous Clustering of Point Defects during Irradiation. *Phys. Status Solidi A* **1979**, 54, 171-178.

[474] Dunn, A.; Capolungo, L. Simulating Radiation Damage Accumulation in α-Fe: A Spatially Resolved Stochastic Cluster Dynamics Approach. *Compu. Mater. Sci.* **2015**, 102, 314–326.

[475] Dunn, A.; Dingreville, R.; Martínez, E.; Capolungo, L. Synchronous Parallel Spatially Resolved Stochastic Cluster Dynamics, *Compu. Mater. Sci.* **2016**, 120, 43-52.

[476] Wirth, B. D.; Hu, X.; Kohnert, A.; Xu, D. Modeling Defect Cluster Evolution in Irradiated Structural Materials: Focus on Comparing to High-Resolution Experimental Characterization Studies. *J. Mater. Res.* **2015**, 30, 1440-1455.

[477] Blondel, S.; Bernholdt, D. E.; Hammond, K. D.; Wirth, B. D. Continuum-Scale Modeling of Helium Bubble Bursting Under Plasma-Exposed Tungsten Surfaces. *Nucl. Fusion.* **2018**, 58, 126034.

[478] Matthews, C.; Perriot, R.; Cooper, M. W. D.; Stanek, C. R.; Andersson, D. A. Cluster Dynamics Simulation of Uranium Self-Diffusion During Irradiation in $UO_2$. *J. Nucl. Mater.* **2019**, 527, 151787.

[479] Matthews, C.; Perriot, R.; Cooper, M. W. D.; Stanek, C. R.; Andersson, D. A. Cluster Dynamics Simulation of Xenon Diffusion During Irradiation in $UO_2$. *J. Nucl. Mater.* **2020**, 540, 152326.

[480] Cooper, M. W. D.; Pastore, G.; Che, Y.; Matthews, C.; Forslund, A.; Stanek, C. R.; Shirvan, K.; Tverberg, T.; Gamble, K. A.; Mays, B.; et al. Fission Gas Diffusion and Release for $Cr_2O_3$-Doped $UO_2$: From the Atomic to the Engineering Scale. *J. Nucl. Mater.* **2020**, 545, 152590.

[481] Gokhman, A.; Bergner. F. Cluster Dynamics Simulation of Point Defect Clusters in Neutron Irradiated Pure Iron. *Radiat. Eff. Defects Solids* **2010**, 165, 216–226.

[482] Pokor, C.; Brechet, Y.; Dubuisson, P.; Massoud, J. P.; Barbu, A. Irradiation Damage in 304 and 316 Stainless Steels: Experimental Investigation and Modeling. Part I: Evolution of the Microstructure. *J. Nucl. Mater.* **2004**, 326, 19–29.

[483] Brimbal, C.; Fournier, L.; Barbu. A. Cluster Dynamics Modeling of the Effect of High Dose Irradiation and Helium on the Microstructure of Austenitic Stainless Steels. *J. Nucl. Mater.* **2016**, 468, 124–139.

[484] Christien, F.; Barbu, A. Effect of Self-Interstitial Diffusion Anisotropy in Electron-Irradiated Zirconium: A Cluster Dynamics Modelling. *J. Nucl. Mater.* **2005**, 346, 272–281.

[485] Clouet, E.; Barbu, A.; Laé, L.; Martin, G. Precipitation Kinetics of $Al_3Zr$ and $Al_3Sc$ in Aluminum Alloys Modeled with Cluster Dynamics. *Acta Mater.* **2005**, 53, 2313–2325.

[486] Bai, X. M.; Ke, H.; Zhang, Y.; Spencer, B. W. Modeling Copper Precipitation Hardening and Embrittlement in a Dilute Fe-0.3 at.% Cu Alloy Under Neutron Irradiation. *J. Nucl. Mater.* **2017**, 495, 442–454.

[487] Skorek, R.; Maillard, S.; Michel, A.; Carlot, G.; Gilabert, E.; Jourdan. T. Modelling Fission Gas Bubble Distribution in $UO_2$. *Defect and Diffusion Forum* **2012**, 323–325, 209–214.

[488] Khalil, S.; Allen, T. R.; El-Azab. A. Off-stoichiometric Defect Clustering in Irradiated Oxides. *Chem. Phys.* **2017**, 487, 1–10.

[489] Kohnert, A. A.; Wirth, B. D.; Capolungo, L. Modeling Microstructural Evolution in Irradiated Materials with Cluster Dynamics Methods: A review. *Compu. Mater. Sci.* **2018**, 149, 442–45.

[490] Vincent-Aublant, E.; Delaye, J.-M.; Van Brutzel, L. Self-Diffusion Near Symmetrical Tilt Grain Boundaries in $UO_2$ Matrix: A Molecular Dynamics Simulation Study. *J. Nucl. Mater.* **2009**, 392,114-120.

[491] Symington, A. R.; Molinari, M.; Brincat, N. A.; Williams, N. R.; Parker, S. C. Defect Segregation Facilitates Oxygen Transport at Fluorite $UO_2$ Grain Boundaries. *Phil. Trans. Royal Soc. A* **2019**, 377, 20190026.

[492] Perriot, R.; Dholabhai, P. P.; B.P. Uberuaga, B. P. Disorder-Induced Transition from Grain Boundary to Bulk Dominated Ionic Diffusion in Pyrochlores. *Nanoscale* **2017**, 9, 6826-6836.

[493] Williams, N. R.; Molinari, M.; Parker, S. C.; Storr, M. T. Atomistic Investigation of the Structure and Transport Properties of Tilt Grain Boundaries of $UO_2$. *J. Nucl. Mater.* **2015**, 458, 45-55.

[494] Murphy, S. T.; Jay, E. E.; Grimes, R. W. Pipe Diffusion at Dislocations in $UO_2$. *J. Nucl. Mater.* **2014**, 447, 143-149.

[495] Galvin, C. O.; Cooper, M. W. D.; Fossati, P. C. M.; Stanek, C. R.; Grimes, R. W.; Andersson, D. A. Pipe and Grain Boundary Diffusion of He in $UO_2$. *J. Phys. Condens. Matter* **2016**, 28, 405002.

[496] Andersson, D. A.; Tonks, M. R.; Casillas, L.; Vyas, S.; Nerikar, P.; Uberuaga, B. P.; Stanek, C. R. Multiscale Simulation of Xenon Diffusion and Grain Boundary Segregation in $UO_2$. *J. Nucl. Mater.* **2015**, 462, 15-25.

[497] Nerikar, P. V.; Parfitt, D. C.; Casillas Trujillo, L. A.; Andersson, D. A.; Unal, C.; Sinnott, S. B.; Grimes, R. W.; Uberuaga, B. P.; Stanek C. R. Segregation of Xenon to Dislocations and Grain Boundaries in Uranium Dioxide *Phys. Rev. B* **2011**, 84, 174105.

[498] Vincent-Aublant, E.; Delaye, J. M.; Van Brutzel, L. Self-diffusion Near Symmetrical Tilt Grain Boundaries in $UO_2$ Matrix: A Molecular Dynamics Simulation Study. *J. Nucl. Mater.* **2009**, 392, 114-120.

[499] Valderrama, B.; He, L.; Henderson, H. B.; Pakarinen, J.; Jaques, B.; Gan, J.; Butt, D. P.; Allen, T. R.; Manuel, M. V. Effect of Grain Boundaries on Krypton Segregation Behavior in Irradiated Uranium Dioxide. *JOM* **2014**, 66, 2562-2568.

[500] Zacharie, I.; Lansiart, S.; Combette, P.; Trotabas, M.; Coster, M.; Groos, M. Microstructural Analysis and Modelling of Intergranular Swelling of Irradiated $UO_2$ Fuel Treated at High Temperature. *J. Nuc. Mat.* **1998** 255, 92-104.

[501] Millett, P. C.; Tonks, M. R. Meso-Scale Modeling of the Influence of Intergranular Gas Bubbles on Effective Thermal Conductivity. *J. Nucl. Mater.* **2011**, 412, 281-286.

[502] Millett, P. C.; Zhang, Y.; Andersson, D. A.; Tonks, M. R.; Biner, S. B. Random-Walk Monte Carlo Simulation of Intergranular Gas Bubble Nucleation in $UO_2$ Fuel. *J. Nucl. Mater.* **2012**, 430, 44-49.  50

[503] Millett, P. C.; Tonks, M. R.; Biner, S. B.; Zhang, L.; Chockalingam, K.; Zhang, Y. Phase-Field Simulation of Intergranular Bubble Growth and Percolation in Bicrystals. *J. Nucl. Mater.* **2012**, 425, 130-135.

504 Dholabhai, P. P.; Aguiar, J. A.; Wu, L.; Holesinger, T. G.; Aoki, T.; Castro, R. H. R.; Uberuaga, B. P. Structure and Segregation of Dopant–Defect Complexes at Grain Boundaries in Nanocrystalline Doped Ceria. *Phys. Chem. Chem. Phys.* **2015**, 17, 15375-15385.

505 Li, F.; Ohkubo, T.; Chen, Y. M.; Kodzuka, M.; Hono, K. Quantitative Atom Probe Analyses of Rare-Earth-Doped Ceria by Femtosecond Pulsed Laser *Ultramicroscopy*, **2011**, 111, 589-594.

506 Li, F.; Ohkubo, T.; Chen, Y. M.; Kozdzuko, M.; Ye, F.; Ou, D. R.; Mori, T.; Hono, K. Laser-Assisted Three-Dimensional Atom Probe Analysis of Dopantdistribution in Gd-doped $CeO_2$. *Scripta Mater.* **2010**, 63, 332-335.

507 Sun, L.; Marrocchelli, D.; Yildiz, B. Edge Dislocation Slows Down Oxide Ion Diffusion in Doped $CeO_2$ by Segregation of Charged Defects. *Nat. Commun.* **2015**, 6, 6294.

508 Barney, W. K.; Wemple, B. D. *Metallography of Irradiated $UO_2$ Containing Fuel Elements*; KAPL-1836; Knolls Atomic Power Laboratory, Schenectady, **1958.**

509 Belle, J. *Uranium Dioxide: Properties and Nuclear Applications;* Naval Reactors Handbooks; U.S. Atomic Energy Commission, Washington, D.C., **1961**.

510 Matzke, Hj.; Turos, A.; Linker, G. Polygonization of Single Crystals of the Fluorite-Type Oxide $UO_2$ Due to High Dose Ion Implantation, *Nucl. Instrum. Methods Phys. Res. B* **1994**, 91, 294-300.

511 Spino, J.; Papaioannou, D. Lattice Parameter Changes Associated with the Rim-Structure Formation in High Burn-Up $UO_2$ Fuels by Micro X-ray Diffraction *J. Nucl. Mat.* **2000**, 281, 146-162.

512 Bai, X. -M.; Tonks, M. R.; Zhang, Y.; Hales, J. D. Multiscale Modeling of Thermal Conductivity of High Burnup Structures in $UO_2$ Fuels *J. Nucl. Mat.* **2016**, 470, 208-215.

513 Wiss, T.; Rondinella, V. V.; Konings, R. J. M.; Staicu, D.; Papaioanou, D.; Bremier, S.; Poml, P.; Benes, O.; Colle, J. -Y.; Van Uffelen, P.; et al. Properties of the High Burnup Structure in Nuclear Light Water Reactor Fuel. *Radiochim. Acta* **2017**, 105, 893-906.

514 Matzke, Hj.; Wang, L. M. High Resolution Transmission Electron Microscopy of Ion Irradiated Uranium Oxide *J. Nucl. Mat.* **1996**, 231, 155-158.

515 Hayashi, K.; Kikuchi, H.; Fukuda, K. Radiation Damage of $UO_2$ by High Energy Heavy Ions *J. Nucl. Mat.* **1997**, 248, 191-195.

516 Sonoda, T.; Kinoshita, M.; Ishikawa, N.; Sataka, M.; Iwase, A.; Yasunaga, K. Clarification of High Density Electronic Excitation Effects on the Microstructural Evolution of $UO_2$. *Nucl. Instrum. Methods Phys. Res. B* **2010**, 268, 3277-3281.

517 Baranov, V. G.; Lunev, A. V.; Reutov, V. F.; Tenishev, A. V.; Isaenkova, M. G.; Khlunov, A. V. An Attempt to Reproduce High Burn-Up Structure by Ion Irradiation of SIMFUEL. *J. Nucl. Mat.* **2014**, 452, 147-157.

518 Miao, Y.; Yao, T.; Lian, J.; Zhu, S.; Bhattacharya, S.; Oaks, A.; Yacout, A. M.; Mo, K. Nano-Crystallization Induced by High-Energy Heavy Ion Irradiation in $UO_2$. *Scripta Mater.* **2018**, 155, 169-174.

519 Gerczak, T. J.; Parhish, C. M.; Edmondson, P. D.; Baldwin, C. A.; Terrani, K. A. Restructuring in High Burnup $UO_2$ Studied Using Modern Electron Microscopy. *J. Nucl. Mat.* **2018**, 509, 245-259.

520 Ferry, S. E.; Dennett, C. A.; Woller, K. B.; Short, M. P. Inferring Radiation-Induced Microstructural Evolution in Single-Crystal Niobium Through Changes in Thermal Transport. *J. Nucl. Mater.* **2019,** 523, 378-382.

521 Cinbiz, M. N.; Wiesenack, W.; Yagnik, S.; Terrani, K. A. In-Pile Thermal Conductivity of Uranium Dioxide at Low Burnup. *J. Nucl. Mater.* **2020,** 538, 152210.

522 Dennett, C. A.; Buller, D. L.; Hattar, K.; Short, M. P. Real-Time Thermomechanical Property monitoring During Ion Beam Irradiation Using In Situ Transient Grating Spectroscopy. *Nucl. Instrum. Meth. Phys. Res. B* **2019,** 440, 126-138.

523 Wang, Y. Z.; Sha, G. F.; Harlow, C.; Yazbeck, M.; Khafizov, M. Impact of Nuclear Reactor Radiation on the Performance of AlN/Sapphire Surface Acoustic Wave Devices. *Nucl. Instrum. Meth. Phys. Res.* **2020,** 481, 35-41.

524 Schley, R. S.; Hua, Z.; Hurley, D. H. *Status Report: Design and Fabricate Installation-Ready Test Capsule for the In-Situ Measurement of Elastic Properties Using an Optical Fiber-Based Measurement Technique*; INL/EXT-19-53123; Idaho National Laboratory, Idaho Falls, **2019**.

525 Guimbretiere, G.; Desgranges, L.; Canizares, A.; Caraballo, R.; Duval, F.; Raimboux, N.; Omnee, R.; Ammar, M. R.; Jegou, C.; Simon, P. In Situ Raman Monitoring of $He^{2+}$ Irradiation Induced Damage in a $UO_2$ Ceramic. *Appl. Phys. Lett.* **2013,** 103, 041904.

526 Petrie, C. M.; Blue, T. E., In Situ Reactor Radiation-Induced Attenuation in Sapphire Optical Fibers Heated up to 1000 degrees C. *Nucl. Instrum. Meth. Phys. Res. B* **2015**, 342, 91-97.

527 Pena-Rodriguez, O.; Crespillo, M. L.; Diaz-Nunez, P.; Perlado, J. M.; Rivera, A.; Olivares, J. In Situ Monitoring the Optical Properties of Dielectric Materials During Ion Irradiation. *Opt. Mater. Express* **2016**, 6, 734-742.

528 Dennett, C. A.; Dacus, B. R.; Barr, C. M.; Clark, T.; Bei, H.; Zhang, Y.; Short, M. P.; Hattar, K. The Dynamic Evolution of Swelling in Nickel Concentrated Solid Solution Alloys Through In Situ Property Monitoring. *Appl. Mater. Today* **2021**, 25, 101187.

529 Khafizov, M.; Chauhan, V; Wang, Y.; Riyad, F.; Hang, N.; Hurley, D. H. Investigation of Thermal Transport in Composites and Ion Beam Irradiated Materials for Nuclear Energy Applications. *J. Mater. Res.* **2017**, 32, 204-216.

530 Lucuta, P. G.; Matzke, Hj.; Hastings, I. J. A Pragmatic Approach to Modelling Thermal Conductivity of Irradiated $UO_2$ Fuel: Review and Recommendations. *J. Nucl. Mater.* **1996,** 232, 166-180.

531 Geelhood, K.; Luscher, W. G.; Beyer, C. E. *FRAPCON-3.4: a Computer Code for the Calculation of Steady State Thermal-Mechanical Behavior of Oxide Fuel Rods for High Burnup;* NRC: Vol. NUREG-CR-7022; US Nuclear Regulatory Commission, Richland, **2011**.

532 Lassmann, K. TRANSURANUS - A Fuel-Rod Analysis Code Ready For Use. *J. Nucl. Mat.* **1992**, 188, 295-302.

533 Van Uffelen, P.; Hales, J.; Li, W.; Rossiter, G.; Williamson, R. A Review of Fuel Performance Modeling. *J. Nucl. Mater.* **2019**, 516, 373-412.

534 Williamson, R.; Hales, J.; Novascone, S.; Tonks, M.; Gaston, D.; Permann, C.; Andrs, D.; Martineau, R. Multidimensional Multiphysics Simulation of Nuclear Fuel Behavior. *J. Nucl. Mat.* **2012**, 423, 149-163.

535 Williamson, R. L.; Hales, J. D.; Novascone S. R.; Pastore G.; Gamble K. A.; Spencer B. W.; Jiang, W.; Pitts, S. A.; Casagranda, A.; Schwen, D.; et al. BISON: A Flexible Code for Advanced Simulation of the Performance of Multiple Nuclear Fuel Forms. *Nucl. Tech.* **2021**, 207, 954-980.

536 Klemens, P. G. Thermal Conductivity and Lattice Vibrational Modes. *Solid State Phys.* **1958,** 7, 1–98.

537 Berman, R. *Thermal Conduction in Solids*; Clarendon Press: Oxford, **1976**, p 11.

538 Staicu, D. Thermal Properties of Irradiated $UO_2$ and MOX. In, *Comprehensive Nuclear Materials;* Elsevier: The Netherlands, **2012**; p 439.

[539] Wiesenack, W.; Tverberg, T. The OECD Halden Reactor Project Fuels Testing Programme: Methods, Selected Results and Plans. *Nucl. Eng. Des.* **2001**, 207, 189-197.

[540] Fujii, H.; Teshima, H.; Kanasugi, K.; Sendo, T. MOX Fuel Performance Experiment Specified for Japanese PWR Utilisation in the HBWR. *J. Nucl. Sci. Tech.* **2006**, 43, 998-1005.

[541] Nakamura, J.; Amaya, M.; Nagase, F.; Fuketa, T. Thermal Conductivity Change in High Burnup MOX Fuel Pellet. *J. Nucl. Sci. Tech.* **2009**, 46, 944-952.

[542] Nakae, N.; Akiyama, H.; Miura, H.; Baba, T.; Kamimura, K.; Kurematsu, K.; Kosaka, Y.; Yoshino, A.; Kitagawa, T. Thermal Property Change of MOX and $UO_2$ Irradiated up to High Burnup of 74 GWd/t. *J. Nucl. Mater.* **2013**, 440, 515-523.

[543] Thermal Conductivities of Irradiated $UO_2$ and $(U,Gd)O_2$. *J. Nucl. Mater.* **2001**, 288, 57-65.

[544] Fink, J. K. Thermophysical Properties of Uranium Dioxide. *J. Nucl. Mater.* **2000,** 279, 1-18.

[545] Bakker, K.; Cordfunke, E. H. P.; Konings, R. J. M.; Schram, R. P. C. Critical Evaluation of the Thermal Properties of $ThO_2$ and $Th_{1-y}U_yO_2$ and a Survey of the Literature Data on $Th_{1-y}Pu_yO_2$. *J. Nucl. Mater.* **1997,** 250, 1-12.

[546] Pastore, G.; Swiler, L. P.; Hales, J. D.; Novascone, S. R.; Perez, D. M.; Spencer, B. W.; Luzzi, L.; Van Uffelen, P.; Williamson, R. L. Uncertainty and Sensitivity Analysis of Fission Gas Behavior in Engineering-Scale Fuel Modeling. *J. Nucl. Mater.* **2015**, 456, 398-408.

[547] Tonks, M. R.; Andersson, D.; Phillpot, S. R.; Zhang, Y.; Williamson, R.; Stanek, C. R.; B.P. Uberuaga, B. L.; Hayes, S. L. Mechanistic Materials Modeling for Nuclear Fuel Performance. *Ann. Nucl. Energy* **2017**, 105, 11–24.

[548] Chockalingam, K.; Millett, P. C.; Tonks, M. R. Effects of Intergranular Gas Bubbles on Thermal Conductivitiy. *J. Nucl. Mater.* **2012**, 430, 166-170.

[549] Biancheria, A. Effect of Porosity on Thermal Conductivity of Ceramic Bodies. *Trans. Am. Nucl. Soc.* **1966**, 9, 1.

[550] Tonks, M. R.; Bhave, C.; Wu, X.; Zhang, Y. 10 – Uncertainty Quantification of Mesoscale Models of Porous Uranium Dioxide. In, *Uncertainty quantification of multiscale materials modeling*. Woodhead Publishing. **2020**, 329-354.

[551] White, J. T.; Nelson, A. T. Thermal Conductivity of $UO_{2+x}$ and $U_4O_{9-y}$. *J. Nucl. Mater.* **2013,** 443, 342-350.

[552] Watanabe, T.; Sinnott, S. B.; Tulenko, J. S.; Grimes, R. W.; Schelling, P. K.; Phillpot, S. R. Thermal Transport Properties of Uranium Dioxide by Molecular Dynamics Simulations. *J. Nucl. Mater.* **2008**, 375, 388-396.

[553] Nichenko, S.; Staicu, D. Thermal Conductivity of Porous $UO_2$: Molecular Dynamics Study. *J. Nucl. Mater.* **2014**, 454, 315–322.

[554] Chen, W.; Bai, X.-M. Unified Effect of Dispersed Xe on the Thermal Conductivity of $UO_2$ Predicted by Three Interatomic Potentials. *JOM* **2020**, 72, 1710-1718.

[555] Deskins, W. R.; Hamed, A.; Kumagai, T.; Dennett, C. A.; Peng, J.; Khafizov, M.; Hurley, D. H.; El-Azab, A. Thermal Conductivity of $ThO_2$ : Effect of Point Defect Disorder. *J. Appl. Phys.* **2021**, 129, 075102.

[556] Lucuta, P. G.; Matzke, H.; Verrall, R. A.; Tasman, H. A. Thermal-Conductivity of SIMFUEL. *J. Nucl. Mater.* **1992**, 188, 198-204.

[557] Ishimoto, S.; Hirai, M.; Ito, I.; Korei, Y. Effects of Soluble Fission-Products on Thermal-Conductivities of Nuclear-Fuel Pellets. *J. Nucl. Sci. Tech.* **1994**, 31, 796-802.

[558] Gibby, R. L., Effect of Plutonium Content on Thermal Conductivity of $(U, Pu)O_2$ Solid Solutions. *J. Nucl. Mater.* **1971**, 38, 163-177.

[559] Hirai, M.; Ishimoto, S. Thermal Diffusivities and Thermal-Conductivities of $UO_2$-$Gd_2O_3$. *J. Nucl. Sci. Tech.* **1991**, 28, 995-1000.

[560] Lucuta, P. G.; Matzke, H.; Verrall, R. A., Thermal Conductivity of Hyperstoichiometric SIMFUEL. *J. Nucl. Mater.* **1995**, 223, 51-60.

[561] Cozzo, C.; Staicu, D.; Somers, J.; Fernandez, A.; Konings, R. J. M., Thermal Diffusivity and Conductivity of Thorium-Plutonium Mixed Oxides. *J. Nucl. Mater.* **2011**, 416, 135-141.

[562] Duriez, C.; Alessandri, J. P.; Gervais, T.; Philipponneau, Y. Thermal Conductivity of Hypostoichiometric Low Pu Content $(U,Pu)O_{2-x}$ Mixed Oxide. *J. Nucl. Mater.* **2000**, 277, 143-158.

[563] Suzuki, K.; Kato, M.; Sunaoshi, T.; Uno, H.; Carvajal-Nunez, U.; Nelson, A. T.; McClellan, K. J. Thermal and Mechanical Properties of $CeO_2$. *J. Am. Ceram. Soc.* **2019,** 102, 1994-2008.

[564] Dennett, C. A.; Choens, R. C.; Taylor, C. A.; Heckman, N. M.; Ingraham, M. D.; Robinson, D.; Boyce, B. L.; Short, M. P.; Hattar, K. Listening to Radiation Damage In Situ: Passive and Active Acoustic Techniques. *JOM* **2020**, 72, 392-402.

[565] Hofmann, F.; Short, M. P.; Dennett, C. A. Transient Grating Spectroscopy: An Ultrarapid, Nondestructive Materials Evaluation Technique. *MRS Bull.* **2019**, 44, 392-402.

[566] Hofmann, F.; Mason, D. R.; Eliason, J. K.; Maznev, A. A.; Nelson, K. A.; Dudarev, S. L. Non-Contact Measurement of Thermal Diffusivity in Ion-Implanted Nuclear Materials. *Sci. Rep.* **2015**, 5, 16042.

[567] Reza, A.; Yu, H.; Mizohata, K.; Hofmann, F. Thermal Diffusivity Degradation and Point Defect Density in Self-Ion Implanted Tungsten. *Acta Mater.* **2020**, 193, 270-279.

[568] Capinskia, W. S.; Maris, H. J.; Bauser, E.; Silier, I.; Asen-Palmer, M.; Ruf, T.; Cardona, M.; Gmelin, E. Thermal Conductivity of Isotopically Enriched Si. *Appl. Phys. Lett.* **1997**, 71, 2109-2111.

[569] Braun, J. L.; Hopkins, P. E. Upper Limit to the Thermal Penetration Depth During Modulated Heating of Multilayer Thin Films with Pulsed and Continuous Wave Lasers: a Numerical Study. *J. Appl. Phys.* **2017**, 121, 175107.

[570] Weisensee, P. B.; Feser, J. P.; Cahill, D. G. Effect of Ion Irradiation on the Thermal Conductivity of $UO_2$ and $U_3O_8$ Epitaxial Layers, *J. Nucl. Mater.* **2014**, 443, 212-217.

[571] Cheaito, R. C.; Gorham, C. S.; Misra, A.; Hattar, K.; Hopkins, P. E. Thermal Conductivity Measurements via Time-Domain Thermoreflectance for the Characterization of Radiation-Induced Damage. *J. Mater. Res.* **2015**, 30, 1403-1412.

[572] Chauhan, V.; Riyad, M. F.; Du, X. P.; Wei, C. D.; Tyburska-Puschel, B.; Zhao, J. C.; Khafizov, M. Thermal Conductivity Degradation and Microstructural Damage Characterization in Low-Dose Ion Beam-Irradiated 3C-SiC. *Metall. Mater. Trans. E* **2017**, 4, 61-69.

[573] Scott, E. A.; Hattar, K.; Rost, C. M.; Gaskins, J. T.; Fazli, M.; Ganski, C.; Li, C.; Bai, T.; Wang, Y.; Esfarjani, K.; et al. Phonon Scattering Effects from Point and Extended Defects on Thermal Conductivity Studied via Ion Irradiation of Crystals with Self-Imputities. *Phys. Rev. Mater.* **2018**, 2, 095001.

[574] Abdullaev, A.; Chauhan, V. S.; Muminov, B.; O'Connell, J.; Skuratov, V. A.; Khafizov, M.; Utegulov, Z. N., Thermal Transport Across Nanoscale Damage Profile in Sapphire Irradiated by Swift Heavy Ions. *J. Appl. Phys.* **2020**, 127, 035108.

[575] Scott, E. A.; Hattar, K.; Braun, J. L.; Rost, C. M.; Gaskins, J. T.; Bai, T.; Wang, Y.; Ganski, C.; Goorsky, M.; Hopkins, P. E. Orders of Magnitude Reduction in the Thermal Conductivity of Polycrystalline Diamond Through Carbon, Nitrogen, and Oxygen Ion Implantation. *Carbon* **2020**, 157, 97-105.

[576] Hurley, D. H.; Telschow, K. L. Simultaneous Microscopic Imaging of Elastic and Thermal Anisotropy *Phys. Rev. B* **2005**, 71, 241410.

[577] Hurley, D. H.; Wright, O. B.; Matsuda, O.; Shinde, S. L. Time Resolved Imaging of Carrier and Thermal Transport in Silicon *J. Appl. Phys.* **2010**, 107, 023521.

[578] Khafizov, M.; Hurley, D. H. Measurement of Thermal Transport Using Time-Resolved Thermal Wave Microscopy. *J. Appl. Phys.* **2011**, 110, 083525.

[579] Feser, J. P.; Cahill, D. G. Probing Anisotropic Heat Transport Using Time-Domain Thermoreflectance With Offset Laser Spots. *Rev. Sci. Instrum.* **2012,** 83, 104901.

[580] Fournier, D.; Marangolo, M.; Fretigny, C. Measurement of Thermal Properties of Bulk Materials and Thin Films by Modulated Thermoreflectance (MTR). *J. Appl. Phys.* **2020**, 128, 241101.

[581] Tang, L.; Dames, C. Anisotropic Thermal Conductivity Tensor Measurements Using Beam-Offset Frequency Domain Thermoreflectance (BO-FDTR) for Materials Lacking In-Plane Symmetry. *Int. J. Heat Mass Transf.* **2021,** 164, 120600.

[582] Hurley, D. H.; Schley, R. S.; Khafizov, M.; Wendt, B. L. Local Measurements of Thermal Conductivity and Diffusivity. *Rev. Sci. Instrum.* **2015**, 86, 123901.

[583] Khafizov, M.; Yablinsky, C.; Allen, T. R.; Hurley, D. H.; Measurement of Thermal Conductivity in Proton Irradiation Silicon. *Nucl. Instrum. Meth. Phys. Res. B* **2014**, 325, 11-14.

[584] Chauhan, V. S.; Abdullaev, A.; Utegulov, Z. N.; O'Connell, J.; Skuratov, V.; Khafizov, M., Simultaneous Characterization of Cross- and In-Plane Thermal Transport in Insulator Patterned by Directionally Aligned Nano-Channels. *AIP Advances* **2020,** 10, 015304.

[585] Hua, Z.; Fleming, A.; Ban, H. The Study of Using Multi-Layered Model to Extract Thermal Property Profiles of Ion-Irradiated Materials. *Int. J. Heat Mass Transf.* **2019**, 131, 206-216.

[586] Riyad, M. F.; Chauhan, V.; Khafizov, M. Implementation of a Multilayer Model for Measurement of Thermal Conductivity in Ion Beam Irradiated Samples Using a Modulated Thermoreflectance Approach. *J. Nucl. Mater.* **2018**, 509, 134-144.

[587] Adkins, C. A.; Pavlov, T. R.; Middlemas, S. C.; Schley, R. S.; Marshall, M. C.; and Cole, M. R. *Demonstrate Thermal Property Measurement on Irradiated Fuel in IMCL;* INL/EXT-19-55902; Idaho National Laboratory, Idaho Falls, **2019**.

[588] Togo, A.; Chaput, L.; Tanaka, I. Distributions of Phonon Lifetimes in Brillouin Zones. *Phys. Rev. B* **2015,** 91, 094306.

[589] Ohashi, K. Scattering of Lattice Waves by Dislocations. *J. Phys. Soc. Japan* **1968,** 24, 437-445.

[590] Klemens, P. G. Some Scattering Problems in Conduction Theory. *Can. J. Phys.* **1957**, 35, 441-450.

[591] Turk, L. A.; Klemens, P. G. Phonon Scattering by Impurity Platelet Precipitates in Diamond. *Phys. Rev. B* **1974**, 9, 4422-4428.

[592] Snead, L. L.; Zinkle, S. J.; White, D. P. Thermal Conductivity Degradation of Ceramic Materials Due to Low Temperature, Low Dose Neutron Irradiation. *J. Nucl. Mater.* **2005,** 340, 187-202.

[593] Morelli, D. T.; Perry, T. A.; Farmer, J. W. Phonon-Scattering in Lightly Neutron-Irradiated Diamond. *Phys. Rev. B* **1993**, 47, 131-139.

[594] Kim, W.; Majumdar, A. Phonon Scattering Cross Section of Polydispersed Spherical Nanoparticles. *J. Appl. Phys.* **2006,** 99, 084306.

[595] Fukushima, S.; Ohmichi, T.; Maeda, A.; Handa, M. Thermal Conductivity of Stoichiometric (Pu,Nd)$O_2$ and (Pu,Y)$O_2$ Solids Solutions. *J. Nucl. Mater.* **1983,** 114, 260-266.

[596] Gurunathan, R.; Hanus, R.; Dylla, M.; Katre, A.; Snyder, G. J. Analytical Models of Phonon-Point-Defect Scattering. *Phys. Rev. Appl.* **2020**, 13, 034011.

[597] Tamura, S. Isotope Scattering of Dispersive Phonons in Ge. *Phys. Rev. B* **1983,** 27, 858-866.

[598] Callaway, J. Model for Lattice Thermal Conductivity at Low Temperatures. *Phys. Rev.* **1959**, 113, 1046-1051.

[599] Cahill, D. G.; Braun, P. V.; Chen, G.; Clarke, D. R.; Fan, S.; Goodson, K. E.; Keblinski, P.; King, W. P.; Mahan, G. D.; Majumdar, A.; et al. Nanoscale Thermal Transport. II. 2003-2012. *Appl. Phys. Rev.* **2014**, 1, 011305.

[600] Srivastava, G. P. *The physics of phonons;* Taylor and Francis: New York, **1990**; p 122.

[601] Reissland, J. A. *The physics of phonons;* Weiley: London, **1973**; p 83.

[602] Torres, E.; CheikNjifon, I.; Kaloni, T. P.; Pencer, J. A. Comparative Analysis of the Phonon Properties in $UO_2$ Using the Boltzmann Transport Equation Coupled with DFT Plus U and Empirical Potentials. *Comput. Mater. Sci.* **2020,** 177, 109594.

[603] Cooper, M. W. D.; Middleburgh, S. C.; Grimes, R. W. Modelling the Thermal Conductivity of $(U_xTh_{1-x})O_2$ and $(U_xPu_{1-x})O_2$. *J. Nucl. Mater.* **2015,** 466, 29-35.

[604] Muta, H.; Murakami, Y.; Uno, M.; Kurosaki, K.; Yamanaka, S. Thermophysical Properties of $Th_{1-x}U_xO_2$ Pellets Prepared by Spark Plasma Sintering Technique. *J. Nucl. Sci. Tech.* **2013**, 50, 181-187.

[605] Yue, S.-Y.; Zhang, X.; Zin, G.; Phillpot, S. R.; Hu, M. Metric for Strong Intrinsic Four-Order Phonon Anharmonicity. *Phys. Rev. B* **2017**, 95, 195203.

[606] Verlet, L. Computer 'Experiments' on Classical Fluids. I. Thermodynamical Properties of Lennard-Jones Molecules. *Phys. Rev.* **1967**, 159, 98–103.

[607] Schelling, P. K.; Phillpot, S. R.; Keblinski, P. Comparison of Atomic-Level Simulation Methods for Computing Thermal Conductivity. *Phys. Rev. B* **2002**, 65, 144306.

[608] Müller-plathe, F. A Simple Nonequilibrium Molecular Dynamics Method for Calculating the Thermal Conductivity. *J. Chem. Phys.* **1997**, 106, 6082.

[609] Poetzsch, R. H. H.; Bottger, H. Interplay of Disorder and Anharmonicity in Heat Conduction: Molecular-Dynamics Study. *Phys. Rev. B* **1994**, 50, 757–764.

[610] Oligschleger, C.; Schon, J. C. Simulation of Thermal Conductivity and Heat Transport in Solids. *Phys. Rev. B* **1999**, 59, 4125–4133.

[611] Sellan, D. P.; Landry, E. S.; Turney, J. E.; Mcgaughey, A. J. H.; Amon, C. H. Size Effects in Molecular Dynamics Thermal Conductivity Predictions. *Phys. Rev. B* **2010**, 81, 214305.

[612] Buckingham, R. A. The Classical Equation of State of Gaseous Helium, Neon and Argon. *Proc. Math. Phys. Eng. Sci.* **1938**, 168, 264-283.

[613] Morse, P. M. Diatomic Molecules According to the Wave Mechanics. II. Vibrational Levels. *Phys. Rev.* **1929**, 34, 57-64.

[614] Basak, C. B.; Sengupta, A. K.; Kamath, H. S. Classical Molecular Dyanamics Simulation of $UO_2$ to Predict Thermophysical Properties. *J. Alloys Compd.*, **2003**, 360, 210-216.

[615] Yakub, E.; Ronchi, C.; Staicu, D. Molecular Dynamics Simulation of Premelting and Melting Phase Transitions in Stoichiometric Uranium Dioxide. *J. Chem. Phys.* **2007**, 127, 094508.

[616] Morelon, N. D.; Ghaleb, D.; Delhaye, J.-M.; Van Brutzel, L. A New Empirical Potential for Simulating the Formation of Defects and Their Mobility in Uranium Dioxde. *Philos. Mag.* **2003**, 83,1533-1550.

[617] Jackson, R. A.; Catlow, C. R. A. Trapping and Solution of Fission Xe in UO$_2$. *J. Nucl. Mater*. **1984**, 127, 161-166.

[618] Govers, K.; Lemehov, S.; Hou, M.; Verwerft, M. Comparison of Interatomic Potentials for UO$_2$, *J. Nucl. Mater*. **2007**, 366, 161-177.

[619] Soulie, A.; Crocombette, J.-P.; Kraych, A.; Garrido, F.l Sattonnay, G.; Clouet, E. Atomistically-Informed Thermal Glide Model for Edge Dislocations in Uranium Dioxide. *Acta Mater*. **2018**, 150, 248-261.

[620] Sattonnay, G.; Tetot, R. Bulk, Surface and Point Defect Properties of UO$_2$ From a Tight-Binding Variable-Charge Model. *J. Phys. Condens. Matter*, **2013**, 25, 125403

[621] Li, Y.; Liang, T.; Sinnott, S. B.; Phillpot, S. R. A Charge-Optimized Many-Body Potential for the U-UO$_2$-O$_2$ System. *J. Phys. Condens. Matter* **2013**, 25, 505401.

[622] Govers, K.; Lemehov, S.; Hou, M.; Verwerft, M. Comparison of Interatomic Potentials for UO$_2$: Part II: Molecular Dynamics Simulations. *J. Nucl. Mater*. **2008**, 376, 66-77.

[623] Dacus, B.; Beeler, B.; Schwen, D. Calculation of Threshold Displacement Energies in UO$_2$ *J. Nucl. Mater.* **2019**, 520, 152-164.

[624] Murphy, S. T.; Rushton, M. J. D.; Grimes, R. W. A Comparison of Empirical Potential Models for the Simulation of Dislocations in Uranium Dioxide. *Prog. Nucl. Energy*, **2014**, 72, 27-32.

[625] Zhang, Y.; Millett, P. C.; Tonks, M. R.; Bai, X-M.; Biner, S. B. Molecular Dynamics Simulations of Intergranular Fracture in UO$_2$ with Nine Empirical Interatomic Potentials *J. Nucl. Mater*. **2014**, 452, 296-303.

[626] Arima, T.; Yamasaki, S.; Inagaki, Y.; Idemitsu, K. Evaluation of Thermal Properties of UO$_2$ and PuO$_2$ by Equilibrium Molecular Dynamics Simulations from 300 to 2000K. *J. Alloys Compd.* **2005**, 400, 43–50.

[627] Maxwell, C. I.; Pencer, J. Molecular Dynamics Modelling of the Thermal Conductivity of Off-Stoichiometric UO$_{2\pm x}$ and (U$_y$Pu$_{1-y}$)O$_{2\pm x}$ Using Equilibrium Molecular Dynamics. *Ann. Nucl. Energy* **2019**, 131, 317–324.

[628] Cooper, M. W. D.; Murphy, S. T.; Rushton, M. J. D.; Grimes, R. W. Thermophysical Properties and Oxygen Transport in the (U$_x$,Pu$_{1-x}$)O$_2$ Lattice. *J. Nucl. Mater*. **2015**, 461, 206-214.

[629] Lunev, A. V.; Tarasov, B. A. A Classical Molecular Dynamics Study of the Correlation Between the Bredig Transition and Thermal Conductivity of Stoichiometric Uranium Dioxide. *J. Nucl. Mater*. **2011**, 415, 217–221.

[630] Park, J.; Farfan, E. B.; Enriquez, C. Thermal Transport in Thorium Dioxide. *Nucl. Eng. Technol*. **2018**, 50, 731–737.

[631] Behera, R. K.; Deo, C. S. Atomistic Models to Investigate Thorium Dioxide (ThO$_2$). *J. Phys. Condens. Matter* **2012**, 24, 215405.

[632] Chernatynskiy, A.; Flint, C.; Sinnott, S. B.; Phillpot, S. R. Critical Assessment of UO$_2$ Classical Potentials for Thermal Conductivity Calculations. *J. Mater. Sci.* **2012**, 47, 7693–7702.

[633] Hyland, G. J. Thermal Conductivity of Solid UO$_2$: Critique and Reccommendation. *J. Nucl. Mater*. **1983**, 113, 125–132.

[634] Arima, T.; Yamasaki, S.; Idemitsu, K.; Inagaki, Y. Equilibrium and Nonequilibrium Molecular Dynamics Simulations of Heat Conduction in Uranium Oxide and Mixed Uranium – Plutonium Oxide. *J. Nucl. Mater*. **2008**, 376, 139–145.

[635] M. J. Qin, M. J.; Middleburgh, S. C.; Cooper, M. W. D.; Rushton, M. J. D.; Puide, M.; Kuo, E. Y.; Grimes, R. W.; Lumpkin, G. R. Thermal Conductivity Variation in Uranium Dioxide with Gadolinia Additions. *J. Nucl. Mater*. **2020**, 540, 152258.

[636] Watanabe, T.; Srivilliputhur, S. G.; Schelling, P. K.; Tulenko, J. S.; Sinnott, S. B.; Phillpot, S. R. Thermal Transport in Off-Stoichiometric Uranium Dioxide by Atomic Level Simulation. *J. Am. Ceram. Soc*. **2009**, 92, 850–856.

[637] Deng, B.; Cherntynskiy, A.; Shukla, P.; Sinnott, S. B.; Phillpot, S. R.; Effects of Edge Dislocations on Thermal Transport in UO$_2$. *J. Nucl. Mat*., **2013**, 434, 203-209.

[638] Phillpot, S. R.; El-Azab, A.; Chernatynskiy, A.; Tulenko, J. S. Thermal Conductivity of UO$_2$ Fuel: Predicting Fuel Performance from Simulation. *JOM.* **2011**, 63, 73–79.

[639] Watanabe, T.; Sinnott, S. B.; Tulenko, J. S.; Grimes, R. W.; Schelling, P. K.; Phillpot, S. R., Thermal Transport Properties of Uranium Dioxide by Molecular Dynamics Simulations. *J. Nucl. Mater.* **2008**, 375, 388-396.

[640] Chen, T.; Chen, D.; Sencer, B. H.; Shao, L. Molecular Dynamic Simulation of Grain Boundary Thermal Resistance in UO$_2$. *J. Nucl. Mater.* **2014**, 452, 364-369.

[641] Deng, B.; Chernatynskiy, A.; Sinnott, S. B.; Phillpot, S. R. Thermal Transport at (001) Twist Grain Boundary in UO$_2$. *J. Nucl. Mater*. **2016**, 479, 167-173.

[642] Chen, W.; Cooper, M. W. D.; Xiao, Z.; Andersson, D. A.; Bai, X. Effect of Xe Bubble Size and Pressure on the Thermal Conductivity of UO$_2$ - A Molecular Dynamics Study. *J. Mater. Res*. **2019**, 34, 2295–2305.

[643] Lee, C.; Chernatynskiy, A.; Shukla, P.; Stoller, R. E.; Sinnott, S. B.; Phillpot, S. R. Effect of Pores and He Bubbles on the Thermal Transport Properties of UO$_2$ by Molecular Dynamics Simulation. *J. Nucl. Mater*. **2015**, 456, 253–259.

[644] Swartz, E. T.; Pohl, R. O. Thermal Boundary Resistance. *Rev. Mod. Phys.* **1989**, 61, 605-668.

[645] Tonks, M. R.; Millett, P. C.; Nerikar, P.; Du, S.; Andersson, D.; Stanek, C. R.; Gaston D.; Andrs, D.; Williamson, R. Multiscale Development of a Fission Gas Thermal Conductivity Model: Coupling Atomic, Meso, and Continuum Level Simulations. *J. Nuc. Mater.* **2013**, 440, 193-200.

[646] Zhu, X.; Gong, H.; Zhao, Y-F. Lin, D-Y. Han, G. Liu, T.; Song, H. Effect of Xe Bubbles on the Thermal Conductivity of UO$_2$: Mechanisms and Model Establishment. *J. Nuc. Mat*. **2020**, 533, 152080.

[647] Hua, Z.; Spackman, J.; Ban, H. Characterization of Kapitza Resistances of Natural Grain Boundaries in Cerium Oxide. *Materialia* **2019**, 5, 100230.

[648] Hurley, D. H.; Khafizov, M.; Shinde, S. L. Measurement of the Kapitza Resistance Across a Bicrystal Interface. *J. Appl. Phys.* **2011**, 109, 083504

[649] Deng, B.; Chernatynskiy, A.; Khafizov, M.; Hurley, D. H.; Phillpot, S. R. Kaptiza Resistance of Si/SiO$_2$ Interface. *J. Appl. Phys*. **2014**, 115, 084910.

[650] Shrestha, K.; Yao, T.; Lian, J.; Antonio, D.; Sessim, M.; Tonks, M. R.; Gofryk, K. The Grain-Size Effect on Thermal Conductivity of Uranium Dioxide. *J. Appl. Phys*. **2019**, 126, 125116.

[651] Katre, A.; Carrete, J.; Dongre, B.; Madsen, G. K. H.; Mingo, N. Exceptionally Strong Phonon Scattering by B Substitution in Cubic SiC. *Phys. Rev. Lett. 2017*, 119, 075902.

[652] Polanco, C. A.; Lindsay, L. Thermal Conductivity of InN with Point Defects from First Principles. *Phys. Rev. B* **2018**, 98, 014306.

[653] Kundu, A.; Mingo, N.; Broido, D. A.; Stewart, D. A. Role of Light and Heavy Embedded Nanoparticles on the Thermal Conductivity of SiGe alloys. *Phys. Rev. B* **2011**, 84, 125426.

[654] Wang, T.; Carrete, J.; van Roekeghem, A.; Mingo, N.; Madsen, G. K. H. Ab Initio Phonon Scattering by Dislocations. *Phys. Rev. B* **2017**, *95,* 245304.